\documentclass[%
reprint,
superscriptaddress,
amsmath,amssymb,
longbibliography,
aps,
]{revtex4-2}

\usepackage{amssymb,amsmath}
\DeclareMathAlphabet\mathbfcal{OMS}{cmsy}{b}{n}
\usepackage{cmap}
\usepackage{graphicx} 
\usepackage{bm}% bold math
\usepackage{color}
\usepackage{blindtext}
\usepackage{hyperref}
\usepackage{listings}
\usepackage{booktabs}
\usepackage{multirow}
\usepackage{amsmath}
\usepackage{mathrsfs}
\usepackage{braket}
\usepackage{mathtools}

\usepackage{dsfont}
\usepackage{tabularx}

\renewcommand{\Re}{\operatorname{Re}}
\renewcommand{\Im}{\operatorname{Im}}

\begin{document}

%\articletype{ARTICLE TEMPLATE}% Specify the article type or omit as appropriate

\title{\large{Momentum, spin, and orbital angular momentum \\ of electromagnetic, acoustic, and water waves}}

%\author{
%\name{Konstantin Y. Bliokh\textsuperscript{a,b,c}\thanks{CONTACT Konstantin Y. Bliokh. Email: konstantin.bliokh@dipc.org}}
%\affil{\textsuperscript{a}Donostia International Physics Center (DIPC), Donostia-San Sebasti\'{a}n 20018, Spain; \textsuperscript{b}IKERBASQUE, Basque Foundation for Science, Bilbao 48009, Spain; \textsuperscript{c}Centre of Excellence ENSEMBLE3 Sp. z o.o., 01-919 Warsaw, Poland}
%}

\author{{Konstantin Y. Bliokh}}
\affiliation{Donostia International Physics Center (DIPC), Donostia-San Sebasti\'{a}n 20018, Spain}
\affiliation{IKERBASQUE, Basque Foundation for Science, Bilbao 48009, Spain}
\affiliation{Centre of Excellence ENSEMBLE3 Sp. z o.o., 01-919 Warsaw, Poland}

\begin{abstract}
Waves of various types carry {\it momentum}, which is associated with their propagation direction, i.e., the phase gradient. The circulation of the wave momentum density gives rise to {\it orbital angular momentum} (AM). Additionally, for waves described by vector fields, local rotation of the wavefield produces {\it spin AM} (or simply, spin). These dynamical wave properties become particularly significant in structured (i.e., inhomogeneous) wavefields. Here we provide an introduction and overview of the momentum and AM properties for a variety of classical waves: electromagnetic, sound, elastic, plasma waves, and water surface waves. A unified field-theory approach, based on Noether's theorem, offers a general framework to describe these diverse physical systems, encompassing longitudinal, transverse, and mixed waves with different dispersion characteristics. We also discuss observable manifestations of the wave momentum and AM providing clear physical interpretations of the derived quantities.
\end{abstract}

\maketitle

%\begin{abbreviations}
%AM - angular momentum;
%OAM - orbital angular momentum;
%SAM - spin angular momentum
%\end{abbreviations}

%\begin{keywords}
%Wave momentum; angular momentum; spin; electromagnetic waves; acoustic waves; water waves
%\end{keywords}

%\twocolumn

%%%%%%%%%%%%%%%%%%%%%%%%%%%%%%
\section{Introduction}
%%%%%%%%%%%%%%%%%%%%%%%%%%%%%%

Electromagnetic, acoustic, and water waves have accompanied humanity since the dawn of civilization. Light and sound waves are fundamental to two of our five basic senses, although recognizing them as waves requires scientific insight. In turn, water surface waves are immediately perceivable as visualized wave phenomena. 

\begin{figure}
\centering
\includegraphics[width=0.7\linewidth]{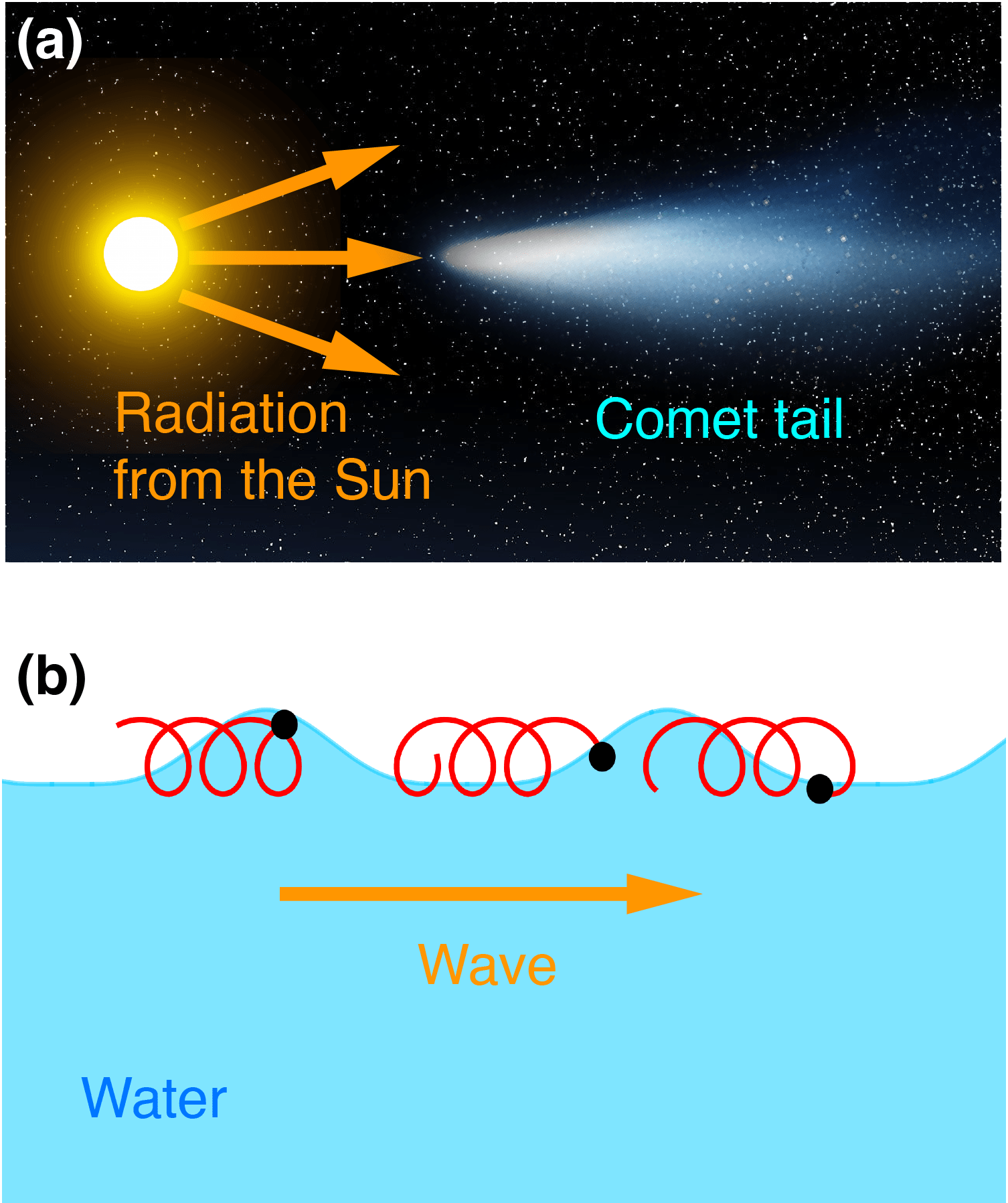}
\caption{(a) As Keppler suggested in the 17th century, a comet's tail points away from the sun due to radiation pressure \cite{Jones1953}. This evidences the electromagnetic wave momentum. (b) The propagation of a water-surface wave induces a drift of water particles in the wave propagation direction, known as the Stokes drift \cite{Falkovich_book}. This is a manifestation of the wave momentum. Here trajectories of drifting water-surface particles are shown in red. In addition, the circular-like local motion of the particles reveals the presence of transverse spin angular momentum in water waves.} 
\label{Fig_Intro}
\end{figure}

The idea that waves carry {\it momentum} dates back to the 17th century, when Kepler attributed the origin of a comet's tail to a `light-wind' emanating from the sun, Fig.~\ref{Fig_Intro}(a). 
In the 18th century, Euler discussed similar effects for light and sound waves, and at the end of the 19th / beginning of the 20th century, stimulated by the appearance of Maxwell's electromagnetic field theory, the electromagnetic and sound wave momenta were extensively studied both theoretically and experimentally \cite{Jones1953, Loudon2012, Beyer1978, Sarvazyan2010UMB}. Furthermore, in the mid-19th century, Stokes described a drift of water-surface particles induced by the propagation of water waves, Fig.~\ref{Fig_Intro}(b). This phenomenon is now known as the {\it Stokes drift} \cite{Falkovich_book, Bremer2018}; it occurs in all media with freely moving particles and corresponds to the canonical momentum carried by waves in such media \cite{Mcintyre1981, Bliokh2022SA, Bliokh2022PRA}.

%FFFFFFFFFFFFFFFFFFFFFFFFFFFFFFFFFFFFFFFFFFFFFF
%FFFFFFFFFFFFFFFFFFFFFFFFFFFFFFFFFFFFFFFFFFFFFF

\begin{figure}
\centering
\includegraphics[width=0.95\linewidth]{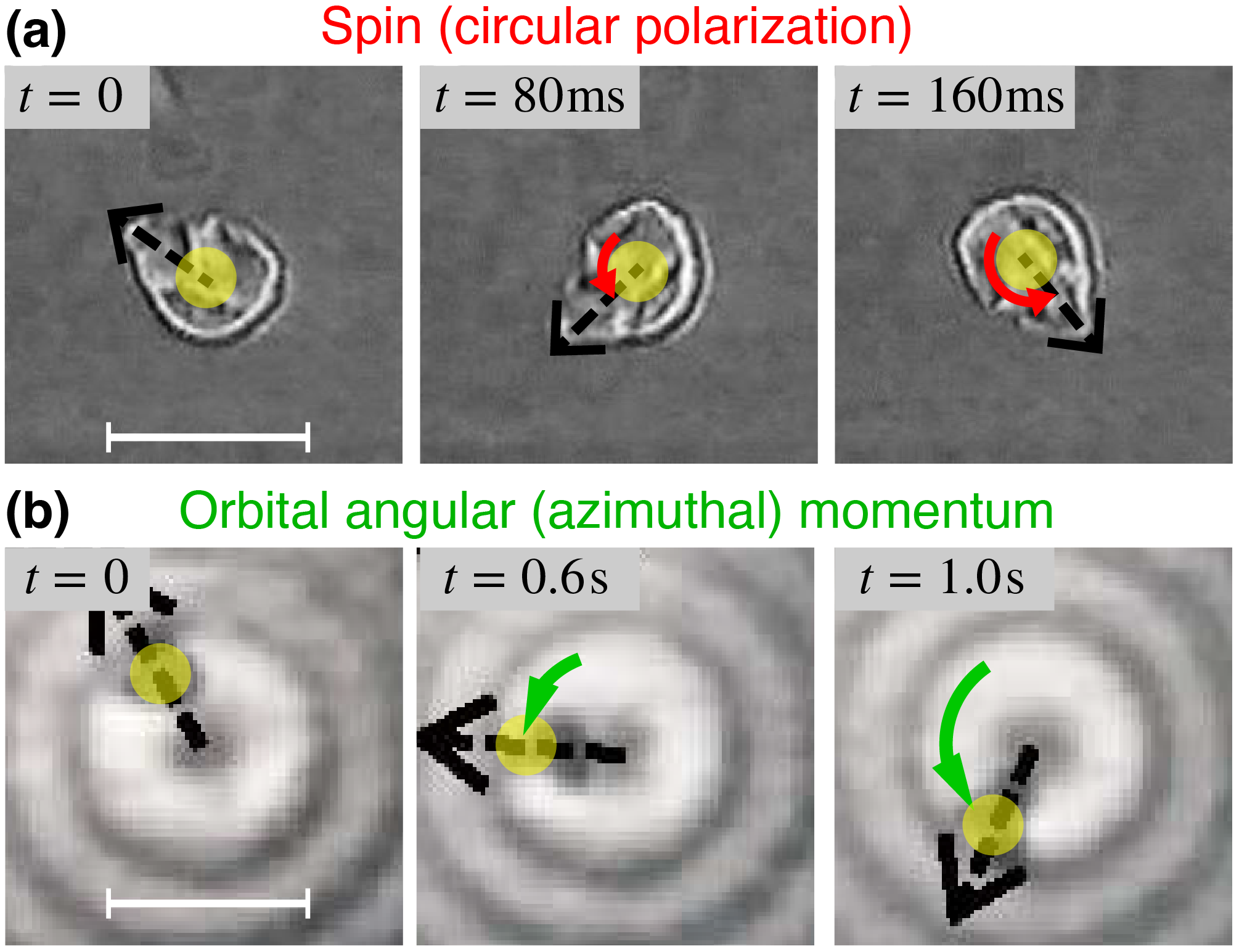}
\caption{Mechanical action of optical spin and OAM on matter. (a) The spin of circularly-polarized light induces rotation of an absorbing or birefringent particle around its center (highlighted in yellow). Adapted from \cite{Friese1998Nat} with permission from Springer Nature. (b) The OAM of optical vortices, characterized by radially-varying intensity (shown in grayscale) and azimuthally-growing phase, generates orbital rotation of an absorbing particle (radially trapped due to the intensity-gradient force) around the vortex center. Adapted from \cite{Garces-Chavez2003PRL} [Copyright (2003) by the American Physical Society]. The OAM and orbital rotation are produced by the azimuthal momentum density and the corresponding azimuthal radiation-pressure force. The scalebars correspond to $10 \lambda \simeq 10\,{\rm \mu m}$, where $\lambda$ is the wavelength of light.} 
\label{Fig_torques}
\end{figure}

It was not until the 20th century that scientists recognized that waves can also carry {\it angular momentum} (AM) \cite{Allen_book, Andrews_book, Bliokh2015PR}. First, in 1899 and 1909, Sadowsky and Poynting theoretically found that the rotation of the electric field in circularly-polarized light generates {\it spin AM} (or simply, spin) \cite{Poynting1909}. The optical spin AM was subsequently measured via the optical torque on matter \cite{Beth1935, Friese1998Nat}, Fig.~\ref{Fig_torques}(a). Second, in 1992, Allen {\it et al}. found that {\it wave vortices}, i.e., azimuthally propagating waves, can carry {\it orbital AM} (OAM) \cite{Allen1992PRA}, also subsequently measured via optical torques \cite{Garces-Chavez2003PRL}, Fig.~\ref{Fig_torques}(b). 

%FFFFFFFFFFFFFFFFFFFFFFFFFFFFFFFFFFFFFFFFFFFFFF
%FFFFFFFFFFFFFFFFFFFFFFFFFFFFFFFFFFFFFFFFFFFFFF

\begin{figure}
\centering
\includegraphics[width=0.75\linewidth]{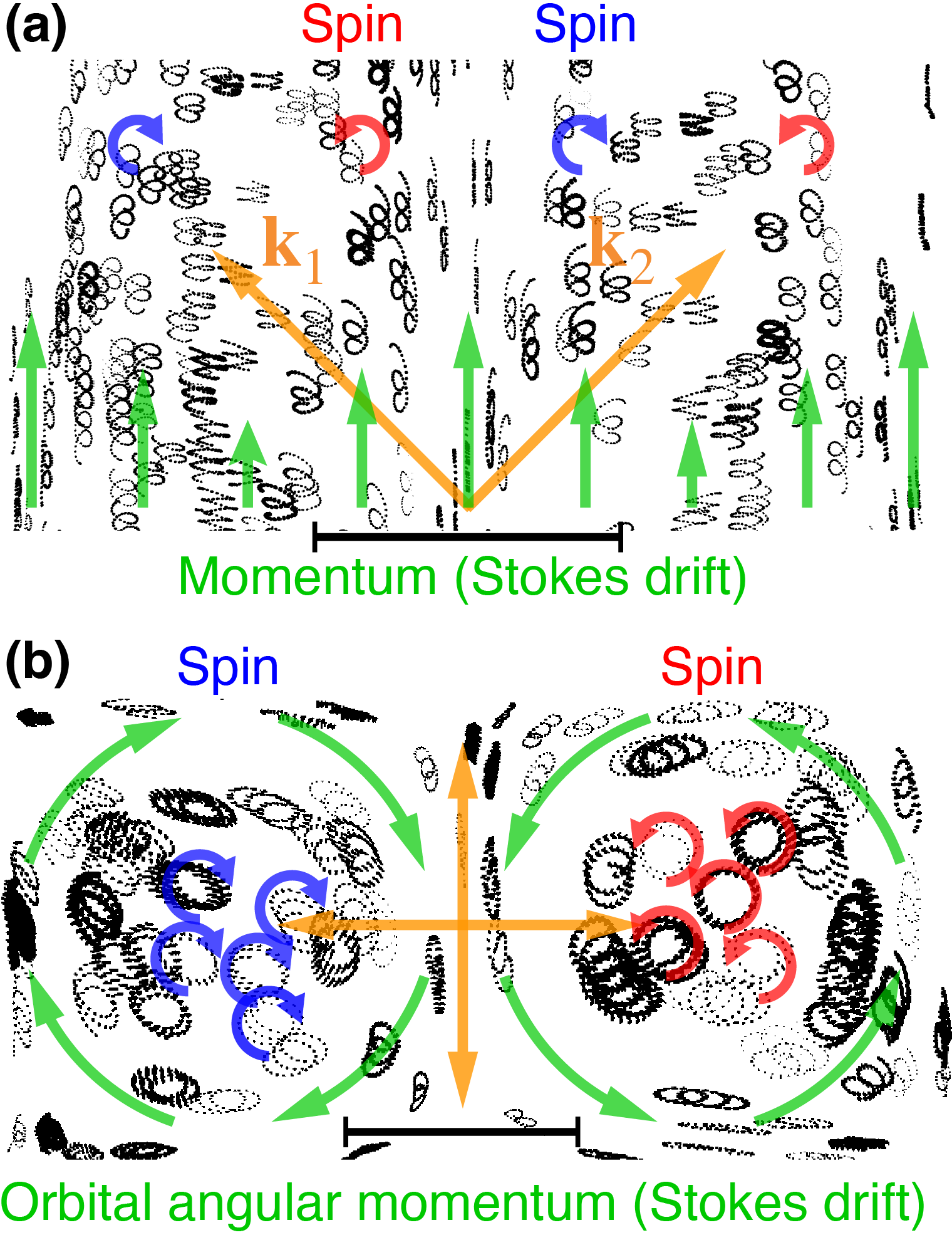}
\caption{Spin, momentum, and OAM in structured water waves visualized through the local circular motion and Stokes drift of water-surface particles (top view) \cite{Bliokh2022SA}. The experimentally observed particle trajectories are shown in black, their circular motions (spin) are highlighted by red and blue arrows, whereas the Stokes drift is highlighted by green arrows. (a) Interference of two plane water waves with equal amplitudes, frequencies, but different wavevectors ${\bf k}_{1,2}$. The scalebar corresponds to $\lambda/2 \simeq 2.2\,{\rm cm}$, where $\lambda$ is the water-wave wavelength. (b) Interference of two orthogonal standing waves with equal amplitudes and frequencies, phase-shifted by $\pi/2$ with respect to each other, generates alternating wave vortices of opposite signs. The circulating azimuthal Stokes drift produces the OAM in these vortices. The scalebar corresponds to $\lambda/4 \simeq 1.4\,{\rm cm}$. Adapted with permission from \cite{Bliokh2022SA}. [Copyright (2022) The Authors] CC BY-NC 4.0.} 
\label{Fig_intro_water}
\end{figure}

Initially, wave momentum and AM were studied for electromagnetic waves, but they were later recognized as universal properties of various kinds of waves, including acoustic and water waves \cite{Jones1973, Mcintyre1981, Peierls1991, Peskin2010, Hefner1999JASA, Nakane2018PRB, Shi2019, Bliokh2019b, Ren2022CPL, Bliokh2022SA, Chaplain2022CP, Bliokh2022PRL, Bliokh2022PRE, Smirnova2024PRL}.
In mechanical (sound, elastic, water) waves, spin can be associated with the local circular motion of medium particles \cite{Jones1973, Nakane2018PRB, Shi2019, Bliokh2019b, Ren2022CPL, Bliokh2022SA, Bliokh2022PRL, Bliokh2022PRE, Smirnova2024PRL}, Figs.~\ref{Fig_Intro}(b) and \ref{Fig_intro_water}. In turn, the OAM of any waves is associated with the azimuthal momentum density, i.e., azimuthal phase gradient, Figs.~\ref{Fig_torques}(b) and \ref{Fig_intro_water}(b). 

Modern research focuses on {\it structured} (i.e., inhomogeneous, non-plane-wave) wavefields \cite{Rubinsztein-Dunlop2016JO, Bliokh2023JO}. Such fields consist of multiple interfering plane waves and exhibit {\it local} space-varying momentum and AM densities. For instance, Fig.~\ref{Fig_intro_water} shows the motion of water-surface particles in the interference of two non-collinear plane water waves and in wave vortices formed by the interference of four plane waves \cite{Bliokh2022SA}. The local circular motion of the particles and their Stokes drift reveal the distributions of the spin and momentum densities.

Since momentum and AM are originally {\it mechanical} concepts, their most direct manifestations are wave-induced {\it forces} and {\it torques} on matter \cite{Toftul2024}. These effects underlie optical and acoustic trapping and manipulations of small particles, from individual atoms to biological organisms, Fig.~\ref{Fig_torques}, as well as numerous optomechanical and acoustomechanical systems \cite{Dholakia2020NRP, Toftul2024}. 

Furthermore, the concepts of wave momentum and AM are fundamental to quantum mechanics, which describes matter as a wave with mechanical properties. In particular, the de Broglie relation linking momentum to wavevector, ${\bf p} = \hbar {\bf k}$, is universal for any quantum particles or quasi-particles. Furthermore, the spin and OAM of electromagnetic and acoustic waves correspond directly to the spin and OAM of photons and phonons \cite{Akhiezer_book, Allen_book, Andrews_book, Nakane2018PRB, Ren2022CPL}.

%FFFFFFFFFFFFFFFFFFFFFFFFFFFFFFFFFFFFFFFFFFFFFF
%FFFFFFFFFFFFFFFFFFFFFFFFFFFFFFFFFFFFFFFFFFFFFF

Thus, understanding the momentum and AM properties of various waves is crucial to modern physics. In this review, we provide the basic theory of the momentum, spin, and OAM in the main types of classical waves: sound, electromagnetic, elastic, plasma, and water-surface waves. For simplicity, we focus on linear monochromatic (single-frequency) waves in homogeneous isotropic media.
Our approach is based on a universal field-theory formalism incorporating Noether's theorem, canonical conserved quantities, and the Belinfante-Rosenfeld relation between the canonical momentum, spin, and energy-flux densities. 
Despite significant differences in the systems under consideration, we demonstrate that this formalism applies equally well to all the wave types.
Moreover, we show that the quantities derived within the  field-theory formalism allow for clear physical interpretations in terms of observable phenomena.

%%%%%%%%%%%%%%%%%%%%%%%%%%%%%%
\section{General approach}
\label{General}
%%%%%%%%%%%%%%%%%%%%%%%%%%%%%%
%%%%%%%%%%%%%%%%%%%%%%%%%%%%%%
\subsection{Types of waves}
\label{Waves}
%%%%%%%%%%%%%%%%%%%%%%%%%%%%%%

Throughout this review, we consider linear waves described by real-valued vector wavefields ${\mathbfcal{F}}({\bf r},t)$ that vary in time and space. The simplest plane wave can be represented by the following complex form:
\begin{equation}
{\mathbfcal{F}}({\bf r},t) = \Re\!\left({\bf F}_0 e^{i{\bf k}\cdot {\bf r} - i\omega t}\right).
\label{planewave}
\end{equation}
Here ${\bf F}_0$ is the complex wave amplitude that includes the {\it polarization} (i.e., the direction of the field), ${\bf k}$ is the wave vector, and $\omega$ is the wave frequency. 
The wave \eqref{planewave} represents a sinusoidal perturbation propagating with velocity $(\omega/k)\, {\bm \kappa}$, where ${\bm \kappa} = {\bf k}/k$. 
For electromagnetic waves, ${\mathbfcal{F}}$ corresponds to the electric field $\mathbfcal{E}$ or magnetic vector-potential $\mathbfcal{A}$, while for all other (mechanical) waves considered in this review, it can be associated with the local wave-induced velocity $\mathbfcal{V}$ or displacement $\mathbfcal{R}$ of the medium particles.

\begin{figure}
\centering
\includegraphics[width=0.7\linewidth]{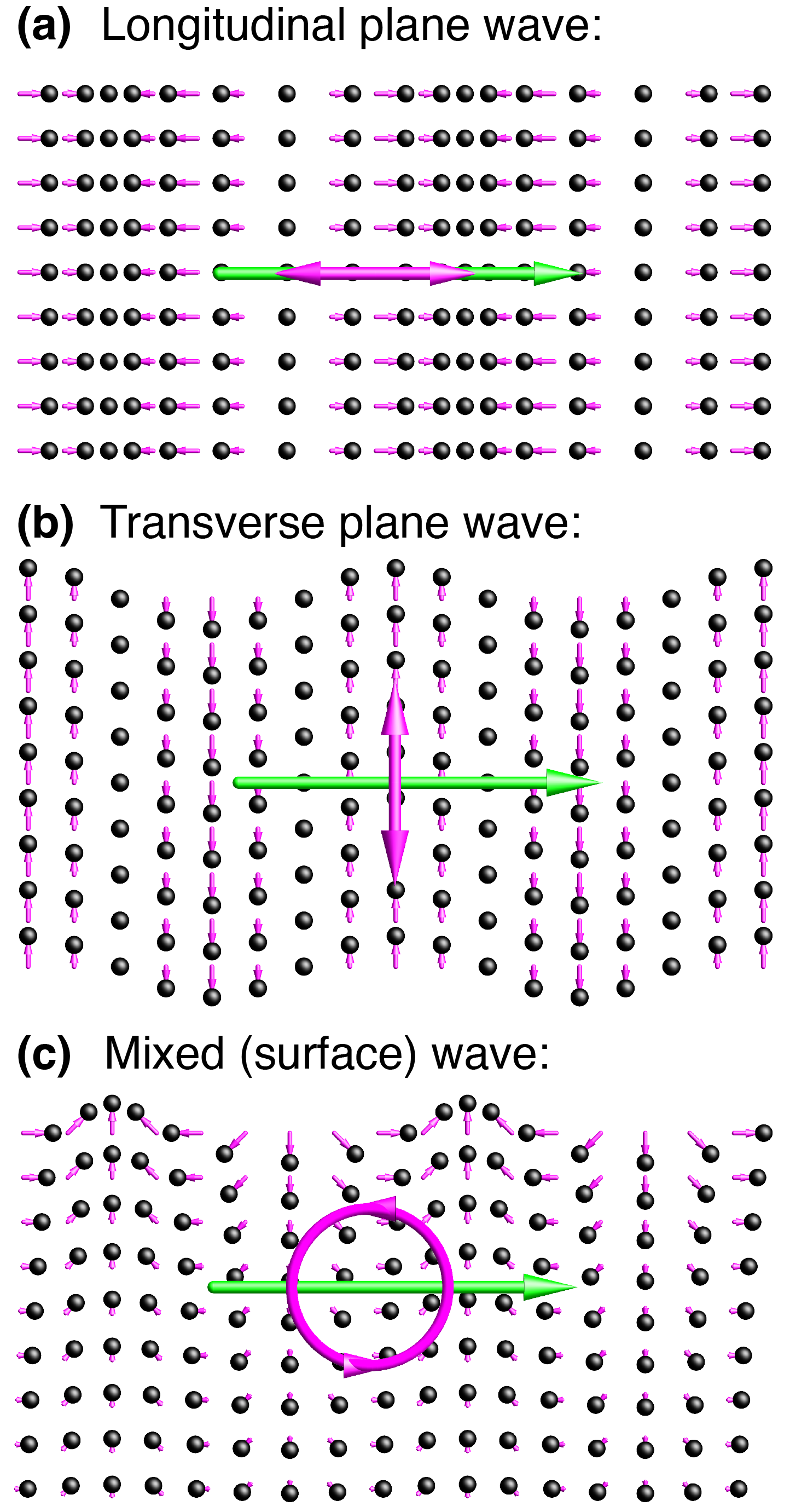}
\caption{Types of waves illustrated by the wave-induced instantaneous displacements $\mathbfcal{R}({\bf r},0)$ (magenta arrows) of particles (balck dots) in a medium. (a) Longitudinal plane wave: the displacements are aligned with the wavevector ${\bf k}$ (green arrow). (b) Transverse wave: the displacements are perpendicular to the wavevector. (c) Mixed wave: the displacements have both longitudinal and transverse components. Here a surface wave is shown, with the wave amplitude decaying in the vertical direction. In this case, the wave is mixed with respect to the {\it real part} of the wavevector, responsible for the propagation of the wave (see Section~\ref{Water}).} 
\label{Fig_wave_types}
\end{figure}

There are three primary types of waves, as shown in  Fig.~\ref{Fig_wave_types}:
\begin{itemize}
\item Longitudinal: ${\bf k} \times {\bf F}_0 = {\bf 0}$;
\item Transverse: ${\bf k} \cdot {\bf F}_0 =0$;
\item Mixed: ${\bf F}_0 = {\bf F}_{0\parallel} + {\bf F}_{0\perp}$, \\
where ${\bf k} \times {\bf F}_{0\parallel} = {\bf 0}$, ${\bf k} \cdot {\bf F}_{0\perp} =0$.
\end{itemize}
For example, sound waves in gases or fluids are longitudinal (the medium particles oscillate along the wave propagation direction), electromagnetic waves are transverse (the electric and magnetic fields are perpendicular to the wave propagation), and water surface waves are mixed (the water-surface particles oscillate in both longitudinal and transverse directions tracing elliptical trajectories). 

Here we should remark that water surface waves actually satisfy both the longitudinal and transverse wave conditions with the complex wavevector ${\bf k}$ describing surface propagation ($\Re{\bf k}$) and orthogonal exponential decay ($\Im{\bf k}$) of the wave (see Section~\ref{Water}). Nonetheless, we classify them as `mixed' with respect to the propagation direction, i.e., $\Re{\bf k}$.

%FFFFFFFFFFFFFFFFFFFFFFFFFFFFFFFFFFFFFFFFFFFFFF
%FFFFFFFFFFFFFFFFFFFFFFFFFFFFFFFFFFFFFFFFFFFFFF

In many problems, a single plane wave is too simplistic, and one needs to consider more complex wavefields formed by superpositions of multiple plane waves \eqref{planewave} with different amplitudes and wavevectors. In the monochromatic case (the frequency $\omega$ is fixed), an arbitrarily structured wavefield can be expressed as 
\begin{equation}
{\mathbfcal{F}}({\bf r},t) = \Re\!\left[ {\bf F}({\bf r}) e^{- i\omega t} \right].
\label{structuredwave}
\end{equation}
Here the inhomogeneous complex field ${\bf F}({\bf r})$ must satisfy the following differential constraints depending on the wave type:
\begin{itemize}
\item Longitudinal: ${\bm \nabla}\times {\bf F} = {\bf 0}$;
\item Transverse: ${\bm \nabla}\cdot {\bf F} = 0$;
\item Mixed: ${\bf F} = {\bf F}_\parallel + {\bf F}_\perp$, \\ 
where ${\bm \nabla}\times {\bf F}_\parallel = {\bf 0}$, ${\bm \nabla}\cdot {\bf F}_\perp = 0$.
\end{itemize}

\begin{figure}[t!]
\centering
\includegraphics[width=0.55\linewidth]{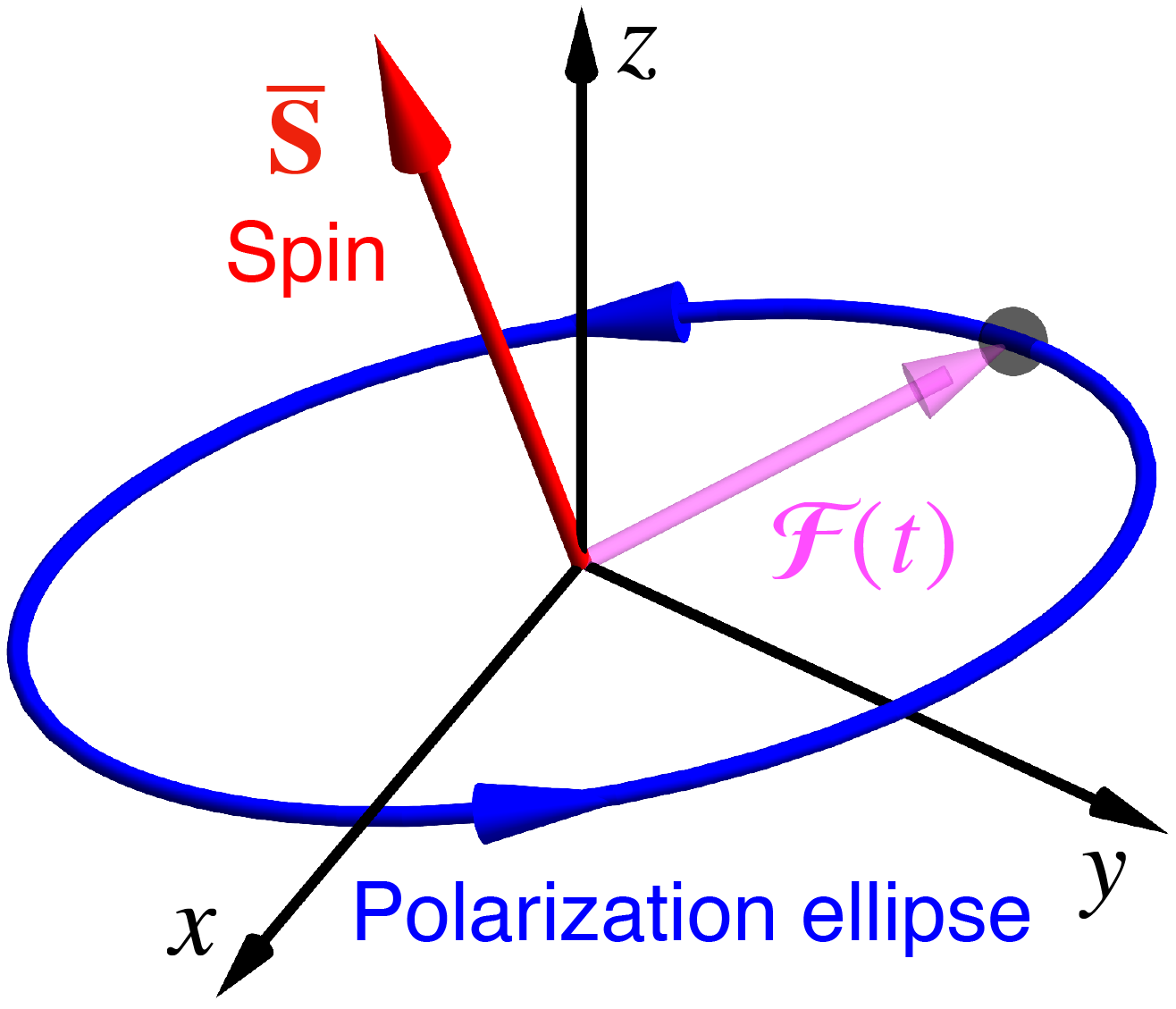}
\caption{Polarization ellipse and spin. A generic monochromatic vector field $\mathbfcal{F}({\bf r},t) = \Re[{\bf F}({\bf r})e^{-i\omega t}]$ traces the polarization ellipse at each point of space. 
The time-averaged spin angular momentum density $\overline{\bf S} \propto \overline{\mathbfcal{F} \times \dot{\mathbfcal{F}}} = \Im({\bf F}^* \!\times {\bf F})/2$ is directed normally to this ellipse and is proportional to its area.} 
\label{Fig_ellipse}
\end{figure}

In this review, we deal with structured waves \eqref{structuredwave} of various kinds. The complex field ${\bf F}({\bf r})$ can be characterized by the amplitudes $|F_i|$ and phases ${\rm Arg}(F_i)$ of its components $i=x,y,z$, or, alternatively, presented as a product of the amplitude, global phase factor, and unit polarization vector: ${\bf F} = |{\bf F}|  e^{i\alpha}\, {\bf f}$, where ${\bf f}^*\!\cdot {\bf f}=1$. Here the phase $\alpha$ 
is chosen such that $\Re{\bf f} \cdot \Im{\bf f} =0$ and $|\Re{\bf f}| \geq |\Im{\bf f}|$ \cite{BornWolf}. 
The polarization vector ${\bf f}$ determines the normalized {\it polarization ellipse} with the major and minor semiaxes $\Re{\bf f}$ and $\Im{\bf f}$, which is traced by the normalized real field, 
$\mathbfcal{F}/|{\bf F}|=\left[ \Re{\bf f}\,\cos(\omega t -\alpha) + \Im{\bf f}\,\sin(\omega t - \alpha) \right]$, over the wave period $2\pi/\omega$ at a given point of space, ${\bf r}$, Fig.~\ref{Fig_ellipse}. The polarization is circular when $|\Re{\bf f}| = |\Im{\bf f}| = 1/\sqrt{2}$ and linear when $|\Re{\bf f}| = 1$ and $|\Im{\bf f}| = 0$.
Importantly, although the concept of polarization was originally introduced for electromagnetic waves, it applies equally to any wave described by a vector field.

\begin{figure}[t!]
\centering
\includegraphics[width=0.85\linewidth]{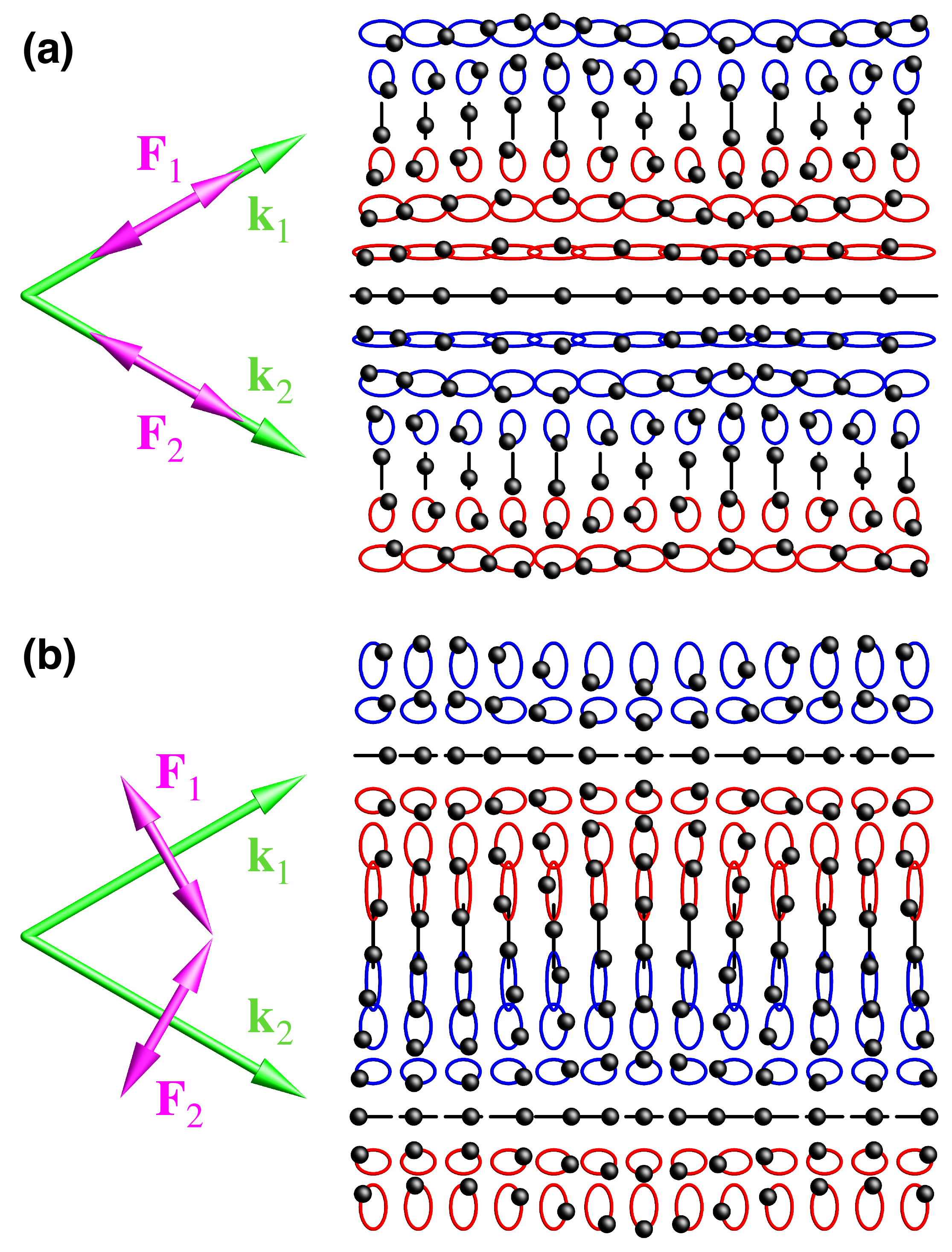}
\caption{Interference of two longitudinal (a) and transverse (b) waves with non-collinear wavevectors ${\bf k}_{1,2}$, equal amplitudes and frequencies. The resulting wavefield is shown via the instantaneous displacements of the effective medium particles and polarizations (elliptical particle trajectories). The counter-clockwise, clockwise, and linear polarizations are shown in red, blue, and black, respectively. Despite the difference in wave types, the polarization distributions and the spin AM density (orthogonal to the plane) appear quite similar in (a) and (b). See also Fig.~\ref{Fig_intro_water}(a) for an analogous water-wave example.} 
\label{Fig_two_wave}
\end{figure}

Polarization is closely related to the wave type only in the simplest {\it plane-wave} case \eqref{planewave}. Indeed, for longitudinal waves, the field has only one nonzero component aligned with the wavevector ${\bf k}$, and thus the polarization is purely linear. For transverse waves, the field generally has two nonzero components in the 2D plane orthogonal to ${\bf k}$, so the polarization can take any elliptical form in this plane, as in polarization optics \cite{BornWolf}. Finally, for mixed waves, the wave field has both longitudinal and transverse components, and the polarization can be elliptical in the plane containing the wavevector ${\bf k}$. 
However, these constraints become largely irrelevant for generic {\it structured} waves, as shown in Fig.~\ref{Fig_two_wave}. Indeed, the interference of two (three) plane longitudinal waves propagating in different directions produces two (three) nonzero field components, which can generate any elliptical polarization in 2D (3D) space. Similarly, the interference of two transverse waves can produce three nonzero field components and a generic polarization ellipse in 3D space. Thus, {\it complex interference fields can have generic ${\bf r}$-dependent 3D polarization textures, regardless of the wave's longitudinal/transverse/mixed type}.   

%FFFFFFFFFFFFFFFFFFFFFFFFFFFFFFFFFFFFFFFFFFFFFF
%FFFFFFFFFFFFFFFFFFFFFFFFFFFFFFFFFFFFFFFFFFFFFF

%FFFFFFFFFFFFFFFFFFFFFFFFFFFFFFFFFFFFFFFFFFFFFF
%FFFFFFFFFFFFFFFFFFFFFFFFFFFFFFFFFFFFFFFFFFFFFF

%%%%%%%%%%%%%%%%%%%%%%%%%%%%%%
%\subsection{Translations, rotations, and Noether's theorem}
\subsection{Canonical momentum, angular momentum, and spin from Noether's theorem}
\label{Canonical}
%%%%%%%%%%%%%%%%%%%%%%%%%%%%%%

\begin{figure}[t!]
\centering
\includegraphics[width=\linewidth]{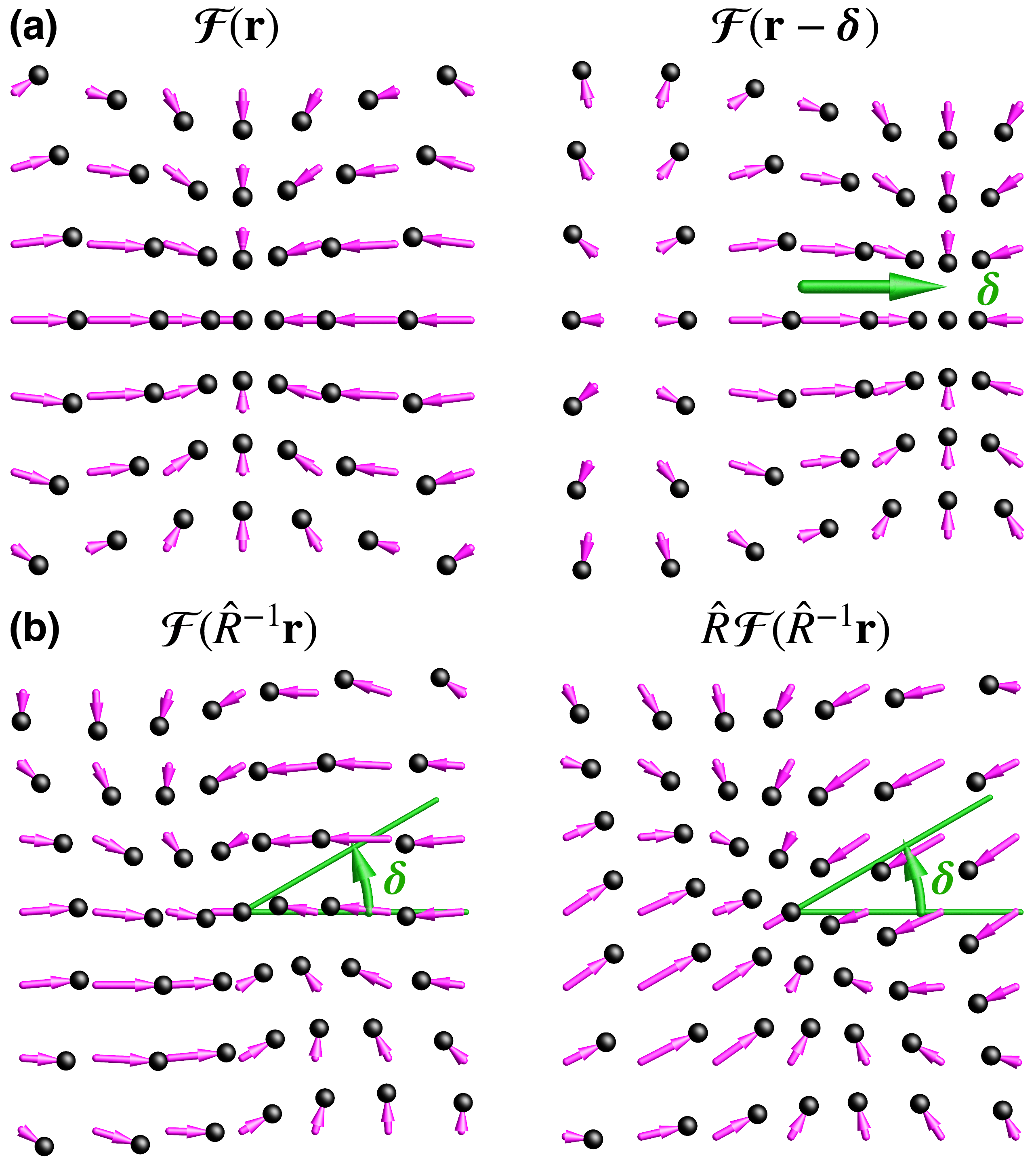}
\caption{(a) An inhomogeneous wavefield $\mathbfcal{F}({\bf r})$ [shown here as a fragment of the two-wave interference, Fig.~\ref{Fig_two_wave}(a)] and its translation $\mathbfcal{F}({\bf r}-{\bm \delta})$, Eq.~\eqref{Translation}. (b) Orbital rotation of the same field, $\mathbfcal{F}(\hat{R}^{-1}{\bf r})= e^{-i{\bm \delta}\cdot\hat{\bf L}}$, (left) and the total (orbital + spin) rotation $\hat{R}\mathbfcal{F}(\hat{R}^{-1}{\bf r}) =e^{-i{\bm \delta}\cdot(\hat{\bf L}+\hat{\bf S})}$, Eq.~\eqref{Rotation} (right). The orbital rotation affects only the distribution of the field without changing its local direction, while the spin rotation modifies the local direction of the field.} 
\label{Fig_trans_rot}
\end{figure}

The momentum and AM properties of wave fields are intimately related to translations and rotations. First, consider an infinitesimal spatial translation of a vector field ${\bf F}({\bf r})$, Fig.~\ref{Fig_trans_rot}(a): 
\begin{equation}
{\bf F}'({\bf r})={\bf F}({\bf r}-{\bm \delta})\simeq{\bf F}({\bf r})-({\bm \delta}\cdot {\bm \nabla}){\bf F}({\bf r})
\simeq e^{-i{\bm \delta}\cdot \hat{\bf P}}\,{\bf F}({\bf r})
 \,.
\label{Translation}
\end{equation}
Here $\hat{\bf P} = -i{\bm\nabla}$ is the canonical momentum operator known from quantum mechanics (we omit the $\hbar$ constant because we are dealing with classical waves). Thus, the canonical momentum operator is the {\it generator of translations} of the wave field. 

Next, consider an infinitesimal rotation of a vector field ${\bf F}({\bf r})$, produced by the operator $\hat{R}=1+{\bm\delta} \times$. (Finite rotations are described by matrices involving cosines and sines of $\delta$; e.g., for ${\bm \delta}$ directed along the $z$-axis: 
$\hat{R}_z = \left( \begin{matrix} \cos\delta & -\sin\delta & 0 \\
\sin\delta & \cos\delta & 0 \\ 
0 & 0 & 1 \end{matrix} \right)$.) 
This rotation should simultaneously rotate the {\it field distribution} in space and its {\it direction} at each point, as shown in Fig.~\ref{Fig_trans_rot}(b), which is described by 
\begin{align}
&{\bf F}'({\bf r})=\hat{R}\,{\bf F}(\hat{R}^{-1}{\bf r})
\simeq \nonumber \\
&{\bf F}({\bf r})-\left[({\bm \delta}\times{\bf r}) \cdot {\bm \nabla}\right]{\bf F}({\bf r}) + {\bm \delta}\times {\bf F}({\bf r}) 
\simeq e^{-i{\bm \delta}\cdot \hat{\bf J}}\,{\bf F}({\bf r})
 \,.
\label{Rotation}
\end{align}
Here $\hat{\bf J} = \hat{\bf L} +\hat{\bf S}$ is the total AM operator, where $\hat{\bf L} = {\bf r}\times \hat{\bf P}$ is the OAM operator, and $-i {\bm \delta}\cdot\hat{\bf S} = {\bm \delta}\times$, i.e., $(\hat{S}_k)_{ij} = -i \epsilon_{ijk}$ ($\epsilon_{ijk}$ is the Levi-Civita symbol) is the vector-matrix operator of spin.
The total AM operator is the {\it generator of rotations} for a 3D vector field, with the orbital and spin parts responsible for the rotations of the spatial distribution and the field directions (polarizations), respectively, Fig.~\ref{Fig_trans_rot}(b). 

In cylindrical coordinates $(r,\varphi,z)$, the $z$-components of the OAM and spin operators take the forms 
\begin{equation}
\hat{L}_z = -i ({\bf r} \times {\bm \nabla})_z = -i\frac{\partial}{\partial \varphi}\,, \quad 
\hat{S}_z = \left( \begin{matrix} 0 & -i & 0 \\
i & 0 & 0 \\ 
0 & 0 & 0 \end{matrix} \right).
\label{J_z}
\end{equation}
In particular, the $\hat{L}_z$ operator has scalar {\it wave-vortex} eigenmodes with azimuthal phase factors $\exp(i\ell\varphi)$, where $\ell = 0,\pm 1,\pm 2,...$ is the `OAM quantum number' or `topological charge' of the vortex \cite{Allen_book, Andrews_book, Allen1992PRA, Bliokh2015PR}. The wave phase acquires a $2\pi\ell$ increment when encircling the $z$-axis. However, circularly-symmetric {\it vector} vortex wavefields ${\bf F}({\bf r})$ are generally eigenmodes of the {\it total} AM operator $\hat{J}_z = \hat{L}_z + \hat{S}_z$ \cite{VanEnk1994JMO, Bliokh2010PRA, Picardi2018O, Bliokh2019b, Bliokh2022PRL, Smirnova2024PRL}, because it is the total AM operator that generates rotations of the vector wavefield (see Section~\ref{Vortex}).

%FFFFFFFFFFFFFFFFFFFFFFFFFFFFFFFFFFFFFFFFFFFFFF
%FFFFFFFFFFFFFFFFFFFFFFFFFFFFFFFFFFFFFFFFFFFFFF

%\blue{[Notes on the transversality, SAM and OAM operators, etc.]}

The canonical OAM and spin operators are well known from the quantum mechanics of photons, i.e., quantum spin-1 particles \cite{Akhiezer_book}. However, it should be emphasized that these AM operators generate SO(3) rotations of {\it any} classical vector fields, independently of the spin of the corresponding quantum particles (e.g., of acoustic displacement fields corresponding to spin-0 phonons).

%Here we should remark that the OAM and SAM operators $\hat{\bf L}$ and $\hat{\bf S}$ separately can be `inconsistent' with the corresponding wave equations. Indeed, a separate orbital or spin/polarization rotation of a transverse (${\bm \nabla}\cdot {\bf F}=0$) or longitudinal (${\bm \nabla}\times {\bf F}={\bf 0}$) wavefield generally makes it non-transverse or non-longitudinal \cite{Akhiezer_book}. Only the total AM operator $\hat{\bf J}$ does not have this problem. For transverse waves, this formal difficulty can be overcome by introducing `corrected' OAM and SAM operators, $\hat{\bf J} = \hat{\bf L}' +\hat{\bf S}'$, consistent with the transversality condition and having {\it the same expectation values} as the canonical $\hat{\bf L}$ and $\hat{\bf S}$ operators \cite{Bliokh2010PRA, VanEnk1994JMO, Barnett2010JMO}. Since here we are interested in the {\it values} of the OAM and spin, rather than operators, this problem does not affect our considerations in what follows. 

Let us now consider the {\it values} of the momentum and AM densities in a wave field. 
According to Noether's theorem, the conserved momentum and AM are related to the symmetry of the Lagrangian with respect to spatial translations and rotations. Let the Lagrangian density of the wave field $\mathbfcal{F}({\bf r},t)$ contain the quadratic `kinetic-energy' term with `velocity' $\dot{\mathbfcal{F}}$ (the overdot denotes the time derivative $\partial/\partial t$):
\begin{equation}
\label{Lagrangian}
{\mathcal{L}}=a\,\dot{\mathbfcal{F}}^2/2 - \Lambda\,, 
\end{equation}
where $a$ is a constant, and the potential energy density $\Lambda$ depends on spatial derivatives of the field $\mathbfcal{F}$. If the field Lagrangian $\int \mathcal{L}\,d^3{\bf r}$ is invariant under translations and rotations, then Noether's theorem results in the following conserved momentum and AM densities \cite{Soper_book, Nakane2018PRB, Bliokh2022PRL, Ren2022CPL, Bliokh2022PRE}:
\begin{equation}
{\bf{P}}=-\frac{\partial \mathcal{L}}{\partial \dot{\mathbfcal{F}}}\cdot ({\bm\nabla}) \mathbfcal{F} 
=-a\, \dot{\mathbfcal{F}} \cdot ({\bm\nabla}) \mathbfcal{F}
\,,
\label{NoetherP}
\end{equation}
\vspace{-0.5cm}
\begin{equation}
{\bf{J}} = -\frac{\partial \mathcal{L}}{\partial \dot{\mathbfcal{F}}}\cdot ({\bf r}\times{\bm\nabla}) \mathbfcal{F} - \frac{\partial \mathcal{L}}{\partial \dot{\mathbfcal{F}}}\times \mathbfcal{F}
=-a\, \dot{\mathbfcal{F}} \cdot ({\bf r}\times{\bm\nabla}) \mathbfcal{F}- a\, \dot{\mathbfcal{F}} \times \mathbfcal{F}.
\label{NoetherJ}
\end{equation}
Here $\dot{\mathbfcal{F}}\cdot ({\bm\nabla}) \mathbfcal{F} \equiv \Sigma_{i} \mathcal{F}_i^* {\bm \nabla} \mathcal{F}_i$, and the two summands in the right-hand side of Eq.~\eqref{NoetherJ} can be associated with the orbital and spin contributions: ${\bf J} = \bf{L} + \bf{S}$.

Equations \eqref{NoetherP} and \eqref{NoetherJ} are quadratic in the wave field and involve structures resembling the canonical momentum and AM operators. Substituting the monochromatic field \eqref{structuredwave}, and performing averaging over the oscillation period $2\pi/\omega$, we obtain the time-averaged momentum, OAM, and spin densities:
\begin{equation}
\overline{\bf P} = \frac{a\omega}{2}\Re[{\bf F}^* \cdot (-i{\bm \nabla}) {\bf F}] 
=\frac{a\omega}{2} \Im[{\bf F}^* \cdot ({\bm \nabla}) {\bf F}]\,,
\label{P_general}
\end{equation}
\vspace{-0.5cm}
\begin{equation}
\overline{\bf L} = \frac{a\omega}{2}
\Im[{\bf F}^* \cdot ({\bf r}\times{\bm \nabla}) {\bf F}]
= {\bf r} \times \overline{\bf P} \,,
\label{L_general}
\end{equation}
\vspace{-0.5cm}
\begin{equation}
\overline{\bf S} = \frac{a\omega}{2}
\Im({\bf F}^* \times {\bf F}) \,.
\label{S_general}
\end{equation}
These are the key equations for our review. 
They have the form of the local expectation values of the operators $\hat{\bf P}$, $\hat{\bf L}$, and $\hat{\bf S}$, where the complex field ${\bf F}({\bf r})$ plays the role of the wavefunction. 
The relation ${\bf L} = {\bf r} \times {\bf P}$ is universal, and we will not repeat it for different kinds of waves, focusing on the momentum and spin densities ${\bf P}$ and ${\bf S}$.
Importantly, Eqs.~\eqref{P_general}--\eqref{S_general} have clear physical interpretations. 

%FFFFFFFFFFFFFFFFFFFFFFFFFFFFFFFFFFFFFFFFFFFFFF
\begin{figure}[t!]
\centering
\includegraphics[width=0.75\linewidth]{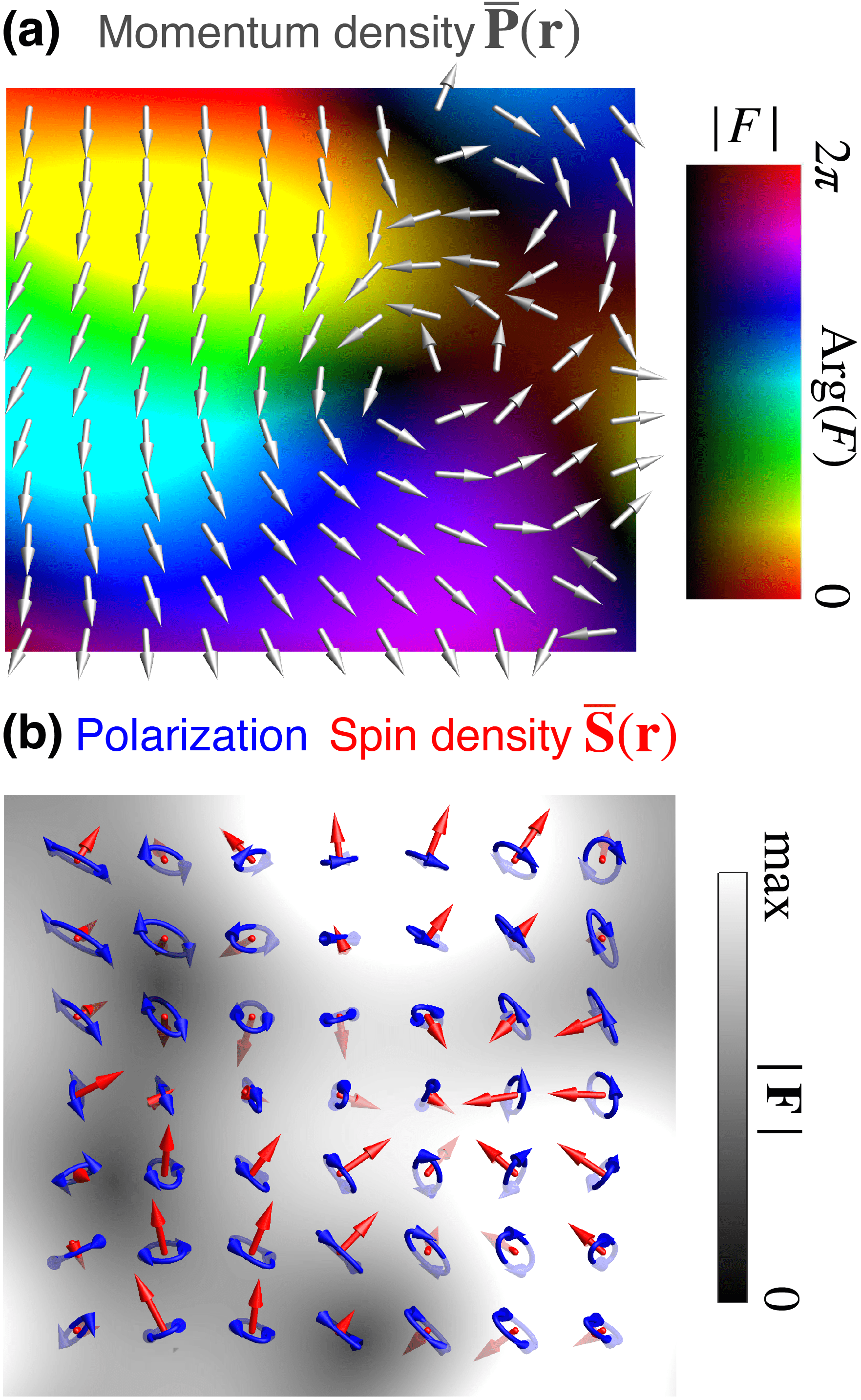}
\caption{(a) A complex scalar wave field $F({\bf r})$ and its canonical momentum density $\overline{\bf P} \propto |F|^2{\bm \nabla}{\rm Arg}(F)$. The phase ${\rm Arg}(F)$ and amplitude $|F|$ are represented by colors and brightness, respectively. For a vector field  ${\bf F}({\bf r})$, the canonical momentum \eqref{P_general} is the sum of such momenta for each of the field components. (b) A vector wave field ${\bf F}({\bf r})$ shown by its amplitude $|{\bf F}|$ (grayscale) and polarization ellipses (blue), along with the corresponding spin density ${\bf S}({\bf r})$ (red) at each point ${\bf r}$ (see Fig.~\ref{Fig_ellipse}. For better visibility, here the polarization ellipses and spin vectors are normalized by the field intensity $|{\bf F}|^2$.} 
\label{Fig_momentum_spin}
\end{figure}
%FFFFFFFFFFFFFFFFFFFFFFFFFFFFFFFFFFFFFFFFFFFFFF

\begin{figure}[t!]
\centering
\includegraphics[width=0.65\linewidth]{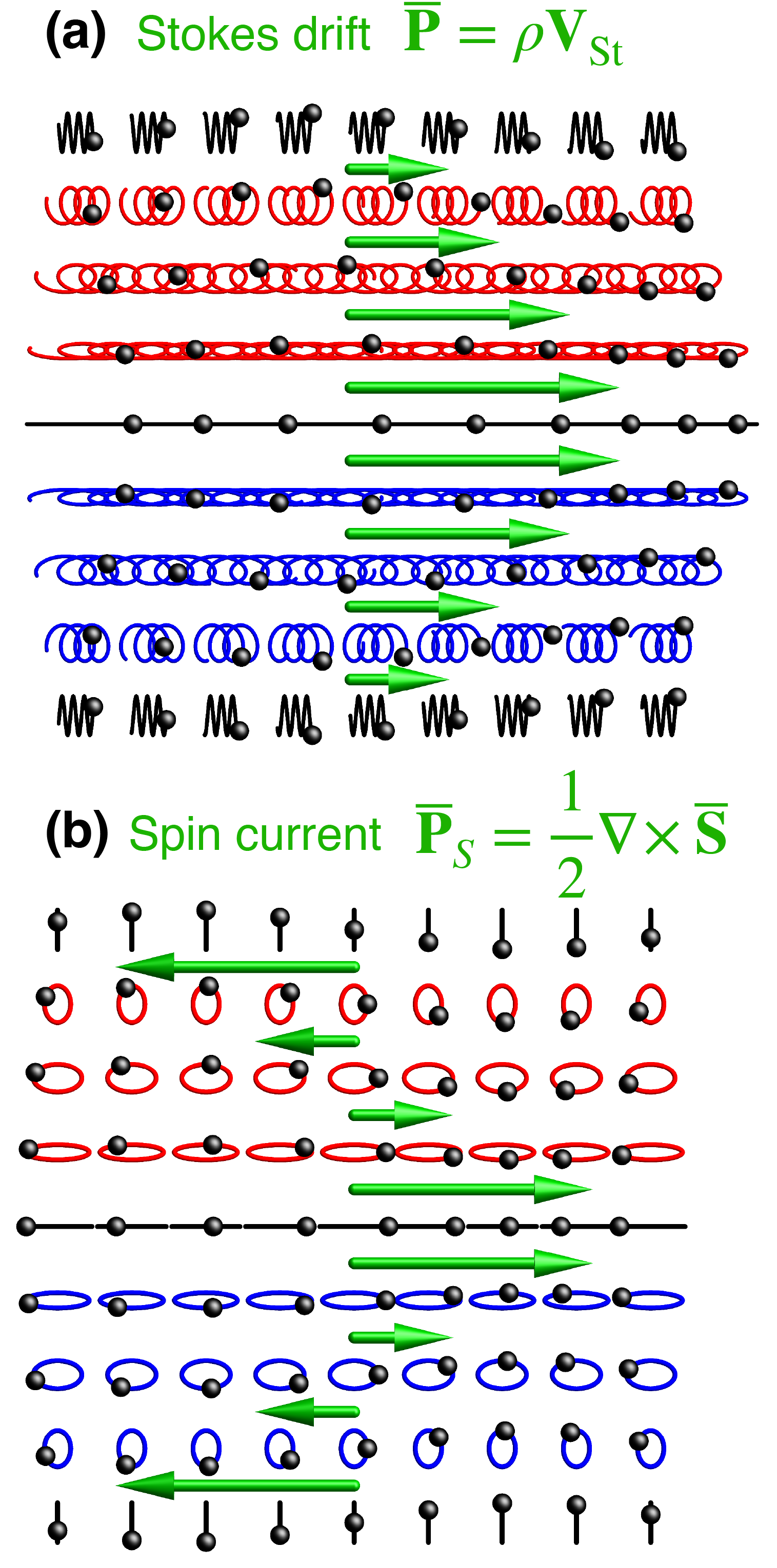}
\caption{Wave momentum in the interference of two longitudinal (e.g., sound) waves, Fig.~\ref{Fig_two_wave}(a). (a) A quadratic correction to the linear oscillatory (elliptical) motion of the medium particles gives rise to the Stokes drift \cite{Bremer2018, Falkovich_book}, see Section~\ref{Stokes} and Fig.~\ref{Fig_intro_water}. Here, the particles' trajectories with the drift correction are shown over three periods of oscillations, $6\pi/\omega$. The time-averaged canonical momentum density corresponds to the momentum density of the drifting medium particles. (b) In the linear approximation, when neighboring particles move along slightly different ellipses, their local velocities/momenta do not fully cancel each other. This results in the spin-induced current and momentum density given by ${\bm \nabla}\times \overline{\bf S}/2$. This spin current is `virtual' or bound, because there is no global transport in this approximation: each particle oscillates around its unperturbed location.} 
\label{Fig_Stokes_spin}
\end{figure}

First, using the amplitude-phase decomposition for each of the field components, $F_i = |F_i|\exp\left[i {\rm Arg}(F_i)\right]$, the momentum density \eqref{P_general} can be associated with the intensity-weighted {\it phase gradient} of the field, averaged over all its components: $\overline{\bf P} \propto \Sigma_i |F_i|^2{\bm \nabla}{\rm Arg}(F_i)$, Fig.~\ref{Fig_momentum_spin}(a). For a scalar field ${\bf F} \to F$, Eq.~\eqref{P_general} takes the form of the {\it probability current} in nonrelativistic quantum mechanics. Remarkably, for waves in media with freely moving particles (such as gases or fluids), the momentum density \eqref{P_general} corresponds to the velocity of the Stokes drift of the medium particles, multiplied by the mass density of the medium: $\overline{\bf P} = \rho \overline{\bf V}_{\rm St}$ \cite{Bliokh2022SA, Bremer2018, Mcintyre1981}, see Figs.~\ref{Fig_intro_water}, \ref{Fig_Stokes_spin}(a), and Section~\ref{Stokes}. 

Second, the OAM density \eqref{L_general} is determined by the vector product of the radius-vector and momentum density, similar to the AM of a point particle in classical mechanics. 
Since the radius-vector is defined with respect to a chosen coordinate origin, the OAM density is an {\it extrinsic} quantity, which depends on the choice of this origin. Indeed, a shift of the origin, ${\bf r} \to {\bf r} +{\bf r}_0$, induces the OAM transformation ${\bf L} \to {\bf L} + {\bf r}_0 \times {\bf P}$.
In cylindrical coordinates $(r,\varphi,z)$, the $z$-component of the OAM density \eqref{L_general} becomes
$\overline{L}_z = r \overline{P}_\varphi = (a\omega/2) \Im({\bf F}^* \partial_\varphi {\bf F})$. Thus, it is the azimuthal component of the momentum density (circulating around the $z$-axis) that produces the OAM in wave vortices with ${\bf F} \propto \exp(i\ell\varphi)$ \cite{Allen_book, Andrews_book, Bliokh2015PR, Allen1992PRA} (see Figs.~\ref{Fig_torques}(b), \ref{Fig_intro_water}(b), and Section~\ref{Vortex}). 

The momentum and OAM densities \eqref{P_general} and \eqref{L_general} can be defined for both vector and scalar wavefields, independently of the number of field components or its polarization properties. In contrast, the spin density \eqref{S_general} essentially involves {\it polarization} of the wavefield.
Namely, it is produced by the {\it rotation} of the real field \eqref{structuredwave} at each point of space ${\bf r}$, Fig.~\ref{Fig_momentum_spin}(b). Using the unit polarization vector introduced in the previous section, the spin density \eqref{S_general} can be written as $\overline{\bf S}= a\omega |{\bf F}|^2 \left( \Re{\bf f} \times \Im{\bf f} \right)$, which shows that the spin is directed normal to the polarization ellipse and is proportional to its area, Fig.~\ref{Fig_ellipse} \cite{Bliokh2015PR, Berry2001}.
For a given wave amplitude $|{\bf F}|$, the magnitude of the spin density is maximized for circular polarizations, and vanishes for linear polarizations.

The spin form \eqref{S_general} has a particularly clear physical meaning for mechanical waves, where the field ${\mathbfcal{F}}\equiv \mathbfcal{R}$ represents the local displacement of the medium particles. The corresponding velocity field is $\mathbfcal{V} = \dot{\mathbfcal{R}}$, so that ${\bf V} = -i\omega {\bf R}$ for the corresponding complex fields. Then, the time-averaged {\it mechanical AM} of the medium particle with respect to its unperturbed position is given by $m\, \overline{\mathbfcal{R} \times \mathbfcal{V}} = m \Re({\bf R}^* \times {\bf V})/2 = m\, \omega \Im({\bf R}^* \times {\bf R})/2$, where $m$ is the mass of the particle \cite{Jones1973, Shi2019, Bliokh2022SA}. For the spin density, the mass $m$  is replaced by the mass density of the medium, $\rho$. 

Evidently, the spin density is an {\it intrinsic} quantity, independent of the choice of coordinate origin. 
Thus, the total AM density of the wave is given by the sum of the orbital (extrinsic) and spin (intrinsic) contributions: ${\bf J} = {\bf L} + {\bf S}$. The OAM density is fully determined by the canonical momentum density ${\bf P}$, while the SAM density ${\bf S}$ is an independent quantity. 

Finally, we should make an important remark on the nature of wave momentum and AM. All the quantities considered here are associated with translations and rotations of the wave field {\it relative to the motionless medium}. Therefore, these canonical momentum and AM are sometimes referred to as {\it pseudo}-momentum and {\it pseudo}-AM \cite{Mcintyre1981, Peierls1991}. This is to distinguish them from quantities associated with translations and rotations of the entire ``wave+medium'' system. 
In many problems involving dynamical properties of waves, the canonical (pseudo) momentum and AM are the relevant quantities. In particular, in various fields, the principal wave-induced radiation force and torque (with respect to the particle's center) on a small absorbing isotropic particle are determined by the canonical momentum and spin densities, respectively \cite{Bliokh2014NC, Toftul2019PRL, Toftul2024, Wang2025N}. This justifies their use as the primary quantities of interest in wave-mechanical systems.

%FFFFFFFFFFFFFFFFFFFFFFFFFFFFFFFFFFFFFFFFFFFFFF
%FFFFFFFFFFFFFFFFFFFFFFFFFFFFFFFFFFFFFFFFFFFFFF

%%%%%%%%%%%%%%%%%%%%%%%%%%%%%%
\subsection{Energy flux, kinetic momentum, and Belinfante-Rosenfeld relation}
\label{Kinetic}
%%%%%%%%%%%%%%%%%%%%%%%%%%%%%%

Another key quantity characterizing wave dynamics is the wave energy density and its flux. Energy conservation is associated with the time-translation invariance of the Lagrangian and the corresponding operator $i\partial/\partial t$. For the field Lagrangian density \eqref{Lagrangian}, Noether's theorem yields the energy density
\begin{equation}
W=\frac{\partial \mathcal{L}}{\partial \dot{\mathbfcal{F}}}\cdot \dot{\mathbfcal{F}} - \mathcal{L}
=\frac{a}{2} \dot{\mathbfcal{F}}^2 + \Lambda\,.
\label{W}
\end{equation}
The local energy conservation law has the form
\begin{equation}
\frac{\partial {W}}{\partial t} 
%\dot{\mathcal{W}}
+ {\bm\nabla} \cdot {\bf U} =0\,,
\label{W_cons}
\end{equation}
where $\bf{U}$ is the {\it energy flux} density. Examples of energy conservation \eqref{W_cons} include the Poynting theorem in electromagnetism \cite{jackson1998ClassicalElectrodynamics} and its analogues for other types of waves \cite{Landau_fluid, Auld_book}. 

According to field theory, the energy, energy flux, and momentum densities are components of the rank-2 {\it energy-momentum tensor} $\mathcal{T}^{\mu\nu}$, $\mu,\nu = 0,1,2,3$ \cite{Soper_book}. Namely, ${W} = \mathcal{T}^{00}$, ${P}_i = \mathcal{T}^{i0}$, whereas the energy flux density is given by 
\begin{equation}
{U}_i=\mathcal{T}^{0i} =\frac{\partial \mathcal{L}}{\partial (\nabla_i{\mathbfcal{F}})}\cdot \dot{\mathbfcal{F}} = -  \frac{\partial \Lambda}{\partial (\nabla_i{\mathbfcal{F}})}\cdot \dot{\mathbfcal{F}}\,.
\label{U}
\end{equation}
One can verify that the energy conservation \eqref{W_cons} is satisfied with Eqs.~\eqref{W} and \eqref{U} using the Euler-Lagrange equation $a \ddot{\mathbfcal{F}} = \Sigma_i \nabla_i \left[\partial \Lambda /\partial (\nabla_i {\mathbfcal{F}})\right]$.

It is tempting to associate the energy flux with the propagation and momentum of the field. Indeed, one can introduce the so-called {\it kinetic momentum} density (noting that the ``canonical-kinetic'' terminology may vary across different contexts):
\begin{equation}
  {\bm\Pi}=\frac{\bf{U}}{c_{ph}c_g}\,.
\label{Pkin}
\end{equation}
Here $c_{ph} = \omega/k$ and $c_g = \partial\omega / \partial k$ are the phase and group velocities of the wave. For example, the kinetic momentum density of an electromagnetic field is given by the Poynting vector (energy flux density) divided by $c^2$, where $c$ is the speed of light \cite{jackson1998ClassicalElectrodynamics}.

Notably, the time-averaged canonical-momentum, kinetic-momentum, and spin densities are related by the fundamental {\it Belinfante-Rosenfeld relation} \cite{Belinfante1940, Soper_book, Berry2009, Bliokh2014NC, Markovich2021PRL, Bliokh2022SA, Bliokh2022PRE}:
\begin{equation}
\overline{\bm \Pi} = \overline{\bf P} + \frac{1}{2} {\bm \nabla}\times \overline{\bf S}\,.
\label{BR}
\end{equation}
Originally derived in relativistic field theory for quantum particles, this relation was used to symmetrize the asymmetric energy-momentum tensor in the presence of spin \cite{Belinfante1940}. Later, it was found to be equally applicable to classical waves of various kinds \cite{Berry2009, Bliokh2014NC, Bliokh2022SA, Bliokh2022PRE}. 

The Belinfante-Rosenfeld relation has a simple interpretation. 
As noted earlier, the momentum density \eqref{P_general} in media with free particles can be associated with the wave-induced current of medium particles, Fig.~\ref{Fig_Stokes_spin}(a). In the presence of spin, i.e., local elliptical motions of the medium particles, there is another, `{\it virtual}' current produced by this motion, Fig.~\ref{Fig_Stokes_spin}(b). If all medium particles move along identical elliptical trajectories, i.e., the spin density \eqref{S_general} is uniform in space, then the local motions of the neighboring particles are mutually canceled, leading to zero net current. However, if the spin density is non-uniform and neighboring particles follow slightly different elliptical orbits, then the local currents from these orbits are not mutually canceled. This results in an additional spin-induced current and the corresponding momentum density proportional to the curl of the spin density \cite{Soper_book, Mita2000AJP, Markovich2021PRL}: $\overline{\bf P}_S = {\bm \nabla}\times \overline{\bf S}/2$. This current is `virtual' because it arises from variations in local rotational motions of the particles and is not associated with any global transport. (A similar bound electric current occurs in solids with non-uniform magnetization produced by local micro-currents.) 

Thus, the kinetic momentum density \eqref{Pkin} and \eqref{BR} is a sum of the canonical momentum density and the virtual spin-induced current. In many cases, the kinetic momentum serves as a fundamental and convenient theoretical quantity, such as the Poynting momentum in electromagnetism \cite{jackson1998ClassicalElectrodynamics}. Its advantage is that it is well-defined for arbitrary fields, including non-monochromatic and static fields. Still, one should remember that for waves, it consists of two contributions with distinct physical properties \cite{Bliokh2014NC, Bliokh2015PR, Antognozzi2016NP, Bliokh2022SA, Ghosh2024JOSA}.

Note that both the time-averaged canonical and kinetic momentum densities satisfy the same stationary continuity (energy conservation) equation and yield the same {\it integral} momentum value for a localized wave packet in a homogeneous medium:
\begin{equation}
{\bm \nabla}\cdot \overline{\bm \Pi} = {\bm \nabla}\cdot \overline{\bf P}  = 0\,, \quad
\int \overline{\bm \Pi}\, d^3{\bf r} = \int \overline{\bf P}\, d^3{\bf r}\,.
\label{Pkin-P}
\end{equation}
However, their {\it local} densities can differ significantly. See, e.g., examples \cite{Bliokh2014NC, Bekshaev2015PRX, Antognozzi2016NP} with an `anomalous' transverse $z$-component of the kinetic momentum in inhomogeneous electromagnetic waves propagating within the $(x,y)$ plane, as well as recent tutorial \cite{Ghosh2024JOSA}.

The kinetic picture also allows for an alternative description of the wave AM density: ${\bf M} = {\bf r} \times {\bm \Pi}$ \cite{jackson1998ClassicalElectrodynamics, Allen_book}. For localized wave fields, the integral value of this kinetic AM matches the integral total AM in the canonical approach:
\begin{equation}
\int \overline{\bf M}\, d^3{\bf r} = 
%\int {\bf J}\, d^3{\bf r} =  
\int \overline{\bf L}\, d^3{\bf r} +  \int \overline{\bf S}\, d^3{\bf r}\,.
\label{Mkin}
\end{equation}
However, locally, the kinetic AM density ${\bf M}$ has limited physical meaning because it is an extrinsic quantity that does not properly characterize the intrinsic spin AM density. For example, the kinetic AM density $M_z$ vanishes in a circularly-polarized electromagnetic plane wave propagating along the $z$-direction. In contrast, the canonical spin density yields the physically meaningful non-zero value $S_z$. To reconcile the integral kinetic and canonical AM values \eqref{Mkin}, one must consider a transversely-confined wave beam rather than an infinite plane wave \cite{Yurchenko2002AJP}.

%%%%%%%%%%%%%%%%%%%%%%%%%%%%%%
\section{Sound waves}
\label{Sound}
%%%%%%%%%%%%%%%%%%%%%%%%%%%%%%

We are now in a position to apply the general approach of Section~\ref{General} to specific wave types. We start with the simplest example of longitudinal waves: sound waves in a homogeneous fluid or gas. The linear wave-induced displacement of medium particles (molecules, neglecting random Brownian motions), ${\mathbfcal{R}}({\bf r},t)$, serves as the vector wave field in this problem. The Lagrangian density of this field is given by
\begin{equation}
\mathcal{L}_s=\frac{\rho}{2}\left[\dot{\mathbfcal{R}}^2-c_s^2 ({\bm\nabla} \cdot \mathbfcal{R})^2 \right]
= \frac{1}{2}\left[\rho {\mathbfcal{V}}^2-\beta {\mathcal{P}}^2\right],
\label{Lsound}
\end{equation}
where $\rho$ is the unperturbed mass density of the medium, $\beta$ is the compressibility of the medium, $c_s = 1/\sqrt{\rho\beta}$ is the speed of sound, ${\mathbfcal{V}}=\dot{\mathbfcal{R}}$ is the local velocity field, and $\mathcal{P}$ is the scalar pressure field (i.e., its wave-induced perturbation). The second equality in Eq.~\eqref{Lsound} follows from $\beta \mathcal{P} = - {\bm\nabla} \cdot \mathbfcal{R}$, which, in turn, follows from the continuity equation $\beta \dot{\mathcal{P}} = - {\bm\nabla} \cdot \mathbfcal{V}$, where $\beta {\mathcal{P}} = \delta\rho/\rho$ represents the dimensionless wave-induced local perturbation of the medium density. The Euler-Lagrange equation for the Lagrangian \eqref{Lsound} yields
\begin{equation}
\ddot{\mathbfcal{R}} = c_s^2 {\bm\nabla}({\bm\nabla} \cdot \mathbfcal{R})\,,
\label{EOM_sound_1}
\end{equation}
which is equivalent to the well-known sound-wave equations \cite{Landau_fluid, Falkovich_book}:
\begin{equation}
\rho \dot{\mathbfcal{V}} = -{\bm\nabla}{\mathcal{P}}\,,\quad
\beta \dot{\mathcal{P}} = - {\bm\nabla} \cdot \mathbfcal{V}\,,
\label{EOM_sound}
\end{equation}
and corresponds to the linear dispersion relation $\omega = kc_s$ with $c_{ph} = c_g = c_s$.

The Lagrangian density \eqref{Lsound} has the assumed form \eqref{Lagrangian} of the kinetic minus potential energy densities. Therefore, we immediately obtain the time-averaged canonical momentum and spin AM densities \eqref{P_general} and \eqref{S_general} in a monochromatic sound-wave field:
\begin{equation}
\overline{\bf P} = 
\frac{\rho\omega}{2} \Im[{\bf R}^* \cdot ({\bm \nabla}) {\bf R}]\,,
\label{P_sound}
\end{equation}
\vspace{-0.5cm}
\begin{equation}
%\overline{\bf L} = {\bf r} \times \overline{\bf P} \,, \quad
\overline{\bf S} = \frac{\rho \omega}{2}
\Im({\bf R}^*\! \times {\bf R}) \,.
\label{S_sound}
\end{equation}
These expressions can also be written in terms of the complex velocity field ${\bf V} = -i\omega {\bf R}$ using the substitution ${\bf R} \to {\bf V}$, $\rho\omega \to \rho/\omega$. 

The time-averaged energy, energy-flux, and kinetic-momentum densities \eqref{W_cons}--\eqref{Pkin} take the forms \cite{Falkovich_book, Landau_fluid}:
\begin{equation}
{W} = 
\frac{\rho}{2}\left[\dot{\mathbfcal{R}}^2+c_s^2 ({\bm\nabla} \cdot {\mathbfcal{R}})^2 \right]
= \frac{1}{2}\left[\rho {\mathbfcal{V}}^2+\beta \mathcal{P}^2\right],
%\overline{W} = 
%\frac{\rho}{4}\left[\omega^2|{\bf R}|^2+c_s^2 ({\bm\nabla} \cdot {\bf R})^2 \right]
%= \frac{1}{4}\left[\rho |{\bf V}|^2+\beta |P|^2\right]\,,
\label{W_sound}
\end{equation}
\vspace{-0.5cm}
\begin{equation}
{\bf U} = - {\rho c_s^2}\, \dot{\mathbfcal{R}} ({\bm\nabla} \cdot {\mathbfcal{R}}) =
\mathcal{P}{\mathbfcal{V}}\,, \quad
{\bm \Pi} = \frac{{\bf U}}{c_s^2}\,.
%\overline{\bf U} = \frac{\rho c_s^2\omega}{2} 
%\Im\!\left[ {\bf R}^* ({\bm\nabla} \cdot {\bf R})\right] =
%\frac{1}{2}\Re(P^*{\bf V})\,, \quad
% \overline{\bm \Pi} = \frac{\overline{\bf U}}{c_s^2}\,.
\label{U_sound}
\end{equation}
Using the sound-wave equations \eqref{EOM_sound_1} or \eqref{EOM_sound}, one can verify that the energy and energy-flux densities \eqref{W_sound} and \eqref{U_sound} satisfy the energy conservation law \eqref{W_cons}, whereas the time-averaged kinetic momentum $\overline{\bm \Pi} = \Re(P^*{\bf V})/2c_s^2$ satisfies the Belinfante-Rosenfeld relation \eqref{BR} with the canonical momentum and spin densities \eqref{P_sound} and \eqref{S_sound}.

For a plane wave \eqref{planewave}, the sound-wave momentum, energy, and spin densities \eqref{P_sound}--\eqref{U_sound} yield  
\begin{equation}
\overline{\bf P} = \overline{\bm \Pi} = \frac{\overline{W}}{\omega}{\bf k}\,, \quad
\overline{\bf S} = {\bf 0}\,.
\label{planewave_s}
\end{equation}
The vanishing spin is a consequence of the longitudinal character of sound waves. Nonetheless, for structured sound waves, starting from the simplest two-wave interference, Fig.~\ref{Fig_two_wave}(a), the spin density $\overline{\bf S}$ is generally nonzero \cite{Jones1973, Shi2019, Bliokh2019b}. As discussed in Section~\ref{Canonical},
this spin density corresponds to the mechanical AM density produced by the local elliptical motion of medium particles. 
Notably, the {\it integral} spin AM of a localized sound wavepacket vanishes: $\int \overline{\bf S}\, d^3{\bf r} = {\bf 0}$. This follows from the identity \cite{Bliokh2019b}: 
\begin{align}
\overline{\bf S} & = \frac{1}{2\rho\omega^3}\Im({\bm \nabla}P^* \times {\bm \nabla}P)=
\frac{1}{2\rho\omega^3}{\bm \nabla}\times \Im(P^* {\bm \nabla}P) \nonumber \\ 
& = \frac{1}{k^2}({\bm \nabla}\times \overline{\bm \Pi})\,.
\end{align}
Thus, the local nonzero spin density complies with the global spin-0 nature of phonons, i.e., spinless quasiparticles (wavepackets) of sound.  

The above consideration illustrates an important distinction between {\it scalar} and {\it longitudinal vector} waves. Sound is often treated as a scalar wave, because it can be described by the scalar pressure field $\mathcal{P}$ satisfying the scalar wave equation $\ddot{\mathcal{P}} - c_s^2 {\bm \nabla}^2 \mathcal{P} =0$, following from Eqs.~\eqref{EOM_sound} and the Klein-Gordon Lagrangian \cite{Bliokh2019PRL}. Such representation perfectly describes the sound-wave field but yields identically vanishing spin density ${\bf S}\equiv {\bf 0}$, as well as identical canonical and kinetic momentum densities ${\bf P}\equiv {\bm \Pi}$ \cite{Burns2020}. In contrast, the vector representation by the displacement field $\mathbfcal{R}$ provides a more complete physical description because it  correctly accounts for the spin AM density associated with the local elliptical motion of medium particles. Thus, the microscopic mechanical properties of the medium select the adequate field-theory representation of the wave.

The vector nature of sound waves, including their spin density, was presented by Jones in 1973 \cite{Jones1973}. However, this concept remained largely unnoticed until it was rediscovered and experimentally confirmed in 2019 by Shi {\it et al}. \cite{Shi2019}. The recognition of the sound-wave spin has led to new insights, such as the explanation of the acoustic torque on small absorbing particles in structured sound waves \cite{Toftul2019PRL, Toftul2024}. 

%%%%%%%%%%%%%%%%%%%%%%%%%%%%%%
\subsection{Stokes drift}
\label{Stokes}
%%%%%%%%%%%%%%%%%%%%%%%%%%%%%%

The Stokes drift phenomenon was originally described for water-surface waves \cite{Bremer2018}, but it is equally present in any waves propagating in media with free particles \cite{Falkovich_book, Bliokh2022PRA, Bliokh2022PRE}. This drift naturally arises as a quadratic correction from the equations of motion \eqref{EOM_sound_1} or \eqref{EOM_sound}. Let us present the main equation of motion as $\rho \dot{\mathbfcal{V}}({\bf r},t)={\mathbfcal{G}}({\bf r},t)$, where ${\mathbfcal{G}}({\bf r},t)$ is the force density. Note that ${\bf r}$ corresponds to the unperturbed position of the particle, while the wave-perturbed particle is displaced to ${\bf r} + \mathbfcal{R}$ and, hence, experiences the modified force ${\mathbfcal{G}}({\bf r}+ \mathbfcal{R},t) \simeq {\mathbfcal{G}}({\bf r},t)+ (\mathbfcal{R}\cdot {\bm \nabla}) {\mathbfcal{G}}({\bf r},t)$. The second term here is quadratic in the wavefield. Using the perturbation theory and denoting the second-order correction to the particle velocity as $\delta\mathbfcal{V}$, we express it in terms of the linear-approximation field: 
\begin{equation}
\delta\dot{\mathbfcal{V}} = (\mathbfcal{R}\cdot {\bm \nabla}) \dot{\mathbfcal{V}} = 
\frac{\partial}{\partial t}\left[ (\mathbfcal{R}\cdot {\bm \nabla}) {\mathbfcal{V}}\right] - \frac{1}{2}{\bm \nabla} (\mathbfcal{V}^2)\, ,
\label{drift}
\end{equation}
where we used ${\bm \nabla} \times \mathbfcal{V} = {\bf 0}$. The second term in Eq.~\eqref{drift} represents the ponderomotive {\it gradient force} \cite{Gaponov1958, Bliokh2022PRA}, which pushes the particle toward regions of lower kinetic energy. In turn, the first term in Eq.~\eqref{drift} corresponds to the time derivative of the Stokes drift velocity. Performing the time averaging for monochromatic fields, the Stokes drift velocity takes the form \cite{Bremer2018, Falkovich_book, Bliokh2022SA, Bliokh2022PRA}: 
\begin{equation}
\overline{\bf V}_{\rm St}=  
\frac{1}{2}\Re\!\left[ ({\bf R}^*\cdot {\bm \nabla}) {\bf{V}}\right] = \frac{\omega}{2}\Im\!\left[ {\bf R}^*\cdot ({\bm \nabla}) {\bf{R}}\right] .
\label{Stokes_drift}
\end{equation}
Thus, the canonical momentum density \eqref{P_sound} is equal to the Stokes drift velocity \eqref{Stokes_drift} multiplied by the mass density of the medium. This provides a clear physical interpretation of the canonical sound-wave momentum, Fig.~\ref{Fig_Stokes_spin}(a). 

%%%%%%%%%%%%%%%%%%%%%%%%%%%%%%
\section{Electromagnetic waves}
\label{Electromagnetic}
%%%%%%%%%%%%%%%%%%%%%%%%%%%%%%

Let us now consider a well-known example of transverse waves: electromagnetic waves in free space. We express the electric ($\mathbfcal{E}$) and magnetic ($\mathbfcal{H}$) fields via the vector-potential $\mathbfcal{A}$ defined by $\mathbfcal{E} = -\dot{\mathbfcal{A}}$ and $\mathbfcal{H} = c {\bm\nabla} \times{\mathbfcal{A}}$, where $c$ is the speed of light. We use Gaussian units, so that $c \mathbfcal{A}$ corresponds to the magnetic vector-potential in the Weyl gauge.    

The electromagnetic Lagrangian density reads \cite{Soper_book, jackson1998ClassicalElectrodynamics}:
\begin{equation}
\mathcal{L}_{EM}=\frac{1}{8\pi}\!\left[\dot{\mathbfcal{A}}^2-c^2({\bm\nabla}\times\mathbfcal{A})^2\right]
= \frac{1}{8\pi}\!\left[ \mathbfcal{E}^2-\mathbfcal{H}^2\right]\,.
\label{LEM}
\end{equation}
Note the similarity with the sound Lagrangian \eqref{Lsound}, where the displacement and velocity fields are substituted with the vector-potential and electric field, whereas the divergence is replaced by the curl. 
The Euler-Lagrange equation for the Lagrangian \eqref{LEM} yields
\begin{equation}
\ddot{\mathbfcal{A}}= - c^2{\bm \nabla}\times({\bm \nabla}\times \mathbfcal{A})\,, 
\label{EOM_EM_1}
\end{equation}
wich, combined with the definition of the vector-potential, leads to Maxwell's equations:
\begin{equation}
\dot{\mathbfcal{E}} = c\,{\bm\nabla}\times{\mathbfcal{H}}\,,\quad
\dot{\mathbfcal{H}} = - c\,{\bm\nabla}\times{\mathbfcal{E}}\,.
\label{EOM_EM}
\end{equation}
These equations describe transverse electromagnetic waves with linear dispersion $\omega = kc$. The remaining Maxwell equations, ${\bm\nabla}\cdot \mathbfcal{E} = {\bm\nabla}\cdot \mathbfcal{H} =0$, follow from Eqs.~\ref{EOM_EM}. 

Since the Lagrangian density \eqref{LEM} has the standard form \eqref{Lagrangian}, the time-averaged momentum and spin densities in a monochromatic electromagnetic field take the forms \eqref{P_general} and \eqref{S_general}:
\begin{equation}
\overline{\bf P} = 
\frac{1}{8\pi\omega} \Im[{\bf E}^* \cdot ({\bm \nabla}) {\bf E}]\,,
\label{P_EM}
\end{equation}
\vspace{-0.5cm}
\begin{equation}
\overline{\bf S} = \frac{1}{8\pi\omega}
\Im({\bf E}^* \times {\bf E}) \,.
\label{S_EM}
\end{equation}
We expressed these quantities in terms of the complex electric field ${\bf E} = -i\omega {\bf A}$. 

The time-averaged electromagnetic energy, energy-flux, and kinetic-momentum densities \eqref{W_cons}--\eqref{Pkin} take the familiar forms: \cite{jackson1998ClassicalElectrodynamics}:
\begin{equation}
{W} = 
\frac{1}{8\pi}\!\left[\dot{\mathbfcal{A}}^2+c^2 ({\bm\nabla} \times {\mathbfcal{A}})^2 \right]
= \frac{1}{8\pi}\!\left[{\mathbfcal{E}}^2 + \mathbfcal{H}^2\right]\,,
\label{W_EM}
\end{equation}
\vspace{-0.5cm}
\begin{equation}
{\bf U} = - \frac{c^2}{4\pi}\,\dot{\mathbfcal{A}}\times({\bm\nabla}\times{\mathbfcal{A}}) = \frac{c}{4\pi}(\mathbfcal{E}\times \mathbfcal{H})\,, \quad
{\bm \Pi} = \frac{{\bf U}}{c^2}\,.
\label{U_EM}
\end{equation}
Here the energy flux density ${\bf U}$ is the Poynting vector. Using Maxwell's equations \eqref{EOM_EM}, one can verify that the energy and energy-flux densities \eqref{W_EM} and \eqref{U_EM} satisfy the energy conservation law \eqref{W_cons} (the Poynting theorem), whereas the time-averaged kinetic momentum $\overline{\bm \Pi} = \Re({\bf E}^*\times {\bf H})/(8\pi c)$ satisfies the Belinfante-Rosenfeld relation \eqref{BR} with the canonical momentum and spin densities \eqref{P_EM} and \eqref{S_EM} \cite{Berry2009, Bliokh2013NJP}.

For a plane wave \eqref{planewave}, the electromagnetic momentum, energy, and spin densities \eqref{P_EM}--\eqref{U_EM} yield
\begin{equation}
\overline{\bf P} = \overline{\bm \Pi} = \frac{\overline{W}}{\omega}{\bf k}\,, \quad
\overline{\bf S} = \frac{\sigma}{k}\,\overline{\bf P}\,,
\label{planewave_EM}
\end{equation}
where ${\bm \kappa} = {\bf k}/k$ and $\sigma = {\bm \kappa} \cdot\Im({\bf E}^*\times {\bf E})/|{\bf E}|^2 \in [-1,1]$ is the polarization {\it helicity}. For a plane electromagnetic wave, the helicity characterizes the degree of circular polarization in the plane orthogonal to the wave vector, with $\sigma=\pm 1$ corresponding to the right- and left-handed circular polarizations. Thus, an  electromagnetic plane wave generally carries nonzero spin, which corresponds to the spin-1 nature photons in quantum electrodynamics \cite{Akhiezer_book}.

For structured electromagnetic waves, starting with the simplest case of two-wave interference Fig.~\ref{Fig_two_wave}(b), the spin density $\overline{\bf S}$ can generally have an arbitrary direction, not attached to the wavevectors and helicities of the interfering waves \cite{Bliokh2015PR, Bekshaev2015PRX, Aiello2015NP}. However, the {\it integral} spin of interfering waves with the same helicity $\sigma=\pm 1$ is proportional to this helicity and integral momentum \cite{Bliokh2015PR}. In this case, the complex electric and magnetic fields satisfy ${\bf H} = -i\sigma {\bf E}$, which yields $\overline{\bf S} = \sigma \overline{\bm \Pi}/ k$ and $\int \overline{\bf S}\,d^3{\bf r} = (\sigma/ k) \int \overline{\bf P}\,d^3{\bf r}$, where we used Eq~\eqref{Pkin-P}. (Note that the kinetic momentum $\overline{\bm \Pi}$ generally depends on the helicity, so that $\overline{\bf S}$ is not strictly proportional to $\sigma$. In contrast, the canonical momentum $\overline{\bf P}$ is independent of $\sigma$.)

Notably, there are alternative ways to define the canonical momentum and spin densities satisfying the Belinfante-Rosenfeld relation \eqref{BR} with the same kinetic-momentum density \eqref{U_EM}. In particular, one can symmetrize expressions \eqref{P_EM} and \eqref{S_EM} between the electric and magnetic field contributions:
\begin{equation}
\overline{\bf P}' = 
\frac{1}{16\pi\omega} \Im[{\bf E}^* \cdot ({\bm \nabla}) {\bf E} + {\bf H}^* \cdot ({\bm \nabla}) {\bf H}]\,,
\label{P_EM_D}
\end{equation}
\vspace{-0.5cm}
\begin{equation}
\overline{\bf S}' = \frac{1}{16\pi\omega}
\Im({\bf E}^* \times {\bf E} + {\bf H}^* \times {\bf H}) \,.
\label{S_EM_D}
\end{equation}
This symmetrization, advocated in a number of studies \cite{Berry2009, Cameron2012, Bliokh2013NJP, Bliokh2014NC, Bliokh2015PR, Aiello2015NP, Bliokh2017NJP}, maintains the {\it dual symmetry} between the electric and magnetic fields in Maxwell's equations \eqref{EOM_EM}, as well as in the electromagnetic energy and energy-flux densities \eqref{W_EM} and \eqref{U_EM}. To derive the dual-symmetric quantities \eqref{P_EM_D} and \eqref{S_EM_D} within the Lagrangian framework, one has to modify the electromagnetic Lagrangian \eqref{LEM} by introducing two vector-potentials and additional constraints \cite{Bliokh2013NJP, Cameron2012NJP_II}. However, this approach is valid only in free space. As we will see in Section~\ref{Plasma}, the presence of electric charges breaks the electric-magnetic symmetry, and is consistent with the electric-biased approach based on the standard Lagrangian \eqref{LEM} and canonical densities \eqref{P_EM} and \eqref{S_EM}. 

Ultimately, the choice between the standard convention \eqref{P_EM} and \eqref{S_EM} and the dual-symmetric approach \eqref{P_EM_D} and \eqref{S_EM_D} depends on the specific problem under consideration. For example, when considering properties of the {\it electric} optical field, or its interaction with small {\it electric}-dipole particles, the standard electric-biased approach is more appropriate \cite{Neugebauer2015PRL, Antognozzi2016NP}. However, for problems essentially involving both electric and magnetic fields, such as interactions with both electric-dipole and magnetic-dipole particles, it is useful to consider electric and magnetic contributions to the momentum and spin densities \cite{Bliokh2014NC, Neugebauer2018PRX, Toftul2024}. Even in such cases, the problem is never perfectly symmetric between the electric and magnetic contributions, and one has to consider the electric and magnetic momentum or spin densities separately.

%%%%%%%%%%%%%%%%%%%%%%%%%%%%%%
\section{Elastic waves}
\label{Elastic}
%%%%%%%%%%%%%%%%%%%%%%%%%%%%%%

\begin{figure}[t!]
\centering
\includegraphics[width=0.65\linewidth]{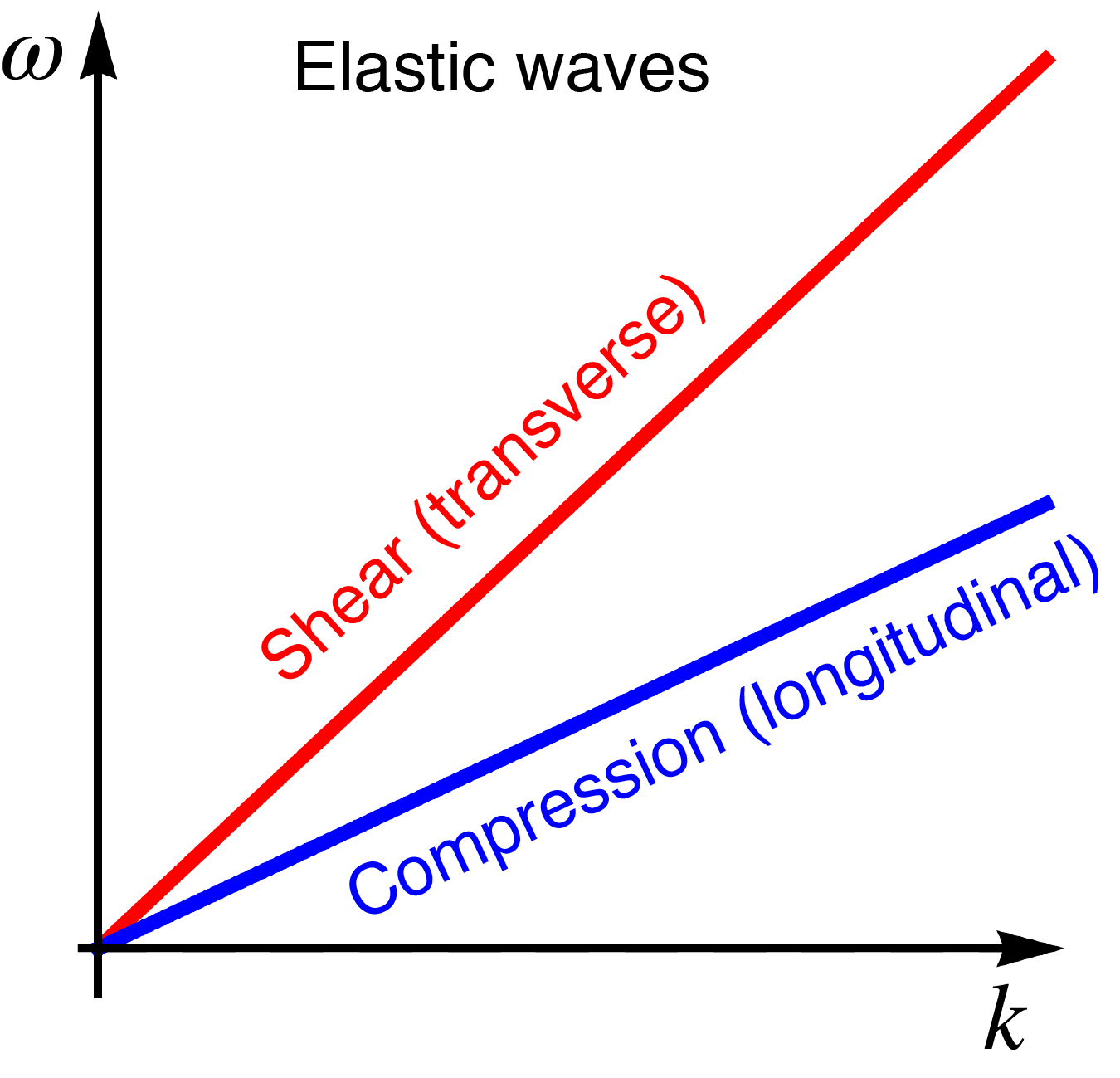}
\caption{Dispersions of elastic compression and shear waves.} 
\label{Fig_elastic}
\end{figure}

Next, we consider more complex systems that exhibit mixed wave fields with both longitudinal and transverse contributions. 
The first such example is linear elastic waves in an isotropic solid \cite{LL_elasticity, Auld_book}. There are two elastic modes: longitudinal compression waves and transverse shear waves. These are mechanical waves in a medium, which are described by the local displacement of the medium particles, $\mathbfcal{R}({\bf r},t)$. The corresponding Lagrangian density is given by
\begin{equation}
\mathcal{L}_e=\frac{\rho}{2} \dot{\mathbfcal{R}}^2
-\frac{\lambda}{2}({\bm\nabla} \cdot \mathbfcal{R})^2
-{\mu}\mathcal{S}_{ij}\mathcal{S}_{ij}
%\left(\nabla_j \mathcal{R}_i + \nabla_i \mathcal{R}_j \right) \left(\nabla_j \mathcal{R}_i + \nabla_i \mathcal{R}_j \right)
\,,
\label{Lelastic}
\end{equation}
where $\mathcal{S}_{ij} = \left(\nabla_j \mathcal{R}_i + \nabla_i \mathcal{R}_j \right)\!/2$, $\lambda$ and $\mu$ are the constant Lam\'{e} coefficients, and summation over repeated indices is assumed.
The Euler-Lagrange equation, corresponding to Eq.~\eqref{Lelastic}, takes the form
\begin{equation}
\rho \ddot{\mathbfcal{R}} = (\lambda + 2\mu) {\bm\nabla}({\bm\nabla} \cdot \mathbfcal{R}) 
- \mu {\bm\nabla}\times ({\bm\nabla}\times \mathbfcal{R})
\,.
\label{EOM_elastic}
\end{equation}
The two terms on the right-hand side of Eq.~\eqref{EOM_elastic} are similar to the corresponding terms for sound and electromagnetic waves, Eqs.~\eqref{EOM_sound_1} and \eqref{EOM_EM_1}. 
For longitudinal compression waves with ${\bm\nabla}\times \mathbfcal{R} = {\bf 0}$, this equation reduces to a sound-like wave equation with dispersion $\omega = k c_l$, $c_l = \sqrt{(\lambda + 2\mu)/\rho}$. For transverse shear waves with ${\bm\nabla}\cdot \mathbfcal{R} = 0$, Eq.~\eqref{EOM_elastic} yields the electromagnetic-like wave equation with dispersion $\omega = k c_t$, $c_t = \sqrt{\mu/\rho}$. These two elastic modes generally coexist but propagate at different speeds, Fig.~\ref{Fig_elastic}. 

%FFFFFFFFFFFFFFFFFFFFFFFFFFFFFFFFFFFFFFFFFFFFFF
%FFFFFFFFFFFFFFFFFFFFFFFFFFFFFFFFFFFFFFFFFFFFFF

Since the Lagrangian density \eqref{Lelastic} has the familiar form \eqref{Lagrangian}, the time-averaged canonical momentum and spin densities in a monochromatic elastic field are obtained from Eqs.~\eqref{P_general} and \eqref{S_general} \cite{Nakane2018PRB, Long2018PNAS, Bliokh2022PRL, Ren2022CPL}:
\begin{equation}
\overline{\bf P} = 
\frac{\rho\omega}{2} \Im[{\bf R}^* \cdot ({\bm \nabla}) {\bf R}]\,,
\label{P_elastic}
\end{equation}
\vspace{-0.5cm}
\begin{equation}
%\overline{\bf L} = {\bf r} \times \overline{\bf P} \,, \quad
\overline{\bf S} = \frac{\rho \omega}{2}
\Im({\bf R}^* \times {\bf R}) \,.
\label{S_elastic}
\end{equation}
These equations are identical to Eqs.~\eqref{P_sound} and \eqref{S_sound} because elastic and sound waves are described by the same kinetic energy density, and their intrinsic spin AM density is determined by the same mechanical AM of the medium particles. 

In turn, the time-averaged energy and energy-flux densities \eqref{W_cons}--\eqref{U} acquire different forms (determined by the potential energy) \cite{Auld_book}:
\begin{equation}
{W} = 
\frac{\rho}{2} \dot{\mathbfcal{R}}^2
+\frac{\lambda}{2}({\bm\nabla} \cdot \mathbfcal{R})^2
+{\mu}\mathcal{S}_{ij}\mathcal{S}_{ij}\,,
\label{W_elastic}
\end{equation}
\vspace{-0.5cm}
\begin{equation}
{\bf U} = - \lambda\, \dot{\mathbfcal{R}} ({\bm\nabla} \cdot {\mathbfcal{R}}) -\mu\! \left[ (\dot{\mathbfcal{R}} \cdot {\bm\nabla}) {\mathbfcal{R}} + \dot{\mathbfcal{R}} \cdot ({\bm\nabla}) {\mathbfcal{R}}\right].
\label{U_elastic}
\end{equation}
One can verify that these quantities satisfy the energy conservation law \eqref{W_cons}.

However, we cannot use Eq.~\eqref{Pkin} for the kinetic energy density, because the longitudinal and transverse elastic modes have different velocities. Instead, we determine the time-averaged kinetic momentum density in a monochromatic field using the Belinfante-Rosenfeld relation \eqref{BR} along with the canonical momentum and spin densities \eqref{P_elastic} and \eqref{S_elastic}. This results in
\begin{equation}
\overline{\bm \Pi} = \frac{\rho\omega}{2} \Im[{\bf R}^* ({\bm \nabla}\cdot  {\bf R}) + 
{\bf R}^* \times ({\bm \nabla}\times  {\bf R})]
\equiv \overline{\bm \Pi}_l + \overline{\bm \Pi}_t\,.
\label{Pkin_elastic}
\end{equation}
This expression consists of two terms corresponding to the longitudinal and transverse elastic modes, similar to the kinetic momentum densities in sound and electromagnetic waves, Eqs.~\eqref{U_sound} and \eqref{U_elastic}. In turn, the time-averaged energy flux density \eqref{U_elastic} can be written, after some algebra, as 
\begin{equation}
\overline{\bf U} = c_l^2 \overline{\bm \Pi}_l + c_t^2 \overline{\bm \Pi}_t - c_t^2 {\bm \nabla}\times \overline {\bf S}\,.
\label{U_elastic_2}
\end{equation}
Here the last term does not contribute to the energy conservation law \eqref{W_cons} because its divergence vanishes. (The energy flux density is generally defined up to the curl of a vector field.) If we ignore it, the energy flux density \eqref{U_elastic_2} becomes a sum of the longitudinal-mode and transverse-mode terms, each associated with the corresponding kinetic-momentum density and velocity coefficient. This provides a generalization of Eq.~\ref{Pkin} to the case of two modes with different velocities. 

For plane compression and shear waves, we obtain the elastic counterparts of Eqs.~\eqref{planewave_s} and \eqref{planewave_EM}:
\begin{align}
{\rm compression:} & \quad \overline{\bf P} = \overline{\bm \Pi} = \frac{\overline{W}}{\omega}{\bf k}\,, \quad
\overline{\bf S} = {\bf 0}\,, \nonumber \\
{\rm shear:} & \quad \overline{\bf P} = \overline{\bm \Pi} = \frac{\overline{W}}{\omega}{\bf k}\,, \quad
\overline{\bf S} = \frac{\sigma}{k}\,\overline{\bf P}\,.
\label{planewave_elastic}
\end{align}
Here $\sigma = {\bm \kappa} \cdot\Im({\bf R}^*\times {\bf R})/|{\bf R}|^2 \in [-1,1]$ is the polarization helicity of transverse elastic waves.

Since compression and shear oscillations coexist in a generic elastic wavefield (e.g., in surface Rayliegh waves \cite{LL_elasticity, Auld_book}), the total displacement field  can be expressed as the sum of two contributions: ${\bf R} = {\bf R}_l + {\bf R}_t$. Substituting this into the canonical momentum and spin densities \eqref{P_elastic} and \eqref{S_elastic}, we find that these quadratic quantities include pure compression, pure shear, and {\it hybrid} contributions \cite{Long2018PNAS, Bliokh2022PRL, Yang2023PRL}: 
\begin{align}
\overline{\bf P} = \overline{\bf P}_l + \overline{\bf P}_t + \overline{\bf P}_h\,, \quad
\overline{\bf S} = \overline{\bf S}_l + \overline{\bf S}_t + \overline{\bf S}_h\,.
\label{hybrid}
\end{align}
Here the hybrid terms involve products of the fields ${\bf R}_l$ and ${\bf R}_t$, leading to effects that are absent in purely longitudinal or purely transverse waves \cite{Yang2023PRL}. 

%%%%%%%%%%%%%%%%%%%%%%%%%%%%%%
\section{Plasma waves}
\label{Plasma}
%%%%%%%%%%%%%%%%%%%%%%%%%%%%%%

So far, we have considered examples with linear dispersion relations, where the phase and group velocities coincide. The next example exhibits both longitudinal and transverse modes with nonlinear dispersion relations. Namely, we examine an isotropic collisionless plasma \cite{Akhiezer_plasma, Krall_book}, where electrons move freely against the background of heavy stationary ions that neutralize the unperturbed electron charge density. As in any gas, sound waves can propagate in the electron gas. These waves cause local perturbations of the electron density and, hence, of the electric charge density. This, in turn, generates oscillating electric fields which obey Maxwell's equations in the presence of charges. 

\begin{figure}[t!]
\centering
\includegraphics[width=0.65\linewidth]{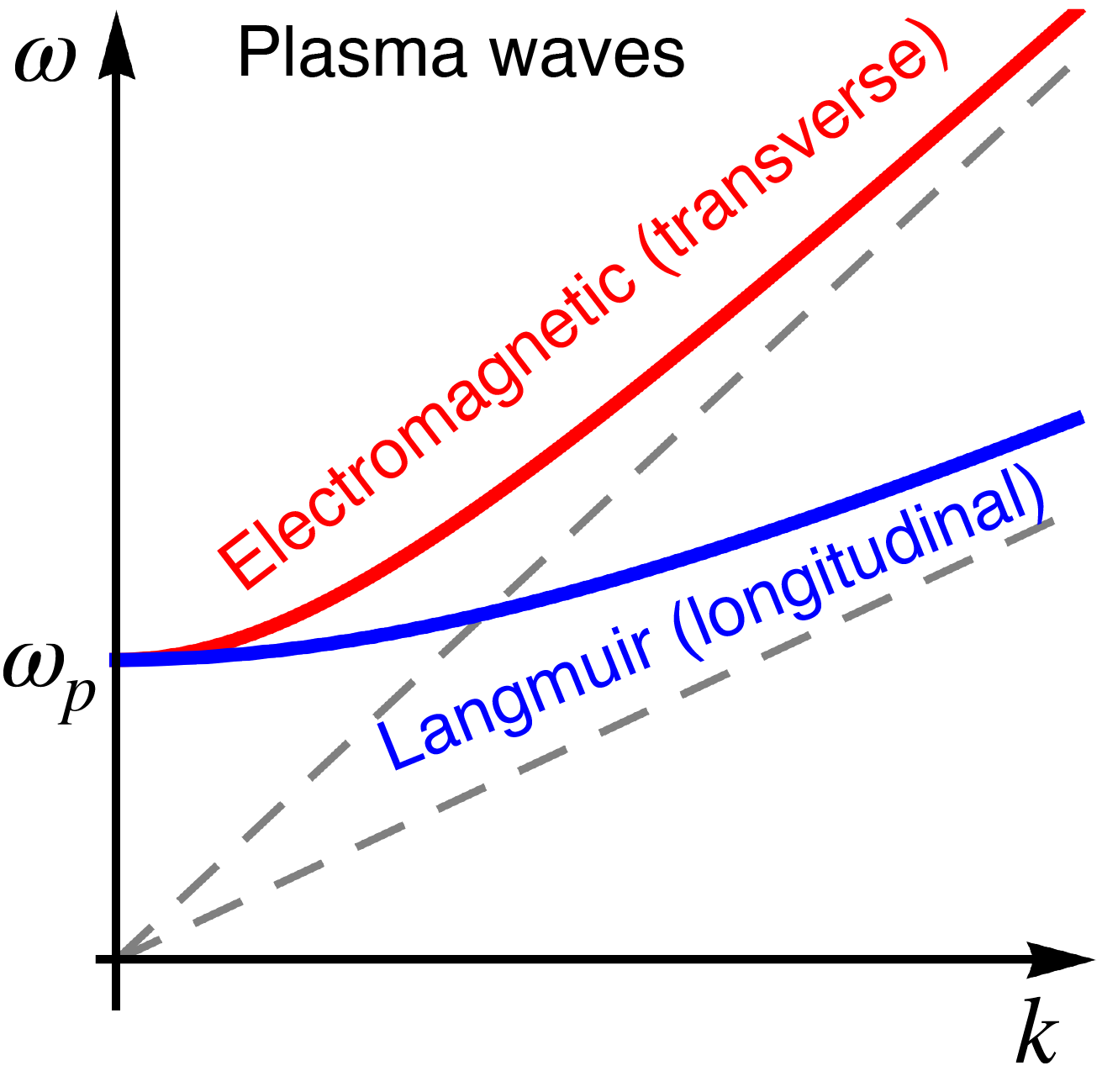}
\caption{Dispersions of electromagnetic and Langmuir waves in an isotropic collisionless electron plasma.} 
\label{Fig_plasma}
\end{figure}

Thus, plasma waves are described by the sound-like Lagrangian \eqref{Lsound} for the electron gas, combined with the electromagnetic-field Lagrangian \eqref{LEM}, and a coupling term between these two parts:   
\begin{equation}
\mathcal{L}_p=\mathcal{L}_s + \mathcal{L}_{EM}
+\frac{e\rho}{m}\dot{\mathbfcal{R}}\cdot \mathbfcal{A}\,,
\label{Lplasma}
\end{equation}
where $m$ and $e<0$ are the electron mass and charge, respectively, $\rho$ is the mass density of the electron gas, whereas the sound-velocity constant in $\mathcal{L}_s$ can be expressed via the temperature $T$ of the electron gas: $c_s=\sqrt{3k_B T/m}$ ($k_B$ is the Boltzmann constant).
The corresponding Euler-Lagrange equations yield two coupled wave equations, analogous to Eqs.~\eqref{EOM_sound_1} and ~\eqref{EOM_EM_1}: 
\begin{align}
\ddot{\mathbfcal{R}}& = c_s^2{\bm \nabla} ({\bm \nabla}\cdot {\mathbfcal{R}}) - \frac{e}{m}\dot {\mathbfcal{A}}\,, \nonumber \\
\ddot{\mathbfcal{A}}& = - c^2{\bm \nabla} \times ({\bm \nabla}\times {\mathbfcal{A}}) + \frac{4\pi e \rho}{m}\dot {\mathbfcal{R}}\,.
\label{EOMplasma}
\end{align}
Using the definitions of the velocity, pressure, electric, and magnetic fields in Sections~\ref{Sound} and \ref{Electromagnetic}, these equations can be rewritten as
\begin{align}
\rho\dot{\mathbfcal{V}}& = -{\bm \nabla} \mathcal{P} + \frac{e\rho}{m}{\mathbfcal{E}}\,, \qquad
\beta \dot{\mathcal{P}} = - {\bm \nabla} \cdot {\mathbfcal{V}}\,, \nonumber \\
\dot{\mathbfcal{E}} &= c\,{\bm \nabla}\times {\mathbfcal{H}} - \frac{4\pi e \rho}{m} {\mathbfcal{V}}\,, \quad
\dot{\mathbfcal{H}} = - c\,{\bm \nabla}\times {\mathbfcal{E}}\,,
\label{EOMplasma_2}
\end{align}
where $\beta= 1/(\rho c_s^2)$. These are the well-known plasma wave equations\cite{Akhiezer_plasma, Krall_book, Bliokh2022PRE}, which support two modes, Fig.~\ref{Fig_plasma}: \\
(i) Longitudinal Langmuir waves, satisfying 
\begin{equation}
{\bm \nabla}\times {\mathbfcal{E}} = {\bm \nabla}\times {\mathbfcal{V}}= {\mathbfcal{H}} = {\bf 0} \quad {\rm and} \quad \omega^2 = \omega_p^2 + c_s^2 k^2\,,
\label{Langmuir_mode}
\end{equation}
where $\omega_p = \sqrt{4\pi\rho e^2/m^2}$ is the plasma frequency. \\
(ii) Transverse electromagnetic waves, satisfying  
\begin{equation}
{\bm \nabla} \cdot {\mathbfcal{V}} = {\bm \nabla} \cdot {\mathbfcal{E}} = \mathcal{P} = 0 \quad {\rm and} \quad \omega^2 = \omega_p^2 + c^2 k^2\,.
\label{Electromagnetic_mode}
\end{equation}
These modes are dispersive: their group and phase velocities differ from each other. Notably, the product of the group and phase velocities, $c_g c_{ph}$, equals $c_s^2$ for Langmuir waves and $c^2$ for electromagnetic waves. 

%FFFFFFFFFFFFFFFFFFFFFFFFFFFFFFFFFFFFFFFFFFFFFF
%FFFFFFFFFFFFFFFFFFFFFFFFFFFFFFFFFFFFFFFFFFFFFF

The Lagrangian density \eqref{Lplasma} does not have the usual ``kinetic-potential'' form \eqref{Lagrangian} because of the additional coupling term. Applying the definitions \eqref{NoetherP} and \eqref{NoetherJ}, we derive the time-averaged canonical momentum and spin densities in plasma waves:
\begin{align}
\overline{\bf P} & = 
\frac{\rho}{2\omega} \Im[{\bf V}^* \cdot ({\bm \nabla}) {\bf V}]
+ \frac{1}{8\pi\omega} \Im[{\bf E}^* \cdot ({\bm \nabla}) {\bf E}] \nonumber \\
& + \frac{e\rho}{2m\omega^2} \Re[{\bf E}^* \cdot ({\bm \nabla}) {\bf V}]\,,
\label{P_plasma}
\end{align}
\vspace{-0.5cm}
\begin{align}
\overline{\bf S} & = \frac{\rho}{2\omega}
\Im({\bf V}^* \times {\bf V}) 
+ \frac{1}{8\pi\omega}
\Im({\bf E}^* \times {\bf E}) \nonumber \\
& + \frac{e\rho}{2m\omega^2} \Re({\bf E}^* \times {\bf V})\,.
\label{S_plasma}
\end{align}
These equations represent the sums of the sound and electromagnetic canonical densities \eqref{P_sound}, \eqref{S_sound} and \eqref{P_EM}, \eqref{S_EM}, plus additional coupling terms involving both the velocity and electric fields. Remarkably, for purely longitudinal Langmuir waves \eqref{Langmuir_mode}, $i \omega {\bf E} = (4\pi e \rho/m) {\bf V}$, the electromagnetic and coupling terms cancel out, reducing Eqs.~\eqref{P_plasma} and \eqref{S_plasma} to sound-wave Eqs.~\eqref{P_sound} and \eqref{S_sound}: $\overline{\bf P} = \overline{\bf P}_s$, $\overline{\bf S} = \overline{\bf S}_s$ (expressed via velocity ${\bf V}= -i\omega {\bf R}$). In turn, for purely transverse electromagnetic waves \eqref{Electromagnetic_mode}, $i \omega {\bf V} = - (e/m) {\bf E}$, the sound and coupling terms cancel out, reducing Eqs.~\eqref{P_plasma} and \eqref{S_plasma} to electromagnetic Eqs.~\eqref{P_EM} and \eqref{S_EM}: $\overline{\bf P} = \overline{\bf P}_{EM}$, $\overline{\bf S} = \overline{\bf S}_{EM}$.

From the Lagrangian density \eqref{Lplasma}, together with Eqs.~\eqref{W} and \eqref{U}, we find that the energy and energy-flux densities of plasma waves represent the sums of the sound and electromagnetic terms \eqref{W_sound}, \eqref{U_sound} and \eqref{W_EM}, \eqref{U_EM}, without additional coupling terms:
\begin{equation}
{W} 
= \frac{1}{2}\left[\rho {\mathbfcal{V}}^2+\beta \mathcal{P}^2\right] + 
\frac{1}{8\pi}\!\left[{\mathbfcal{E}}^2 + \mathbfcal{H}^2\right]\equiv W_s + W_{EM}\,,
\label{W_plasma}
\end{equation}
\vspace{-0.5cm}
\begin{equation}
{\bf U}  = \mathcal{P}{\mathbfcal{V}}+ \frac{c}{4\pi}(\mathbfcal{E}\times \mathbfcal{H}) \equiv {\bf U}_s + {\bf U}_{EM}\,.
\label{U_plasma}
\end{equation}
As in elastic waves, we obtain the time-averaged kinetic-momentum density by substituting Eqs.~\eqref{P_plasma} and \eqref{S_plasma} into the Belinfante-Rosenfeld relation \eqref{BR}.
The resulting general expression is rather cumbersome, but one can verify that for purely longitudinal Langmuir and for purely transverse electromagnetic waves
%, Eqs.~\eqref{Langmuir_mode} and \eqref{Electromagnetic_mode}, with $\overline{\bf U} = \overline{\bf U}_s$ and $\overline{\bf U} = \overline{\bf U}_{EM}$, respectively, 
it satisfies relation \eqref{Pkin} \cite{Bliokh2022PRE}:
\begin{equation}
\overline{\bm \Pi} = \frac{\overline{\bf U}_s}{c_s^2}
\quad {\rm and} \quad
\overline{\bm \Pi} = \frac{\overline{\bf U}_{EM}}{c^2}\,.
\label{Pkin_plasma}
\end{equation}
Here the use of the products of the group and phase velocities is essential.

Finally, the plane-wave values of the momentum and spin densities in Langmuir and electromagnetic waves are given by the sound and electromagnetic Eqs.~\eqref{planewave_s} and \eqref{planewave_EM}, respectively.

%%%%%%%%%%%%%%%%%%%%%%%%%%%%%%
\subsection{The Abraham-Minkowski problem}
\label{AM}
%%%%%%%%%%%%%%%%%%%%%%%%%%%%%%

Electromagnetic waves in plasma provide an example of electromagnetic waves in a dispersive medium. Therefore, the famous Abraham-Minkowski dilemma concerning the momentum of electromagnetic waves in a medium \cite{Pfeifer2007, Milonni2010AOP, Barnett2010PTRS, Partanen2017PRA, Bliokh2017NJP} becomes relevant. In the simplest case of an electromagnetic plane wave in a non-dispersive medium with dielectric permittivity $\varepsilon$, the Abraham momentum is equivalent to the Poynting momentum and is given by $\overline{\bf P}_A = \overline{\bm \Pi} = (\overline{W}/\omega)\, {\bf k}_0/n$, where $\overline{\bm \Pi} = {\rm Re}({\bf E}^* \times {\bf H})/(8\pi c)$, $\overline{W} = (\varepsilon |{\bf E}|^2 + |{\bf H}|^2)/(16\pi\omega)$ is the electromagnetic energy density in the medium, ${\bf k}_0 = {\bm \kappa}\,\omega/c$, and $n=\sqrt{\varepsilon}$ is the refractive index of the medium. In turn, the Minkowski momentum is given by $\overline{\bf P}_M = \varepsilon\, \overline{\bm \Pi} = (\overline{W}/\omega)\, n{\bf k}_0$.

However, plasma is a dispersive medium, where permittivity depends on the frequency: $\varepsilon (\omega) = c^2k^2/\omega^2 = 1 - \omega_p^2/\omega^2$ \cite{Akhiezer_plasma, Krall_book}. In dispersive media, the Abraham-Minkowski dilemma acquires significant corrections \cite{Barnett2010PTRS, Philbin2011PRA, Bliokh2017NJP}. Specifically, the Abraham momentum involves the {\it group} refractive index $n\to n_g = c/c_g$, while the Minkowski momentum is determined by the {\it phase} refractive index $n = n_{ph} = c/c_{ph}$. Remarkably, for electromagnetic waves in plasma, $c_g c_{ph} = c^2$ leads to $n_g = 1/n_{ph}$. As a result, the Abraham and Minkowski plane-wave momenta become {\it identical} in plasma, so that the dilemma disappears \cite{Bliokh2022PRE}.

This intriguing result is unique to plasma-like dispersions and does not hold in other dispersive media, such as those composed of neutral dipole-like atoms. Unfortunately, such media cannot be easily described using an exact microscopic Lagrangian involving electrons and fields, as is possible for plasma. Instead, they require phenomenological macroscopic approaches \cite{Philbin2011PRA, Bliokh2017NJP}.

%%%%%%%%%%%%%%%%%%%%%%%%%%%%%%
\section{Water surface waves}
\label{Water}
%%%%%%%%%%%%%%%%%%%%%%%%%%%%%%

Finally, we turn our attention to the most familiar type of waves: water waves \cite{Falkovich_book, Landau_fluid}. These surface waves exhibit both the longitudinal and transverse field components within a single mode having a nonlinear dispersion. Before analyzing this system, it is worth recalling a famous remark by
Richard P. Feynman \cite{Feynman_I}: ``{\it The next waves of interest, that are easily seen by everyone and which are usually used as an example of waves in elementary courses, are water waves. As we shall soon see, they are the worst possible example, because they are in no respects like sound and light; they have all the complications that waves can have.}'' Despite these complications, we will demonstrate that the momentum and AM properties of water waves share have much in common with those in sound and light.

%FFFFFFFFFFFFFFFFFFFFFFFFFFFFFFFFFFFFFFFFFFFFFF
\begin{figure}[t!]
\centering
\includegraphics[width=0.9\linewidth]{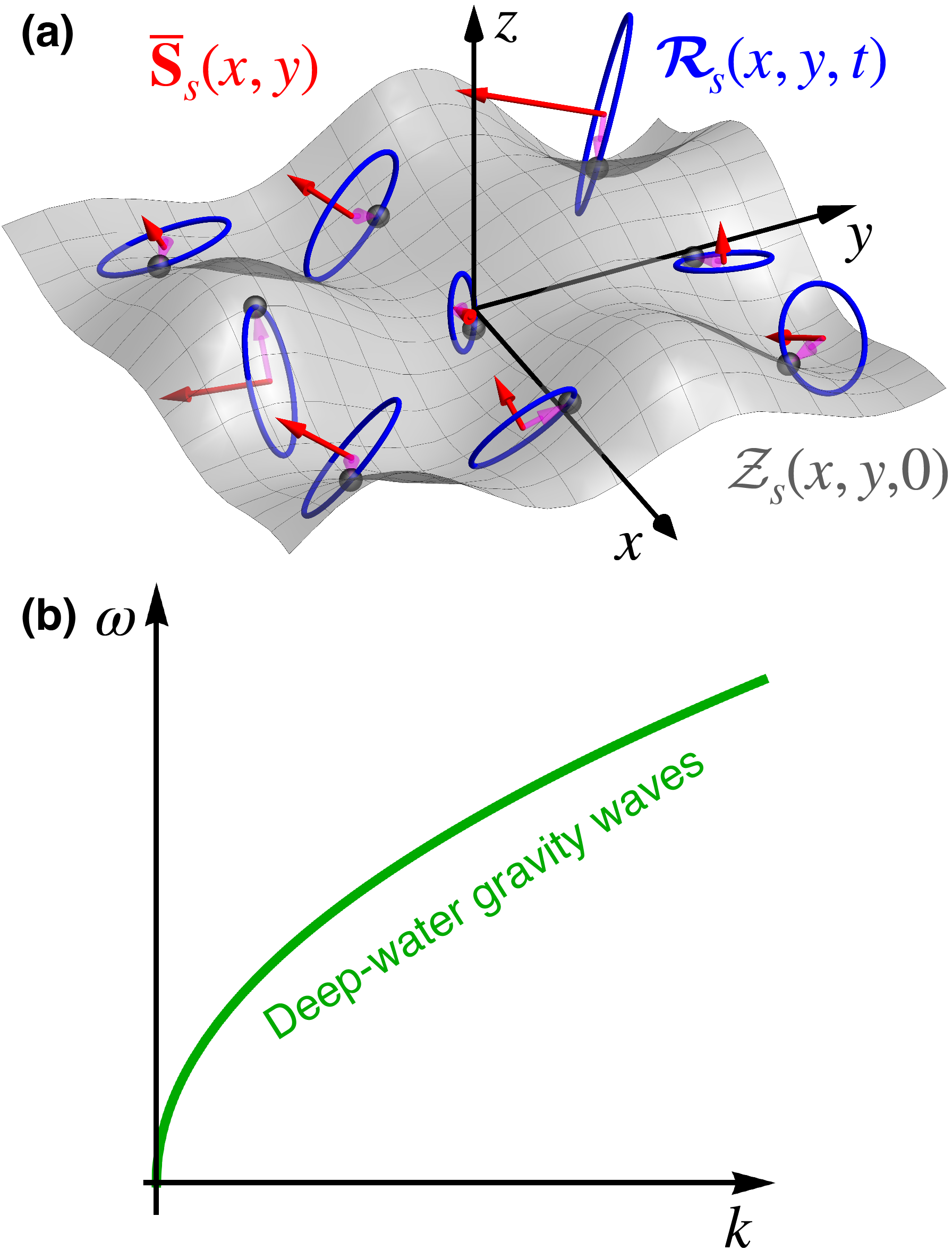}
\caption{(a) Schematics of monochromatic structured water surface waves. The figure illustrates: the instantaneous wave-perturbed water surface $z= {\mathcal{Z}}_s(x,y,t=0)$ (gray), local elliptical trajectories (polarizations) of water-surface particles traced by the 3D displacement $\mathbfcal{R}_s (x,y,t)$ (blue), and the corresponding time-averaged spin AM density $\overline{\bf S}_s (x,y)$ (red). See also Figs.~\ref{Fig_ellipse} and \ref{Fig_momentum_spin}(b).
(b) Dispersion of deep-water gravity water waves.} 
\label{Fig_water}
\end{figure}
%FFFFFFFFFFFFFFFFFFFFFFFFFFFFFFFFFFFFFFFFFFFFFF

Note that surface waves are inherently {\it anisotropic} and cannot be described within the isotropic 3D approach employed in previous sections. The wave momentum lies within the 2D propagation plane $(x,y)$, but the wave field and polarization/spin still have a 3D nature, see Fig.~\ref{Fig_water}(a). Furthermore, the system's symmetries include 2D in-plane translations and SO(2) rotations about the orthogonal $z$-axis, implying that only the $(x,y)$-components of the wave momentum and the $z$-component of the AM are conserved. 

These peculiarities make constructing a natural Lagrangian field-theory formalism for water-surface waves challenging.
Therefore, we will focus on the wave equations of motion, and restrict ourselves to monochromatic deep-water gravity waves. 

Akin to sound and elastic waves, water waves can be described by the 3D displacement of water particles $\mathbfcal{R}({\bf r},t)$.
It is well-known that, in the linear approximation, water-surface particles follow local circular orbits during the propagation of a plane gravity wave. These orbits are based on vertical $z$-oscillations and $\pi/2$-phase-shifted longitudinal (along the wave vector) oscillations, Figs.~\ref{Fig_Intro}(b) and \ref{Fig_wave_types}(c). Thus, the wave field $\mathbfcal{R}({\bf r},t)$ contains both the transverse and longitudinal components with respect to the wave vector. However, for an incompressible fluid, the velocity and displacement fields satisfy the transverse-wave condition ${\bm\nabla} \cdot \mathbfcal{R} = 0$. Moreover, the linearized Euler equation with the potential gravity term yields ${\bm\nabla} \times \mathbfcal{R} = {\bf 0}$. Thus, gravity waves are formally both transverse and longitudinal at the same time \cite{Yang2023PRL}. This apparent contradiction is resolved by noticing that a surface plane wave can be considered as a plane wave with a {\it complex} wave vector, which includes an imaginary vertical component. For a deep-water gravity plane wave propagating along the $x$-axis, the wave field has the form $\mathbfcal{R} = \Re[{\bf R}_0\exp(ik x + kz-i\omega t)] \equiv \Re[{\bf R}_0\exp(i{\bf k}\cdot {\bf r}-i\omega t)]$. Here both the longitudinal and transverse wave conditions are satisfied, ${\bf k} \cdot {\bf R}_0 =0$ and ${\bf k} \times {\bf R}_0 = {\bf 0}$, with the field $R_{0x}=i R_{0z}$ containing both the longitudinal and transverse components. 

For a generic monochromatic water-wave field, representing a superpositions of plane waves with different directions and amplitudes, the $(x,y)$-dependence of the field can be complex, the local particle orbits (polarizations) represent generic 3D ellipses, Fig.~\ref{Fig_water}(a), but the $z$-dependence of the field remains the same: $\mathbfcal{R} = \Re[{\bf R}(x,y)\exp(kz-i\omega t)]$. 
Due to such anisotropic character of the problem, it is useful to separate the in-plane and vertical components of the field. Moreover, since the $z$-dependence is known, we can focus on the water-surface particles at $z=0$. In this manner, we introduce the ``(2+1)-dimensional'' notations:
\begin{align}
{\bf r}_s &= (x,y)\,, \quad {\bm\nabla}_s = (\nabla_x, \nabla_y)\,, 
\nonumber \\
\mathbfcal{R}_s({\bf r}_s,t) &= \left.(\mathcal{X}, \mathcal{Y})\right|_{z=0}\,, \quad \mathcal{Z}_s({\bf r}_s,t) = \left.\mathcal{Z}\right|_{z=0}\,. \nonumber
\end{align}
In these notations, water-surface particles obey the following equations of motion \cite{Bliokh2022SA}:
\begin{equation}
\ddot{\mathbfcal{R}}_s = -g\, {\bm \nabla}_s \mathcal{Z}_s\,, \qquad
\ddot{\mathcal{Z}}_s = g\, {\bm \nabla}_s \cdot {\mathbfcal{R}}_s\,, 
\label{EOM_water}
\end{equation}
where $g$ is the gravitational acceleration. Notably, Eqs.~\eqref{EOM_water} resemble the 2D sound-wave equations \eqref{EOM_sound} with the second time derivatives. For monochromatic waves, $\partial^2/\partial t^2 = - \omega^2$, and Eqs.~\eqref{EOM_water} describe gravity waves with dispersion $\omega^2 = kg$, Fig.~\ref{Fig_water}(b). The phase and group velocities are given by $c_{ph} = \sqrt{g/k}$ and $c_{g} = \sqrt{g/k}/2$, so that $c_{ph} c_{g} = g/2k$.

Since the kinetic properties of water-surface particles are similar to those in sound and elastic waves, we can write the time-averaged canonical momentum and spin densities in the same form as is Eqs.~\eqref{P_sound}, \eqref{S_sound}, \eqref{P_elastic}, and \eqref{S_elastic}. However, to accommodate these densities to our formalism, we use 2D surface densities obtained by the $z$-integration of the 3D volume densities, considering the 2D in-plane momentum and the vertical $z$-component of the AM.  
As a result, we derive \cite{Bliokh2022SA}:  
\begin{equation}
\overline{\bf P}_s = 
\frac{\rho\omega}{4k} \Im[{\bf R}_s^* \cdot ({\bm \nabla}_s) {\bf R}_s + Z_s^* {\bm \nabla}_sZ_s]\,, 
\label{P_water}
\end{equation}
\vspace{-0.2cm}
\begin{equation}
\overline{\bf S}_{sz} = \frac{\rho \omega}{4k}
\Im({\bf R}_s^* \times {\bf R}_s) \,.
\label{S_water}
\end{equation}
where $\rho$ is the water density, and the $z$-integration of the $\exp(2kz)$ dependencies introduces the factors $1/2k$.

Next, we examine the 2D energy and energy-flux densities in gravity water waves. The energy density can be written as the sum of the regular kinetic energy and potential energy due to gravity (caused by the elevation of the water surface) \cite{Peskin2010}: 
\begin{equation}
W_s = \frac{\rho}{4k}\left( \mathbfcal{V}_s^2 + \mathcal{V}_{zs}^2\right) +\frac{1}{2}\rho g \mathcal{Z}_s^2 \,,
\label{W_water}
\end{equation}
where $\mathbfcal{V}_s = \dot{\mathbfcal{R}_s}$ and $\mathcal{V}_{zs} = \dot{\mathcal{Z}}_s$. 
Using Eqs.~\eqref{EOM_water}, one can verify that this expression satisfies the 2D energy conservation equation $\partial W/\partial t + {\bm\nabla}_s \cdot {\bf U}_s =0$, with the energy-flux density and corresponding kinetic-momentum density given by:
\begin{equation}
{\bf U}_s = \frac{\rho g}{2k}\mathcal{Z}_s\mathbfcal{V}_s\,, \quad
{\bm \Pi}_s=\frac{{\bf U}_s}{c_{ph}c_g } = {\rho}\mathcal{Z}_s\mathbfcal{V}_s\,.
\label{U_water}
\end{equation}
Here the kinetic-momentum density takes the natural form of the 2D mass density excess ${\rho}\mathcal{Z}_s$ (due to the water-surface elevation) multiplied by the in-plane surface velocity $\mathbfcal{V}_s$ \cite{Peskin2010}.

Performing time averaging of the kinetic-momentum density and using Eqs.~\eqref{EOM_water}, we find that the canonical momentum and spin densities \eqref{P_water} and \eqref{S_water} satisfy the 2D version of the Belinfante-Rosenfeld relation \eqref{BR}:
\begin{equation}
\overline{\bm \Pi}_s = \overline{\bf P}_s + \frac{1}{2}{\bm\nabla}_s \times \overline{\bf S}_{sz} = 
\frac{\rho}{2} \Re(Z_s^* {\bf V}_s)\,.
\label{Pkin_water}
\end{equation}
Thus, the general relations between the canonical densities, the energy conservation law, and the kinetic momentum density hold even in the nontrivial case of dispersive surface waves.

A significant advantage of water waves, compared to optical and acoustic fields, is that one can directly observe the evolution of water-surface particles, i.e., time-dependent vector wave fields. As shown in Fig.~\ref{Fig_intro_water}, this allows one to directly observe the microscopic mechanical origin of the wave momentum, spin, and OAM \cite{Bliokh2022SA}. 
Moreover, recent studies have shown that the canonical momentum and spin AM densities \eqref{P_water} and \eqref{S_water} in structured water waves manifest as wave-induced forces and torques on small floating particles \cite{Wang2025N}, entirely similar to the optical/acoustic forces and torques in optical/acoustic tweezers, \cite{Toftul2024} (see Section~\ref{Vortex}). Notably, motion of floating particles, related to the Stokes drift and water-wave momentum density, is an important fluid-mechanical problem related to the transport of marine litter and other phenomena \cite{Bremer2018, Huang2011OE, Huang2016JEM, Alsina2020JGRO, Calvert2021JFM}.

%%%%%%%%%%%%%%%%%%%%%%%%%%%%%%
\section{Example: cylindrical wave vortices}
\label{Vortex}
%%%%%%%%%%%%%%%%%%%%%%%%%%%%%%

%FFFFFFFFFFFFFFFFFFFFFFFFFFFFFFFFFFFFFFFFFFFFFF
\begin{figure*}[t!]
\centering
\includegraphics[width=0.95\linewidth]{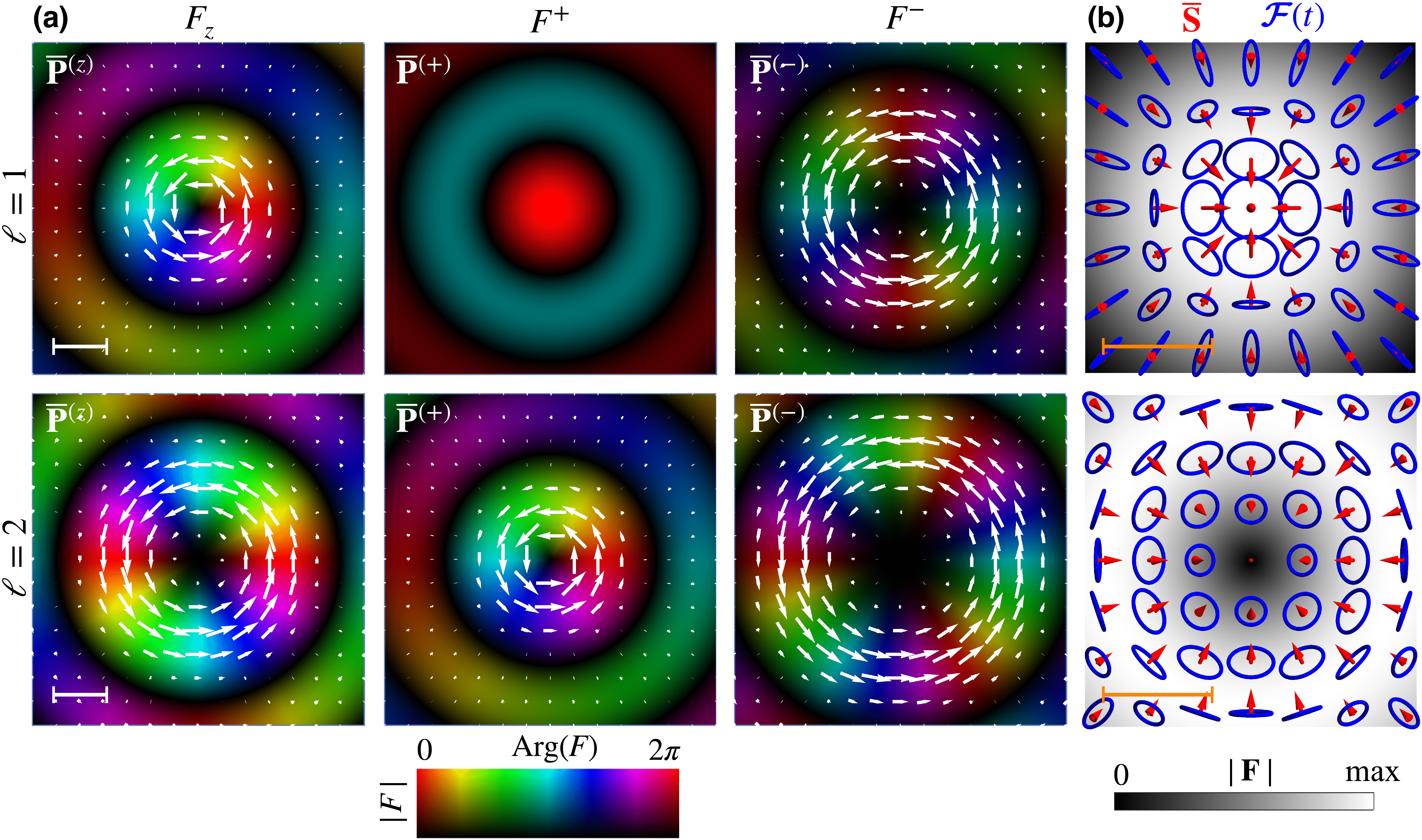}
\caption{The $(x,y)$-plane distributions of the cylindrical vortex field \eqref{F_cylinder} and \eqref{F_circular} with $\ell=1$ (upper row) and $\ell=2$ (lower row), with its azimuthal momentum density (which generates the OAM), polarization ellipses, and the spin density. The radial dependencies $G_z = J_\ell (kr)$ and $G^\pm = \pm J_{\ell\mp 1} (kr)/\sqrt{2}$ (where $J_\ell$ is the Bessel functions of the first kind) correspond to the water-wave vortex solutions \cite{Smirnova2024PRL}. (a) The amplitude-phase distributions of the field components $F_z({\bf r})$, $F^+({\bf r})$, and $F^-({\bf r})$, encoded by the brightness-color (as in Fig.~\ref{Fig_momentum_spin}(a)), along with their contributions to the canonical momentum density: $\overline{\bf P}^{(z)}\propto {\rm Im}(F_z^* {\bm \nabla} F_z)$ and $\overline{\bf P}^{(\pm)}\propto {\rm Im}(F^{\pm*} {\bm \nabla} F^{\pm})$ (white arrows). (b) Distributions of the total field amplitude $|{\bf F}({\bf r})|$ (grayscale), polarization ellipses traced by the real field $\mathbfcal{F}({\bf r},t)$ (blue), and the corresponding spin density $\overline{\bf S}$ (red arrows). All scalebars correspond to $k^{-1} = \lambda/2\pi$.} 
\label{Fig_vortex}
\end{figure*}
%FFFFFFFFFFFFFFFFFFFFFFFFFFFFFFFFFFFFFFFFFFFFFF

We have examined the generic momentum and spin properties of various waves, where the OAM density \eqref{L_general} is fully determined by the canonical momentum density. Let us now consider an important example of azimuthal {\it vortex waves} that carry a well-defined AM component along the $z$-axis (see Figs.~\ref{Fig_torques}(b) and \ref{Fig_intro_water}(b)). Assuming rotational symmetry of the system about the $z$-axis, such waves are natural eigenmodes in cylindrical coordinates $(r,\varphi,z)$. Cylindrical vortex modes with similar momentum and AM properties appear in all kinds of waves discussed in this review: in free-space optics \cite{Bliokh2010PRA} and acoustics \cite{Bliokh2019b}, in optical fibers \cite{Picardi2018O} and elastic waveguides \cite{Bliokh2022PRL, Chaplain2022CP}, as well as in water waves \cite{Smirnova2024PRL, Wang2025N}. Therefore, we will describe the generic properties of such waves characterized by a complex vector wavefield ${\bf F}({\bf r})$. 

The general form of a cylindrical vortex wavefield in cylindrical coordinates is 
\begin{equation}
{\bf F} = {\bf G}\,e^{i\ell\varphi + ik_z z}= \{ G_r (r), G_\varphi (r), G_z (r) \} e^{i\ell\varphi + ik_z z}\,.
\label{F_cylinder}
\end{equation}
Here $\ell=0,\pm 1, \pm 2, ...$ is the azimuthal quantum number, $k_z$ is the wavenumber for propagation along the $z$-axis ($k_z =0$ for surface waves), whereas the radial dependencies of the field components are determined by the equations of motion and boundary conditions in the problem. 
The common azimuthal phase factor $e^{i\ell \varphi}$ in Eq.~\eqref{F_cylinder} determines a wave vortex with topological charge $\ell$, where the phase increases by $2\pi\ell$ upon encircling the $z$-axis \cite{Allen_book, Andrews_book, Bliokh2015PR}. For a scalar wave, this would mean that the wave is an eigenmode of the canonical OAM operator $\hat{L}_z = -i\partial/\partial \varphi$ with eigenvalue $\ell$  \cite{Allen1992PRA}.

However, although the cylindrical components of the vector ${\bf G}$ depend only on $r$, the vector itself depends on both $r$ and $\varphi$. This is because the local orientations of the cylindrical coordinates varies with $\varphi$. Consequently, the vector field \eqref{F_cylinder} is {\it not} an eigenmode of the OAM operator $\hat{L}_z$. Indeed, the Cartesian components of the field ${\bf F}$ are given by $F_x = F_r \cos\varphi - F_\varphi \sin\varphi$ and $F_y = F_r \sin\varphi + F_\varphi \cos\varphi$. Given the circular symmetry of the system, it is convenient to use the basis of circular polarizations in the $(x,y)$ plane: $F^\pm = (F_x \mp i F_y)/\sqrt{2} = (F_r \mp i F_\varphi)e^{\mp i \varphi}/\sqrt{2} \equiv \tilde{F}^\pm e^{\mp i \varphi}$. In this basis, the wavefield \eqref{F_cylinder} becomes
\begin{equation}
{\bf F} = \{{G}^+ (r) e^{-i\varphi},{G}^- (r) e^{i\varphi}, G_z (r) \} e^{i\ell\varphi + ik_z z}\,,
\label{F_circular}
\end{equation}
where ${G}^\pm = (G_r \mp i G_\varphi)/\sqrt{2}$.
This reveals that different field components carry vortices with distinct topological charges $\{\ell-1, \ell+1, \ell \}$, see Fig.~\ref{Fig_vortex}.

To fully characterize the vector field \eqref{F_cylinder} and \eqref{F_circular}, we must consider both the OAM and spin parts of the total AM operator, Eqs.~\eqref{J_z}. These operators were originally defined for the Cartesian field components. In the circular basis \eqref{F_circular}, the OAM operator $\hat{L}_z$ remains the same, while the spin operator becomes diagonal: $\hat{S}_z = {\rm diag}(1,-1,0)$. 
It is easy to see that the cylindrical wave \eqref{F_circular} is an eigenmode of the total AM operator $\hat{J}_z = \hat{L}_z + \hat{S}_z$: $\hat{J}_z {\bf F} = \ell\, {\bf F}$, where $\ell$ is the {\it total} AM quantum number. 
This result is expected, as $\hat{J}_z$ generates rotations of the circularly-symmetric vector field ${\bf F} ({\bf r})$ about the $z$-axis, see Section~\ref{Canonical} and Fig.~\ref{Fig_trans_rot}(b). 

The $z$-components of the time-averaged momentum, OAM, and spin densities in the cylindrical wave \eqref{F_cylinder} and \eqref{F_circular} can be obtained from Eqs.~\eqref{P_general}--\eqref{S_general}: 
\begin{align}
\overline{P}_z & = k_z \frac{2W_{\rm kin}}{\omega}\,, 
\label{P_cylinder} \\
\overline{L}_z & = \left( \ell - \frac{|G^+|^2 - |G^-|^2}{|{\bf G}|^2}\right)\!\frac{2W_{\rm kin}}{\omega}\,, 
\label{L_cylinder} \\
\overline{S}_z & = \frac{|G^+|^2 - |G^-|^2}{|{\bf G}|^2}\frac{2W_{\rm kin}}{\omega}\,. 
\label{S_cylinder}
\end{align}
Here $\overline{W}_{\rm kin} = a\omega^2|{\bf F}|^2/4$ is the time-averaged kinetic energy of the wavefield, and the total AM density takes the natural `quantized' form $\overline{J}_z = \overline{L}_z + \overline{S}_z = \ell\, (2 \overline{W}_{\rm kin}/\omega)$. 

We recall that the OAM \eqref{L_cylinder} is generated by the azimuthal component of the canonical momentum density: $\overline{L}_z = r \overline{P}_\varphi$. Figure~\ref{Fig_vortex} shows an example of the $(x,y)$-plane distributions of the components of the cylindrical field \eqref{F_circular}, along with their contributions to the azimuthal momentum density (and hence the OAM), as well as the polarization ellipses with the corresponding spin density. We used radial dependencies $G_z = J_\ell (kr)$ and $G^\pm = \pm J_{\ell\mp 1} (kr)/\sqrt{2}$ (where $J_\ell$ is the Bessel functions of the first kind), corresponding to the water-wave vortex solutions \cite{Smirnova2024PRL}.

%FFFFFFFFFFFFFFFFFFFFFFFFFFFFFFFFFFFFFFFFFFFFFF
\begin{figure}[t!]
\centering
\includegraphics[width=\linewidth]{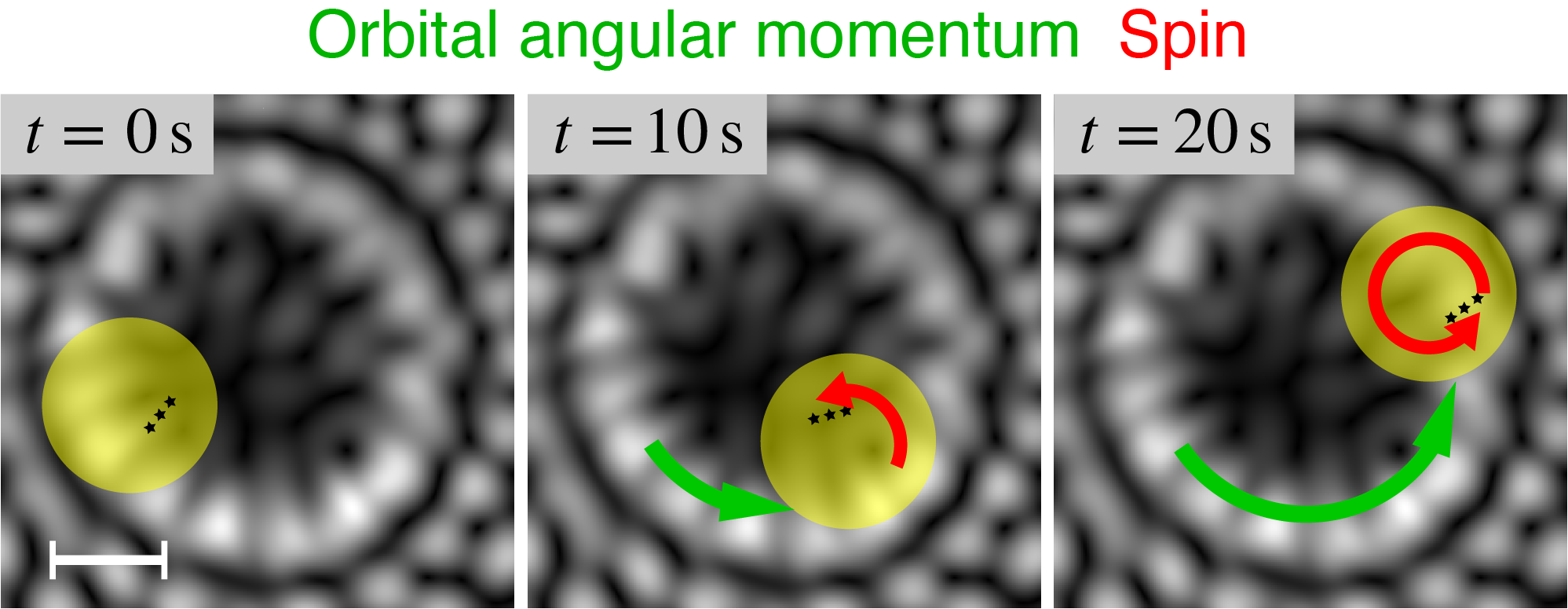}
\caption{Dynamics of a floating ping-pong ball in a water-wave vortex, similar to Fig.~\ref{Fig_vortex} with $\ell=8$ \cite{Wang2025N}. The vertical-field amplitude $|F_z(x,y)|$ is shown in grayscale, and the ball is highlighted in yellow. Similar to the mechanical action of optical spin and OAM, Fig.~\ref{Fig_torques}, the ball undergoes orbital rotation (green arrows) due to the water-wave OAM (or the azimuthal momentum density), Fig.~\ref{Fig_vortex}(a), as well as the spinning rotation (red arrows) due to the local spin density, Fig.~\ref{Fig_vortex}(b). The radial gradient of azimuthal momentum also contributes to the spinning rotation. The scalebar corresponds to the wavelength $\lambda \simeq 2.7\,{\rm cm}$. Adapted from \cite{Wang2025N} with permission from Springer Nature.} 
\label{Fig_ping-pong} 
\end{figure}
%FFFFFFFFFFFFFFFFFFFFFFFFFFFFFFFFFFFFFFFFFFFFFF

Figure~\ref{Fig_ping-pong} shows the experimentally observed dynamics of a floating ping-pong ball in such water-wave vortex with $\ell=8$ \cite{Wang2025N}. Similar to the mechanical action of optical spin and OAM, Fig.~\ref{Fig_torques}, the ball exhibits orbital rotation due to the water-wave OAM $\overline{L}_z$ (or the azimuthal momentum $\overline{P}_\varphi$), as well as the spinning rotation due to the spin AM $\overline{S}_z$. The ball is radially trapped at the inner side of the maximum-intensity ring where $\overline{S}_z>0$, see Fig.~\ref{Fig_vortex}(b) for $\ell=2$. Additionally, the radial gradient of the azimuthal radiation-pressure force, determined by $\partial \overline{P}_\varphi/\partial r$, also contributes to the ball's spinning motion with the same sign \cite{Wang2025N}.

%%%%%%%%%%%%%%%%%%%%%%%%%%%%%%
\section{Conclusions}
%%%%%%%%%%%%%%%%%%%%%%%%%%%%%%

We have described the fundamental momentum and AM properties of various types of linear classical waves: sound, electromagnetic, elastic, plasma waves, and gravity water-surface waves. We focused on the canonical momentum and AM densities, which are associated with translations and rotations of the wavefield relative to the motionless medium. 
This allowed us to employ a universal Lagrangian field-theory formalism and Noether's theorem. In this approach, the time-averaged canonical momentum density in a monochromatic wavefield is given by the sum of the {\it intensity-weighted phase gradients} of all the field components. For any vector waves, the canonical AM density consists of the orbital and spin parts. The OAM density is determined by the vector product of the radius-vector and the canonical momentum density, while the spin AM is produced by the {\it local rotation of the vector wavefield} tracing elliptical trajectories (polarizations) in monochromatic waves. 

Furthermore, we have described the general Belinfante-Rosenfeld relation, which connects the canonical momentum, spin, and kinetic momentum densities. In turn, the kinetic momentum is directly related to the energy flux density appearing in the energy conservation law. 

Remarkably, these general relations hold for different types of waves: longitudinal (curl-less), transverse (divergence-less), or mixed, including dispersive waves with different phase and group velocities. For mechanical waves, the spin AM density corresponds to the mechanical AM generated by the local elliptical motions of the medium particles. Meanwhile, the canonical wave momentum and OAM densities in media with free particles arise from the Stokes drift of the medium particles. 
Moreover, the canonical wave momentum and spin densities produce directly observable radiation forces and torques on small absorbing particles \cite{Toftul2024, Wang2025N}, which play a crucial role in optical and acoustic tweezers.

These features provide a clear physical interpretation of the wave characteristics derived from field-theory Lagrangians. 
This is particularly important for resolving long-standing controversies over the definitions of momentum, spin, and OAM in various wave systems, especially in cases where direct experimental tests are not feasible \cite{Leader2014PR}. 

The local momentum and AM densities provide a powerful framework for the description of {\it structured} waves, restricted by monochromaticity but arbitrarily inhomogeneous in space \cite{Rubinsztein-Dunlop2016JO, Bliokh2023JO}. Such structured waves are ubiquitous in modern optics and acoustics, where these are employed in numerous opto-mechanical and acousto-mechanical applications. In addition, structured quantum-matter waves have found applications in electron microscopy and neutron interferometry \cite{Bliokh2017PR, Larocque2018CP, Bliokh2023JO}. Moreover, there is a rapidly growing interest in structured elastic waves, including `chiral phonons' in solids \cite{Zhang2014PRL, Nakane2018PRB, Zhu2018S, Chaplain2022CP,  Ishito2023NP, Tauchert2022N}, as well as in structured water waves \cite{Rozenman2019F, Zhu2024NRP, Bliokh2022SA, Wang2025N}. The approach presented in this review can be extended to other types of waves, such as, e.g., spin waves in magnetic condensed-matter systems \cite{Valet2025_I, Valet2025_II}. Finally, the increasing interest in space-time-structured waves \cite{Shen2023JO} and complex materials highlights the need to extend the wave momentum and AM theory to encompass polychromatic waves and inhomogeneous, anisotropic media.

%Throughout this review, we mostly assumed the monochromatic character of the wavefields, as well as the homogeneous and isotropic character of the media. These assumptions have allowed us to use Noether's theorem and time-averaged wave properties. 

%Note also recent surge of interest in `chiral phonons', i.e., quantized elastic waves carrying AM (both spin and orbital) and underlying the Einstein–de Haas effect \cite{Zhang2014PRL, Garanin2015PRB, Nakane2018PRB, Zhu2018S, Ishito2023NP, Tauchert2022N, Ueda2023N, Luo2023S, Choi2024NN}.  

%%%%%%%%%%%%%%%%%%%%%%%%%%%%%%
\section*{Acknowledgements} 
%%%%%%%%%%%%%%%%%%%%%%%%%%%%%%
%\section*{Funding}
I acknowledge support from Marie Sk\l{}odowska-Curie COFUND Programme of the European Commission (project HORIZON-MSCA-2022-COFUND-101126600-SmartBRAIN3), 
ENSEMBLE3 Project (MAB/2020/14) which is carried out within the International Research Agendas Programme (IRAP) of the Foundation for Polish Science co-financed by the European Union under the European Regional Development Fund and Teaming Horizon 2020 programme (GA. No. 857543) of the European Commission and the project of the Minister of Science and Higher Education ``Support for the activities of Centers of Excellence established in Poland under the Horizon 2020 program'' (contract MEiN/2023/DIR/3797).

%\bibliographystyle{tfnlm}
%\bibliography{interactnlmsample}
\bibliography{refs}

\begin{thebibliography}{91}%
\makeatletter
\providecommand \@ifxundefined [1]{%
 \@ifx{#1\undefined}
}%
\providecommand \@ifnum [1]{%
 \ifnum #1\expandafter \@firstoftwo
 \else \expandafter \@secondoftwo
 \fi
}%
\providecommand \@ifx [1]{%
 \ifx #1\expandafter \@firstoftwo
 \else \expandafter \@secondoftwo
 \fi
}%
\providecommand \natexlab [1]{#1}%
\providecommand \enquote  [1]{``#1''}%
\providecommand \bibnamefont  [1]{#1}%
\providecommand \bibfnamefont [1]{#1}%
\providecommand \citenamefont [1]{#1}%
\providecommand \href@noop [0]{\@secondoftwo}%
\providecommand \href [0]{\begingroup \@sanitize@url \@href}%
\providecommand \@href[1]{\@@startlink{#1}\@@href}%
\providecommand \@@href[1]{\endgroup#1\@@endlink}%
\providecommand \@sanitize@url [0]{\catcode `\\12\catcode `\$12\catcode
  `\&12\catcode `\#12\catcode `\^12\catcode `\_12\catcode `\%12\relax}%
\providecommand \@@startlink[1]{}%
\providecommand \@@endlink[0]{}%
\providecommand \url  [0]{\begingroup\@sanitize@url \@url }%
\providecommand \@url [1]{\endgroup\@href {#1}{\urlprefix }}%
\providecommand \urlprefix  [0]{URL }%
\providecommand \Eprint [0]{\href }%
\providecommand \doibase [0]{https://doi.org/}%
\providecommand \selectlanguage [0]{\@gobble}%
\providecommand \bibinfo  [0]{\@secondoftwo}%
\providecommand \bibfield  [0]{\@secondoftwo}%
\providecommand \translation [1]{[#1]}%
\providecommand \BibitemOpen [0]{}%
\providecommand \bibitemStop [0]{}%
\providecommand \bibitemNoStop [0]{.\EOS\space}%
\providecommand \EOS [0]{\spacefactor3000\relax}%
\providecommand \BibitemShut  [1]{\csname bibitem#1\endcsname}%
\let\auto@bib@innerbib\@empty
%</preamble>
\bibitem [{\citenamefont {Jones}(1953)}]{Jones1953}%
  \BibitemOpen
  \bibfield  {author} {\bibinfo {author} {\bibfnamefont {R.~V.}\ \bibnamefont
  {Jones}},\ }\bibfield  {title} {\bibinfo {title} {{Pressure of Radiation}},\
  }\href {https://doi.org/10.1038/1711089a0} {\bibfield  {journal} {\bibinfo
  {journal} {Nature}\ }\textbf {\bibinfo {volume} {171}},\ \bibinfo {pages}
  {1089} (\bibinfo {year} {1953})}\BibitemShut {NoStop}%
\bibitem [{\citenamefont {Loudon}\ and\ \citenamefont
  {Baxter}(2012)}]{Loudon2012}%
  \BibitemOpen
  \bibfield  {author} {\bibinfo {author} {\bibfnamefont {R.}~\bibnamefont
  {Loudon}}\ and\ \bibinfo {author} {\bibfnamefont {C.}~\bibnamefont
  {Baxter}},\ }\bibfield  {title} {\bibinfo {title} {{Contributions of John
  Henry Poynting to the understanding of radiation pressure}},\ }\href
  {https://doi.org/10.1098/rspa.2011.0573} {\bibfield  {journal} {\bibinfo
  {journal} {Proc. R. Soc. A.}\ }\textbf {\bibinfo {volume} {468}},\ \bibinfo
  {pages} {1825} (\bibinfo {year} {2012})}\BibitemShut {NoStop}%
\bibitem [{\citenamefont {Beyer}(1978)}]{Beyer1978}%
  \BibitemOpen
  \bibfield  {author} {\bibinfo {author} {\bibfnamefont {R.~T.}\ \bibnamefont
  {Beyer}},\ }\bibfield  {title} {\bibinfo {title} {{Radiation
  pressure{\ifmmode---\else\textemdash\fi}the history of a mislabeled
  tensor}},\ }\href {https://doi.org/10.1121/1.381833} {\bibfield  {journal}
  {\bibinfo  {journal} {J. Acoust. Soc. Am.}\ }\textbf {\bibinfo {volume}
  {63}},\ \bibinfo {pages} {1025} (\bibinfo {year} {1978})}\BibitemShut
  {NoStop}%
\bibitem [{\citenamefont {Sarvazyan}\ \emph {et~al.}(2010)\citenamefont
  {Sarvazyan}, \citenamefont {Rudenko},\ and\ \citenamefont
  {Nyborg}}]{Sarvazyan2010UMB}%
  \BibitemOpen
  \bibfield  {author} {\bibinfo {author} {\bibfnamefont {A.~P.}\ \bibnamefont
  {Sarvazyan}}, \bibinfo {author} {\bibfnamefont {O.~V.}\ \bibnamefont
  {Rudenko}},\ and\ \bibinfo {author} {\bibfnamefont {W.~L.}\ \bibnamefont
  {Nyborg}},\ }\bibfield  {title} {\bibinfo {title} {{Biomedical Applications
  of Radiation Force of Ultrasound: Historical Roots and Physical Basis}},\
  }\href {https://doi.org/10.1016/j.ultrasmedbio.2010.05.015} {\bibfield
  {journal} {\bibinfo  {journal} {Ultrasound Med. Biol.}\ }\textbf {\bibinfo
  {volume} {36}},\ \bibinfo {pages} {1379} (\bibinfo {year}
  {2010})}\BibitemShut {NoStop}%
\bibitem [{\citenamefont {Falkovich}(2018)}]{Falkovich_book}%
  \BibitemOpen
  \bibfield  {author} {\bibinfo {author} {\bibfnamefont {G.}~\bibnamefont
  {Falkovich}},\ }\href {https://doi.org/10.1017/9781316416600} {\emph
  {\bibinfo {title} {{Fluid Mechanics}}}}\ (\bibinfo  {publisher} {Cambridge
  University Press},\ \bibinfo {address} {Cambridge},\ \bibinfo {year}
  {2018})\BibitemShut {NoStop}%
\bibitem [{\citenamefont {van~den Bremer}\ and\ \citenamefont
  {Breivik}(2018)}]{Bremer2018}%
  \BibitemOpen
  \bibfield  {author} {\bibinfo {author} {\bibfnamefont {T.~S.}\ \bibnamefont
  {van~den Bremer}}\ and\ \bibinfo {author} {\bibfnamefont {{\O}.}~\bibnamefont
  {Breivik}},\ }\bibfield  {title} {\bibinfo {title} {{Stokes drift}},\ }\href
  {https://doi.org/10.1098/rsta.2017.0104} {\bibfield  {journal} {\bibinfo
  {journal} {Philos. Trans. Royal Soc. A}\ }\textbf {\bibinfo {volume} {376}},\
  \bibinfo {pages} {20170104} (\bibinfo {year} {2018})}\BibitemShut {NoStop}%
\bibitem [{\citenamefont {Mcintyre}(1981)}]{Mcintyre1981}%
  \BibitemOpen
  \bibfield  {author} {\bibinfo {author} {\bibfnamefont {M.~E.}\ \bibnamefont
  {Mcintyre}},\ }\bibfield  {title} {\bibinfo {title} {{On the {`}wave
  momentum{'} myth}},\ }\href {https://doi.org/10.1017/S0022112081001626}
  {\bibfield  {journal} {\bibinfo  {journal} {J. Fluid Mech.}\ }\textbf
  {\bibinfo {volume} {106}},\ \bibinfo {pages} {331} (\bibinfo {year}
  {1981})}\BibitemShut {NoStop}%
\bibitem [{\citenamefont {Bliokh}\ \emph
  {et~al.}(2022{\natexlab{a}})\citenamefont {Bliokh}, \citenamefont {Punzmann},
  \citenamefont {Xia}, \citenamefont {Nori},\ and\ \citenamefont
  {Shats}}]{Bliokh2022SA}%
  \BibitemOpen
  \bibfield  {author} {\bibinfo {author} {\bibfnamefont {K.~Y.}\ \bibnamefont
  {Bliokh}}, \bibinfo {author} {\bibfnamefont {H.}~\bibnamefont {Punzmann}},
  \bibinfo {author} {\bibfnamefont {H.}~\bibnamefont {Xia}}, \bibinfo {author}
  {\bibfnamefont {F.}~\bibnamefont {Nori}},\ and\ \bibinfo {author}
  {\bibfnamefont {M.}~\bibnamefont {Shats}},\ }\bibfield  {title} {\bibinfo
  {title} {{Field theory spin and momentum in water waves}},\ }\href
  {https://doi.org/10.1126/sciadv.abm1295} {\bibfield  {journal} {\bibinfo
  {journal} {Sci. Adv.}\ }\textbf {\bibinfo {volume} {8}},\ \bibinfo {pages}
  {eabm1295} (\bibinfo {year} {2022}{\natexlab{a}})}\BibitemShut {NoStop}%
\bibitem [{\citenamefont {Bliokh}\ \emph
  {et~al.}(2022{\natexlab{b}})\citenamefont {Bliokh}, \citenamefont {Bliokh},\
  and\ \citenamefont {Nori}}]{Bliokh2022PRA}%
  \BibitemOpen
  \bibfield  {author} {\bibinfo {author} {\bibfnamefont {K.~Y.}\ \bibnamefont
  {Bliokh}}, \bibinfo {author} {\bibfnamefont {Y.~P.}\ \bibnamefont {Bliokh}},\
  and\ \bibinfo {author} {\bibfnamefont {F.}~\bibnamefont {Nori}},\ }\bibfield
  {title} {\bibinfo {title} {{Ponderomotive forces, Stokes drift, and momentum
  in acoustic and electromagnetic waves}},\ }\href
  {https://doi.org/10.1103/PhysRevA.106.L021503} {\bibfield  {journal}
  {\bibinfo  {journal} {Phys. Rev. A}\ }\textbf {\bibinfo {volume} {106}},\
  \bibinfo {pages} {L021503} (\bibinfo {year}
  {2022}{\natexlab{b}})}\BibitemShut {NoStop}%
\bibitem [{\citenamefont {Allen}\ \emph {et~al.}(2003)\citenamefont {Allen},
  \citenamefont {Barnett},\ and\ \citenamefont {Padgett}}]{Allen_book}%
  \BibitemOpen
  \bibfield  {author} {\bibinfo {author} {\bibfnamefont {L.}~\bibnamefont
  {Allen}}, \bibinfo {author} {\bibfnamefont {S.~M.}\ \bibnamefont {Barnett}},\
  and\ \bibinfo {author} {\bibfnamefont {M.~J.}\ \bibnamefont {Padgett}},\
  }\href@noop {} {\emph {\bibinfo {title} {{Optical Angular Momentum}}}}\
  (\bibinfo  {publisher} {{IoP Publishing}},\ \bibinfo {year}
  {2003})\BibitemShut {NoStop}%
\bibitem [{\citenamefont {Andrews}\ and\ \citenamefont
  {Babiker}(2012)}]{Andrews_book}%
  \BibitemOpen
  \bibfield  {author} {\bibinfo {author} {\bibfnamefont {D.~L.}\ \bibnamefont
  {Andrews}}\ and\ \bibinfo {author} {\bibfnamefont {M.}~\bibnamefont
  {Babiker}},\ }\href@noop {} {\emph {\bibinfo {title} {{The Angular Momentum
  of Light}}}}\ (\bibinfo  {publisher} {Cambridge University Press},\ \bibinfo
  {address} {Cambridge},\ \bibinfo {year} {2012})\BibitemShut {NoStop}%
\bibitem [{\citenamefont {Bliokh}\ and\ \citenamefont
  {Nori}(2015)}]{Bliokh2015PR}%
  \BibitemOpen
  \bibfield  {author} {\bibinfo {author} {\bibfnamefont {K.~Y.}\ \bibnamefont
  {Bliokh}}\ and\ \bibinfo {author} {\bibfnamefont {F.}~\bibnamefont {Nori}},\
  }\bibfield  {title} {\bibinfo {title} {Transverse and longitudinal angular
  momenta of light},\ }\href {https://doi.org/10.1016/j.physrep.2015.06.003}
  {\bibfield  {journal} {\bibinfo  {journal} {Phys. Rep.}\ }\textbf {\bibinfo
  {volume} {592}},\ \bibinfo {pages} {1} (\bibinfo {year} {2015})}\BibitemShut
  {NoStop}%
\bibitem [{\citenamefont {Poynting}(1909)}]{Poynting1909}%
  \BibitemOpen
  \bibfield  {author} {\bibinfo {author} {\bibfnamefont {J.~H.}\ \bibnamefont
  {Poynting}},\ }\bibfield  {title} {\bibinfo {title} {{The wave motion of a
  revolving shaft, and a suggestion as to the angular momentum in a beam of
  circularly polarised light}},\ }\href
  {https://doi.org/10.1098/rspa.1909.0060} {\bibfield  {journal} {\bibinfo
  {journal} {Proc. R. Soc. London A}\ }\textbf {\bibinfo {volume} {82}},\
  \bibinfo {pages} {560} (\bibinfo {year} {1909})}\BibitemShut {NoStop}%
\bibitem [{\citenamefont {Beth}(1935)}]{Beth1935}%
  \BibitemOpen
  \bibfield  {author} {\bibinfo {author} {\bibfnamefont {R.~A.}\ \bibnamefont
  {Beth}},\ }\bibfield  {title} {\bibinfo {title} {Direct {{Detection}} of the
  {{Angular Momentum}} of {{Light}}},\ }\href
  {https://doi.org/10.1103/PhysRev.48.471} {\bibfield  {journal} {\bibinfo
  {journal} {Phys. Rev.}\ }\textbf {\bibinfo {volume} {48}},\ \bibinfo {pages}
  {471} (\bibinfo {year} {1935})}\BibitemShut {NoStop}%
\bibitem [{\citenamefont {Friese}\ \emph {et~al.}(1998)\citenamefont {Friese},
  \citenamefont {Nieminen}, \citenamefont {Heckenberg},\ and\ \citenamefont
  {Rubinsztein-Dunlop}}]{Friese1998Nat}%
  \BibitemOpen
  \bibfield  {author} {\bibinfo {author} {\bibfnamefont {M.~E.~J.}\
  \bibnamefont {Friese}}, \bibinfo {author} {\bibfnamefont {T.~A.}\
  \bibnamefont {Nieminen}}, \bibinfo {author} {\bibfnamefont {N.~R.}\
  \bibnamefont {Heckenberg}},\ and\ \bibinfo {author} {\bibfnamefont
  {H.}~\bibnamefont {Rubinsztein-Dunlop}},\ }\bibfield  {title} {\bibinfo
  {title} {{Optical alignment and spinning of laser-trapped microscopic
  particles}},\ }\href {https://doi.org/10.1038/28566} {\bibfield  {journal}
  {\bibinfo  {journal} {Nature}\ }\textbf {\bibinfo {volume} {394}},\ \bibinfo
  {pages} {348} (\bibinfo {year} {1998})}\BibitemShut {NoStop}%
\bibitem [{\citenamefont {Allen}\ \emph {et~al.}(1992)\citenamefont {Allen},
  \citenamefont {Beijersbergen}, \citenamefont {Spereeuw},\ and\ \citenamefont
  {Woerdman}}]{Allen1992PRA}%
  \BibitemOpen
  \bibfield  {author} {\bibinfo {author} {\bibfnamefont {L.}~\bibnamefont
  {Allen}}, \bibinfo {author} {\bibfnamefont {M.~W.}\ \bibnamefont
  {Beijersbergen}}, \bibinfo {author} {\bibfnamefont {R.~J.~C.}\ \bibnamefont
  {Spereeuw}},\ and\ \bibinfo {author} {\bibfnamefont {J.~P.}\ \bibnamefont
  {Woerdman}},\ }\bibfield  {title} {\bibinfo {title} {Orbital angular momentum
  of light and the transformation of {{Laguerre}}-{{Gaussian}} laser modes},\
  }\href@noop {} {\bibfield  {journal} {\bibinfo  {journal} {Phys. Rev. A}\
  }\textbf {\bibinfo {volume} {45}},\ \bibinfo {pages} {8185} (\bibinfo {year}
  {1992})}\BibitemShut {NoStop}%
\bibitem [{\citenamefont {{Garc{\'e}s-Ch{\'a}vez}}\ \emph
  {et~al.}(2003)\citenamefont {{Garc{\'e}s-Ch{\'a}vez}}, \citenamefont
  {McGloin}, \citenamefont {Padgett}, \citenamefont {Dultz}, \citenamefont
  {Schmitzer},\ and\ \citenamefont {Dholakia}}]{Garces-Chavez2003PRL}%
  \BibitemOpen
  \bibfield  {author} {\bibinfo {author} {\bibfnamefont {V.}~\bibnamefont
  {{Garc{\'e}s-Ch{\'a}vez}}}, \bibinfo {author} {\bibfnamefont
  {D.}~\bibnamefont {McGloin}}, \bibinfo {author} {\bibfnamefont {M.~J.}\
  \bibnamefont {Padgett}}, \bibinfo {author} {\bibfnamefont {W.}~\bibnamefont
  {Dultz}}, \bibinfo {author} {\bibfnamefont {H.}~\bibnamefont {Schmitzer}},\
  and\ \bibinfo {author} {\bibfnamefont {K.}~\bibnamefont {Dholakia}},\
  }\bibfield  {title} {\bibinfo {title} {Observation of the {{Transfer}} of the
  {{Local Angular Momentum Density}} of a {{Multiringed Light Beam}} to an
  {{Optically Trapped Particle}}},\ }\href
  {https://doi.org/10.1103/PhysRevLett.91.093602} {\bibfield  {journal}
  {\bibinfo  {journal} {Phys. Rev. Lett.}\ }\textbf {\bibinfo {volume} {91}},\
  \bibinfo {pages} {093602} (\bibinfo {year} {2003})}\BibitemShut {NoStop}%
\bibitem [{\citenamefont {Jones}(1973)}]{Jones1973}%
  \BibitemOpen
  \bibfield  {author} {\bibinfo {author} {\bibfnamefont {W.~L.}\ \bibnamefont
  {Jones}},\ }\bibfield  {title} {\bibinfo {title} {Asymmetric wave-stress
  tensors and wave spin},\ }\href {https://doi.org/10.1017/S0022112073002466}
  {\bibfield  {journal} {\bibinfo  {journal} {J. Fluid Mech.}\ }\textbf
  {\bibinfo {volume} {58}},\ \bibinfo {pages} {737} (\bibinfo {year}
  {1973})}\BibitemShut {NoStop}%
\bibitem [{\citenamefont {Peierls}(1991)}]{Peierls1991}%
  \BibitemOpen
  \bibfield  {author} {\bibinfo {author} {\bibfnamefont {R.}~\bibnamefont
  {Peierls}},\ }\href@noop {} {\emph {\bibinfo {title} {More Surprises in
  Theoretical Physics}}}\ (\bibinfo  {publisher} {Princeton University Press},\
  \bibinfo {year} {1991})\BibitemShut {NoStop}%
\bibitem [{\citenamefont {Peskin}(2010)}]{Peskin2010}%
  \BibitemOpen
  \bibfield  {author} {\bibinfo {author} {\bibfnamefont {C.~S.}\ \bibnamefont
  {Peskin}},\ }\bibfield  {title} {\bibinfo {title} {Wave momentum},\ }\href
  {www.math. nyu.edu/faculty/peskin/papers/wave_momentum.pdf} {\bibfield
  {journal} {\bibinfo  {journal} {The Silver Dialogues}\ } (\bibinfo {year}
  {2010})}\BibitemShut {NoStop}%
\bibitem [{\citenamefont {Hefner}\ and\ \citenamefont
  {Marston}(1999)}]{Hefner1999JASA}%
  \BibitemOpen
  \bibfield  {author} {\bibinfo {author} {\bibfnamefont {B.~T.}\ \bibnamefont
  {Hefner}}\ and\ \bibinfo {author} {\bibfnamefont {P.~L.}\ \bibnamefont
  {Marston}},\ }\bibfield  {title} {\bibinfo {title} {{An acoustical helicoidal
  wave transducer with applications for the alignment of ultrasonic and
  underwater systems}},\ }\href {https://doi.org/10.1121/1.428184} {\bibfield
  {journal} {\bibinfo  {journal} {J. Acoust. Soc. Am.}\ }\textbf {\bibinfo
  {volume} {106}},\ \bibinfo {pages} {3313} (\bibinfo {year}
  {1999})}\BibitemShut {NoStop}%
\bibitem [{\citenamefont {Nakane}\ and\ \citenamefont
  {Kohno}(2018)}]{Nakane2018PRB}%
  \BibitemOpen
  \bibfield  {author} {\bibinfo {author} {\bibfnamefont {J.~J.}\ \bibnamefont
  {Nakane}}\ and\ \bibinfo {author} {\bibfnamefont {H.}~\bibnamefont {Kohno}},\
  }\bibfield  {title} {\bibinfo {title} {{Angular momentum of phonons and its
  application to single-spin relaxation}},\ }\href
  {https://doi.org/10.1103/PhysRevB.97.174403} {\bibfield  {journal} {\bibinfo
  {journal} {Phys. Rev. B}\ }\textbf {\bibinfo {volume} {97}},\ \bibinfo
  {pages} {174403} (\bibinfo {year} {2018})}\BibitemShut {NoStop}%
\bibitem [{\citenamefont {Shi}\ \emph {et~al.}(2019)\citenamefont {Shi},
  \citenamefont {Zhao}, \citenamefont {Long}, \citenamefont {Yang},
  \citenamefont {Wang}, \citenamefont {Chen}, \citenamefont {Ren},\ and\
  \citenamefont {Zhang}}]{Shi2019}%
  \BibitemOpen
  \bibfield  {author} {\bibinfo {author} {\bibfnamefont {C.}~\bibnamefont
  {Shi}}, \bibinfo {author} {\bibfnamefont {R.}~\bibnamefont {Zhao}}, \bibinfo
  {author} {\bibfnamefont {Y.}~\bibnamefont {Long}}, \bibinfo {author}
  {\bibfnamefont {S.}~\bibnamefont {Yang}}, \bibinfo {author} {\bibfnamefont
  {Y.}~\bibnamefont {Wang}}, \bibinfo {author} {\bibfnamefont {H.}~\bibnamefont
  {Chen}}, \bibinfo {author} {\bibfnamefont {J.}~\bibnamefont {Ren}},\ and\
  \bibinfo {author} {\bibfnamefont {X.}~\bibnamefont {Zhang}},\ }\bibfield
  {title} {\bibinfo {title} {Observation of acoustic spin},\ }\href
  {https://doi.org/10.1093/nsr/nwz059} {\bibfield  {journal} {\bibinfo
  {journal} {Natl. Sci. Rev.}\ }\textbf {\bibinfo {volume} {6}},\ \bibinfo
  {pages} {707} (\bibinfo {year} {2019})}\BibitemShut {NoStop}%
\bibitem [{\citenamefont {Bliokh}\ and\ \citenamefont
  {Nori}(2019{\natexlab{a}})}]{Bliokh2019b}%
  \BibitemOpen
  \bibfield  {author} {\bibinfo {author} {\bibfnamefont {K.~Y.}\ \bibnamefont
  {Bliokh}}\ and\ \bibinfo {author} {\bibfnamefont {F.}~\bibnamefont {Nori}},\
  }\bibfield  {title} {\bibinfo {title} {{Spin and orbital angular momenta of
  acoustic beams}},\ }\href {https://doi.org/10.1103/PhysRevB.99.174310}
  {\bibfield  {journal} {\bibinfo  {journal} {Phys. Rev. B}\ }\textbf {\bibinfo
  {volume} {99}},\ \bibinfo {pages} {174310} (\bibinfo {year}
  {2019}{\natexlab{a}})},\ \bibinfo {note} {{Erratum: {Phys. Rev. B} {\bf 105},
  219901(E) (2022)}}\BibitemShut {NoStop}%
\bibitem [{\citenamefont {Ren}(2022)}]{Ren2022CPL}%
  \BibitemOpen
  \bibfield  {author} {\bibinfo {author} {\bibfnamefont {J.}~\bibnamefont
  {Ren}},\ }\bibfield  {title} {\bibinfo {title} {{From Elastic Spin to Phonon
  Spin: Symmetry and Fundamental Relations}},\ }\href
  {https://doi.org/10.1088/0256-307X/39/12/126301} {\bibfield  {journal}
  {\bibinfo  {journal} {Chin. Phys. Lett.}\ }\textbf {\bibinfo {volume} {39}},\
  \bibinfo {pages} {126301} (\bibinfo {year} {2022})}\BibitemShut {NoStop}%
\bibitem [{\citenamefont {Chaplain}\ \emph {et~al.}(2022)\citenamefont
  {Chaplain}, \citenamefont {De~Ponti},\ and\ \citenamefont
  {Starkey}}]{Chaplain2022CP}%
  \BibitemOpen
  \bibfield  {author} {\bibinfo {author} {\bibfnamefont {G.~J.}\ \bibnamefont
  {Chaplain}}, \bibinfo {author} {\bibfnamefont {J.~M.}\ \bibnamefont
  {De~Ponti}},\ and\ \bibinfo {author} {\bibfnamefont {T.~A.}\ \bibnamefont
  {Starkey}},\ }\bibfield  {title} {\bibinfo {title} {{Elastic orbital angular
  momentum transfer from an elastic pipe to a fluid}},\ }\href
  {https://doi.org/10.1038/s42005-022-01057-0} {\bibfield  {journal} {\bibinfo
  {journal} {Commun. Phys.}\ }\textbf {\bibinfo {volume} {5}},\ \bibinfo
  {pages} {279} (\bibinfo {year} {2022})}\BibitemShut {NoStop}%
\bibitem [{\citenamefont {Bliokh}(2022)}]{Bliokh2022PRL}%
  \BibitemOpen
  \bibfield  {author} {\bibinfo {author} {\bibfnamefont {K.~Y.}\ \bibnamefont
  {Bliokh}},\ }\bibfield  {title} {\bibinfo {title} {{Elastic Spin and Orbital
  Angular Momenta}},\ }\href {https://doi.org/10.1103/PhysRevLett.129.204303}
  {\bibfield  {journal} {\bibinfo  {journal} {Phys. Rev. Lett.}\ }\textbf
  {\bibinfo {volume} {129}},\ \bibinfo {pages} {204303} (\bibinfo {year}
  {2022})}\BibitemShut {NoStop}%
\bibitem [{\citenamefont {Bliokh}\ and\ \citenamefont
  {Bliokh}(2022)}]{Bliokh2022PRE}%
  \BibitemOpen
  \bibfield  {author} {\bibinfo {author} {\bibfnamefont {K.~Y.}\ \bibnamefont
  {Bliokh}}\ and\ \bibinfo {author} {\bibfnamefont {Y.~P.}\ \bibnamefont
  {Bliokh}},\ }\bibfield  {title} {\bibinfo {title} {{Momentum, angular
  momentum, and spin of waves in an isotropic collisionless plasma}},\ }\href
  {https://doi.org/10.1103/PhysRevE.105.065208} {\bibfield  {journal} {\bibinfo
   {journal} {Phys. Rev. E}\ }\textbf {\bibinfo {volume} {105}},\ \bibinfo
  {pages} {065208} (\bibinfo {year} {2022})}\BibitemShut {NoStop}%
\bibitem [{\citenamefont {Smirnova}\ \emph {et~al.}(2024)\citenamefont
  {Smirnova}, \citenamefont {Nori},\ and\ \citenamefont
  {Bliokh}}]{Smirnova2024PRL}%
  \BibitemOpen
  \bibfield  {author} {\bibinfo {author} {\bibfnamefont {D.~A.}\ \bibnamefont
  {Smirnova}}, \bibinfo {author} {\bibfnamefont {F.}~\bibnamefont {Nori}},\
  and\ \bibinfo {author} {\bibfnamefont {K.~Y.}\ \bibnamefont {Bliokh}},\
  }\bibfield  {title} {\bibinfo {title} {{Water-Wave Vortices and Skyrmions}},\
  }\href {https://doi.org/10.1103/PhysRevLett.132.054003} {\bibfield  {journal}
  {\bibinfo  {journal} {Phys. Rev. Lett.}\ }\textbf {\bibinfo {volume} {132}},\
  \bibinfo {pages} {054003} (\bibinfo {year} {2024})}\BibitemShut {NoStop}%
\bibitem [{\citenamefont {Rubinsztein-Dunlop}\ \emph
  {et~al.}(2016)\citenamefont {Rubinsztein-Dunlop}, \citenamefont {Forbes},
  \citenamefont {Berry}, \citenamefont {Dennis}, \citenamefont {Andrews},
  \citenamefont {Mansuripur}, \citenamefont {Denz}, \citenamefont {Alpmann},
  \citenamefont {Banzer}, \citenamefont {Bauer}, \citenamefont {Karimi},
  \citenamefont {Marrucci}, \citenamefont {Padgett}, \citenamefont
  {Ritsch-Marte}, \citenamefont {Litchinitser}, \citenamefont {Bigelow},
  \citenamefont {Rosales-Guzm{\ifmmode\acute{a}\else\'{a}\fi}n}, \citenamefont
  {Belmonte}, \citenamefont {Torres}, \citenamefont {Neely}, \citenamefont
  {Baker}, \citenamefont {Gordon}, \citenamefont {Stilgoe}, \citenamefont
  {Romero}, \citenamefont {White}, \citenamefont {Fickler}, \citenamefont
  {Willner}, \citenamefont {Xie}, \citenamefont {McMorran},\ and\ \citenamefont
  {Weiner}}]{Rubinsztein-Dunlop2016JO}%
  \BibitemOpen
  \bibfield  {author} {\bibinfo {author} {\bibfnamefont {H.}~\bibnamefont
  {Rubinsztein-Dunlop}}, \bibinfo {author} {\bibfnamefont {A.}~\bibnamefont
  {Forbes}}, \bibinfo {author} {\bibfnamefont {M.~V.}\ \bibnamefont {Berry}},
  \bibinfo {author} {\bibfnamefont {M.~R.}\ \bibnamefont {Dennis}}, \bibinfo
  {author} {\bibfnamefont {D.~L.}\ \bibnamefont {Andrews}}, \bibinfo {author}
  {\bibfnamefont {M.}~\bibnamefont {Mansuripur}}, \bibinfo {author}
  {\bibfnamefont {C.}~\bibnamefont {Denz}}, \bibinfo {author} {\bibfnamefont
  {C.}~\bibnamefont {Alpmann}}, \bibinfo {author} {\bibfnamefont
  {P.}~\bibnamefont {Banzer}}, \bibinfo {author} {\bibfnamefont
  {T.}~\bibnamefont {Bauer}}, \bibinfo {author} {\bibfnamefont
  {E.}~\bibnamefont {Karimi}}, \bibinfo {author} {\bibfnamefont
  {L.}~\bibnamefont {Marrucci}}, \bibinfo {author} {\bibfnamefont
  {M.}~\bibnamefont {Padgett}}, \bibinfo {author} {\bibfnamefont
  {M.}~\bibnamefont {Ritsch-Marte}}, \bibinfo {author} {\bibfnamefont {N.~M.}\
  \bibnamefont {Litchinitser}}, \bibinfo {author} {\bibfnamefont {N.~P.}\
  \bibnamefont {Bigelow}}, \bibinfo {author} {\bibfnamefont {C.}~\bibnamefont
  {Rosales-Guzm{\ifmmode\acute{a}\else\'{a}\fi}n}}, \bibinfo {author}
  {\bibfnamefont {A.}~\bibnamefont {Belmonte}}, \bibinfo {author}
  {\bibfnamefont {J.~P.}\ \bibnamefont {Torres}}, \bibinfo {author}
  {\bibfnamefont {T.~W.}\ \bibnamefont {Neely}}, \bibinfo {author}
  {\bibfnamefont {M.}~\bibnamefont {Baker}}, \bibinfo {author} {\bibfnamefont
  {R.}~\bibnamefont {Gordon}}, \bibinfo {author} {\bibfnamefont {A.~B.}\
  \bibnamefont {Stilgoe}}, \bibinfo {author} {\bibfnamefont {J.}~\bibnamefont
  {Romero}}, \bibinfo {author} {\bibfnamefont {A.~G.}\ \bibnamefont {White}},
  \bibinfo {author} {\bibfnamefont {R.}~\bibnamefont {Fickler}}, \bibinfo
  {author} {\bibfnamefont {A.~E.}\ \bibnamefont {Willner}}, \bibinfo {author}
  {\bibfnamefont {G.}~\bibnamefont {Xie}}, \bibinfo {author} {\bibfnamefont
  {B.}~\bibnamefont {McMorran}},\ and\ \bibinfo {author} {\bibfnamefont
  {A.~M.}\ \bibnamefont {Weiner}},\ }\bibfield  {title} {\bibinfo {title}
  {{Roadmap on structured light}},\ }\href
  {https://doi.org/10.1088/2040-8978/19/1/013001} {\bibfield  {journal}
  {\bibinfo  {journal} {J. Opt.}\ }\textbf {\bibinfo {volume} {19}},\ \bibinfo
  {pages} {013001} (\bibinfo {year} {2016})}\BibitemShut {NoStop}%
\bibitem [{\citenamefont {Bliokh}\ \emph {et~al.}(2023)\citenamefont {Bliokh},
  \citenamefont {Karimi}, \citenamefont {Padgett}, \citenamefont {Alonso},
  \citenamefont {Dennis}, \citenamefont {Dudley}, \citenamefont {Forbes},
  \citenamefont {Zahedpour}, \citenamefont {Hancock}, \citenamefont
  {Milchberg}, \citenamefont {Rotter}, \citenamefont {Nori}, \citenamefont
  {{\ifmmode\ddot{O}\else\"{O}\fi}zdemir}, \citenamefont {Bender},
  \citenamefont {Cao}, \citenamefont {Corkum}, \citenamefont
  {Hern{\ifmmode\acute{a}\else\'{a}\fi}ndez-Garc{\ifmmode\acute{\imath}\else\'{\i}\fi}a},
  \citenamefont {Ren}, \citenamefont {Kivshar}, \citenamefont {Silveirinha},
  \citenamefont {Engheta}, \citenamefont {Rauschenbeutel}, \citenamefont
  {Schneeweiss}, \citenamefont {Volz}, \citenamefont {Leykam}, \citenamefont
  {Smirnova}, \citenamefont {Rong}, \citenamefont {Wang}, \citenamefont
  {Hasman}, \citenamefont {Picardi}, \citenamefont {Zayats}, \citenamefont
  {Rodr{\ifmmode\acute{\imath}\else\'{\i}\fi}guez-Fortu{\ifmmode\tilde{n}\else\~{n}\fi}o},
  \citenamefont {Yang}, \citenamefont {Ren}, \citenamefont {Khanikaev},
  \citenamefont {Al{\ifmmode\grave{u}\else\`{u}\fi}}, \citenamefont
  {Brasselet}, \citenamefont {Shats}, \citenamefont {Verbeeck}, \citenamefont
  {Schattschneider}, \citenamefont {Sarenac}, \citenamefont {Cory},
  \citenamefont {Pushin}, \citenamefont {Birk}, \citenamefont {Gorlach},
  \citenamefont {Kaminer}, \citenamefont {Cardano}, \citenamefont {Marrucci},
  \citenamefont {Krenn},\ and\ \citenamefont {Marquardt}}]{Bliokh2023JO}%
  \BibitemOpen
  \bibfield  {author} {\bibinfo {author} {\bibfnamefont {K.~Y.}\ \bibnamefont
  {Bliokh}}, \bibinfo {author} {\bibfnamefont {E.}~\bibnamefont {Karimi}},
  \bibinfo {author} {\bibfnamefont {M.~J.}\ \bibnamefont {Padgett}}, \bibinfo
  {author} {\bibfnamefont {M.~A.}\ \bibnamefont {Alonso}}, \bibinfo {author}
  {\bibfnamefont {M.~R.}\ \bibnamefont {Dennis}}, \bibinfo {author}
  {\bibfnamefont {A.}~\bibnamefont {Dudley}}, \bibinfo {author} {\bibfnamefont
  {A.}~\bibnamefont {Forbes}}, \bibinfo {author} {\bibfnamefont
  {S.}~\bibnamefont {Zahedpour}}, \bibinfo {author} {\bibfnamefont {S.~W.}\
  \bibnamefont {Hancock}}, \bibinfo {author} {\bibfnamefont {H.~M.}\
  \bibnamefont {Milchberg}}, \bibinfo {author} {\bibfnamefont {S.}~\bibnamefont
  {Rotter}}, \bibinfo {author} {\bibfnamefont {F.}~\bibnamefont {Nori}},
  \bibinfo {author} {\bibfnamefont {{\ifmmode\mbox{\c{S}}\else\c{S}\fi}.~K.}\
  \bibnamefont {{\ifmmode\ddot{O}\else\"{O}\fi}zdemir}}, \bibinfo {author}
  {\bibfnamefont {N.}~\bibnamefont {Bender}}, \bibinfo {author} {\bibfnamefont
  {H.}~\bibnamefont {Cao}}, \bibinfo {author} {\bibfnamefont {P.~B.}\
  \bibnamefont {Corkum}}, \bibinfo {author} {\bibfnamefont {C.}~\bibnamefont
  {Hern{\ifmmode\acute{a}\else\'{a}\fi}ndez-Garc{\ifmmode\acute{\imath}\else\'{\i}\fi}a}},
  \bibinfo {author} {\bibfnamefont {H.}~\bibnamefont {Ren}}, \bibinfo {author}
  {\bibfnamefont {Y.}~\bibnamefont {Kivshar}}, \bibinfo {author} {\bibfnamefont
  {M.~G.}\ \bibnamefont {Silveirinha}}, \bibinfo {author} {\bibfnamefont
  {N.}~\bibnamefont {Engheta}}, \bibinfo {author} {\bibfnamefont
  {A.}~\bibnamefont {Rauschenbeutel}}, \bibinfo {author} {\bibfnamefont
  {P.}~\bibnamefont {Schneeweiss}}, \bibinfo {author} {\bibfnamefont
  {J.}~\bibnamefont {Volz}}, \bibinfo {author} {\bibfnamefont {D.}~\bibnamefont
  {Leykam}}, \bibinfo {author} {\bibfnamefont {D.~A.}\ \bibnamefont
  {Smirnova}}, \bibinfo {author} {\bibfnamefont {K.}~\bibnamefont {Rong}},
  \bibinfo {author} {\bibfnamefont {B.}~\bibnamefont {Wang}}, \bibinfo {author}
  {\bibfnamefont {E.}~\bibnamefont {Hasman}}, \bibinfo {author} {\bibfnamefont
  {M.~F.}\ \bibnamefont {Picardi}}, \bibinfo {author} {\bibfnamefont {A.~V.}\
  \bibnamefont {Zayats}}, \bibinfo {author} {\bibfnamefont {F.~J.}\
  \bibnamefont
  {Rodr{\ifmmode\acute{\imath}\else\'{\i}\fi}guez-Fortu{\ifmmode\tilde{n}\else\~{n}\fi}o}},
  \bibinfo {author} {\bibfnamefont {C.}~\bibnamefont {Yang}}, \bibinfo {author}
  {\bibfnamefont {J.}~\bibnamefont {Ren}}, \bibinfo {author} {\bibfnamefont
  {A.~B.}\ \bibnamefont {Khanikaev}}, \bibinfo {author} {\bibfnamefont
  {A.}~\bibnamefont {Al{\ifmmode\grave{u}\else\`{u}\fi}}}, \bibinfo {author}
  {\bibfnamefont {E.}~\bibnamefont {Brasselet}}, \bibinfo {author}
  {\bibfnamefont {M.}~\bibnamefont {Shats}}, \bibinfo {author} {\bibfnamefont
  {J.}~\bibnamefont {Verbeeck}}, \bibinfo {author} {\bibfnamefont
  {P.}~\bibnamefont {Schattschneider}}, \bibinfo {author} {\bibfnamefont
  {D.}~\bibnamefont {Sarenac}}, \bibinfo {author} {\bibfnamefont {D.~G.}\
  \bibnamefont {Cory}}, \bibinfo {author} {\bibfnamefont {D.~A.}\ \bibnamefont
  {Pushin}}, \bibinfo {author} {\bibfnamefont {M.}~\bibnamefont {Birk}},
  \bibinfo {author} {\bibfnamefont {A.}~\bibnamefont {Gorlach}}, \bibinfo
  {author} {\bibfnamefont {I.}~\bibnamefont {Kaminer}}, \bibinfo {author}
  {\bibfnamefont {F.}~\bibnamefont {Cardano}}, \bibinfo {author} {\bibfnamefont
  {L.}~\bibnamefont {Marrucci}}, \bibinfo {author} {\bibfnamefont
  {M.}~\bibnamefont {Krenn}},\ and\ \bibinfo {author} {\bibfnamefont
  {F.}~\bibnamefont {Marquardt}},\ }\bibfield  {title} {\bibinfo {title}
  {{Roadmap on structured waves}},\ }\href
  {https://doi.org/10.1088/2040-8986/acea92} {\bibfield  {journal} {\bibinfo
  {journal} {J. Opt.}\ }\textbf {\bibinfo {volume} {25}},\ \bibinfo {pages}
  {103001} (\bibinfo {year} {2023})}\BibitemShut {NoStop}%
\bibitem [{\citenamefont {Toftul}\ \emph {et~al.}(2024)\citenamefont {Toftul},
  \citenamefont {Golat}, \citenamefont
  {Rodr{\ifmmode\acute{\imath}\else\'{\i}\fi}guez-Fortu{\ifmmode\tilde{n}\else\~{n}\fi}o},
  \citenamefont {Nori}, \citenamefont {Kivshar},\ and\ \citenamefont
  {Bliokh}}]{Toftul2024}%
  \BibitemOpen
  \bibfield  {author} {\bibinfo {author} {\bibfnamefont {I.}~\bibnamefont
  {Toftul}}, \bibinfo {author} {\bibfnamefont {S.}~\bibnamefont {Golat}},
  \bibinfo {author} {\bibfnamefont {F.~J.}\ \bibnamefont
  {Rodr{\ifmmode\acute{\imath}\else\'{\i}\fi}guez-Fortu{\ifmmode\tilde{n}\else\~{n}\fi}o}},
  \bibinfo {author} {\bibfnamefont {F.}~\bibnamefont {Nori}}, \bibinfo {author}
  {\bibfnamefont {Y.}~\bibnamefont {Kivshar}},\ and\ \bibinfo {author}
  {\bibfnamefont {K.~Y.}\ \bibnamefont {Bliokh}},\ }\bibfield  {title}
  {\bibinfo {title} {{Radiation forces and torques in optics and acoustics}},\
  }\bibfield  {journal} {\bibinfo  {journal} {arXiv}\ }\href
  {https://doi.org/10.48550/arXiv.2410.23670} {10.48550/arXiv.2410.23670}
  (\bibinfo {year} {2024}),\ \Eprint {https://arxiv.org/abs/2410.23670}
  {2410.23670} \BibitemShut {NoStop}%
\bibitem [{\citenamefont {Dholakia}\ \emph {et~al.}(2020)\citenamefont
  {Dholakia}, \citenamefont {Drinkwater},\ and\ \citenamefont
  {Ritsch-Marte}}]{Dholakia2020NRP}%
  \BibitemOpen
  \bibfield  {author} {\bibinfo {author} {\bibfnamefont {K.}~\bibnamefont
  {Dholakia}}, \bibinfo {author} {\bibfnamefont {B.~W.}\ \bibnamefont
  {Drinkwater}},\ and\ \bibinfo {author} {\bibfnamefont {M.}~\bibnamefont
  {Ritsch-Marte}},\ }\bibfield  {title} {\bibinfo {title} {{Comparing acoustic
  and optical forces for biomedical research}},\ }\href
  {https://doi.org/10.1038/s42254-020-0215-3} {\bibfield  {journal} {\bibinfo
  {journal} {Nat. Rev. Phys.}\ }\textbf {\bibinfo {volume} {2}},\ \bibinfo
  {pages} {480} (\bibinfo {year} {2020})}\BibitemShut {NoStop}%
\bibitem [{\citenamefont {Akhiezer}\ and\ \citenamefont
  {Berestetsky}(1965)}]{Akhiezer_book}%
  \BibitemOpen
  \bibfield  {author} {\bibinfo {author} {\bibfnamefont {A.~I.}\ \bibnamefont
  {Akhiezer}}\ and\ \bibinfo {author} {\bibfnamefont {V.~B.}\ \bibnamefont
  {Berestetsky}},\ }\href@noop {} {\emph {\bibinfo {title} {{Quantum
  Electrodynamics}}}}\ (\bibinfo  {publisher} {Interscience},\ \bibinfo
  {address} {New York},\ \bibinfo {year} {1965})\BibitemShut {NoStop}%
\bibitem [{\citenamefont {Born}\ and\ \citenamefont {Wolf}(1999)}]{BornWolf}%
  \BibitemOpen
  \bibfield  {author} {\bibinfo {author} {\bibfnamefont {M.}~\bibnamefont
  {Born}}\ and\ \bibinfo {author} {\bibfnamefont {E.}~\bibnamefont {Wolf}},\
  }\href@noop {} {\emph {\bibinfo {title} {{Principles of Optics}}}}\ (\bibinfo
   {publisher} {Cambridge University Press},\ \bibinfo {address} {Cambridge},\
  \bibinfo {year} {1999})\BibitemShut {NoStop}%
\bibitem [{\citenamefont {Van~Enk}\ and\ \citenamefont
  {Nienhuis}(1994)}]{VanEnk1994JMO}%
  \BibitemOpen
  \bibfield  {author} {\bibinfo {author} {\bibfnamefont {S.~J.}\ \bibnamefont
  {Van~Enk}}\ and\ \bibinfo {author} {\bibfnamefont {G.}~\bibnamefont
  {Nienhuis}},\ }\bibfield  {title} {\bibinfo {title} {{Commutation Rules and
  Eigenvalues of Spin and Orbital Angular Momentum of Radiation Fields}},\
  }\href {https://doi.org/10.1080/09500349414550911} {\bibfield  {journal}
  {\bibinfo  {journal} {J. Mod. Opt.}\ }\textbf {\bibinfo {volume} {41}},\
  \bibinfo {pages} {963} (\bibinfo {year} {1994})}\BibitemShut {NoStop}%
\bibitem [{\citenamefont {Bliokh}\ \emph {et~al.}(2010)\citenamefont {Bliokh},
  \citenamefont {Alonso}, \citenamefont {Ostrovskaya},\ and\ \citenamefont
  {Aiello}}]{Bliokh2010PRA}%
  \BibitemOpen
  \bibfield  {author} {\bibinfo {author} {\bibfnamefont {K.~Y.}\ \bibnamefont
  {Bliokh}}, \bibinfo {author} {\bibfnamefont {M.~A.}\ \bibnamefont {Alonso}},
  \bibinfo {author} {\bibfnamefont {E.~A.}\ \bibnamefont {Ostrovskaya}},\ and\
  \bibinfo {author} {\bibfnamefont {A.}~\bibnamefont {Aiello}},\ }\bibfield
  {title} {\bibinfo {title} {{Angular momenta and spin-orbit interaction of
  nonparaxial light in free space}},\ }\href
  {https://doi.org/10.1103/PhysRevA.82.063825} {\bibfield  {journal} {\bibinfo
  {journal} {Phys. Rev. A}\ }\textbf {\bibinfo {volume} {82}},\ \bibinfo
  {pages} {063825} (\bibinfo {year} {2010})}\BibitemShut {NoStop}%
\bibitem [{\citenamefont {Picardi}\ \emph {et~al.}(2018)\citenamefont
  {Picardi}, \citenamefont {Bliokh}, \citenamefont
  {Rodr{\ifmmode\acute{\imath}\else\'{\i}\fi}guez-Fortu{\ifmmode\tilde{n}\else\~{n}\fi}o},
  \citenamefont {Alpeggiani},\ and\ \citenamefont {Nori}}]{Picardi2018O}%
  \BibitemOpen
  \bibfield  {author} {\bibinfo {author} {\bibfnamefont {M.~F.}\ \bibnamefont
  {Picardi}}, \bibinfo {author} {\bibfnamefont {K.~Y.}\ \bibnamefont {Bliokh}},
  \bibinfo {author} {\bibfnamefont {F.~J.}\ \bibnamefont
  {Rodr{\ifmmode\acute{\imath}\else\'{\i}\fi}guez-Fortu{\ifmmode\tilde{n}\else\~{n}\fi}o}},
  \bibinfo {author} {\bibfnamefont {F.}~\bibnamefont {Alpeggiani}},\ and\
  \bibinfo {author} {\bibfnamefont {F.}~\bibnamefont {Nori}},\ }\bibfield
  {title} {\bibinfo {title} {{Angular momenta, helicity, and other properties
  of dielectric-fiber and metallic-wire modes}},\ }\href
  {https://doi.org/10.1364/OPTICA.5.001016} {\bibfield  {journal} {\bibinfo
  {journal} {Optica}\ }\textbf {\bibinfo {volume} {5}},\ \bibinfo {pages}
  {1016} (\bibinfo {year} {2018})}\BibitemShut {NoStop}%
\bibitem [{\citenamefont {Soper}(1976)}]{Soper_book}%
  \BibitemOpen
  \bibfield  {author} {\bibinfo {author} {\bibfnamefont {D.~E.}\ \bibnamefont
  {Soper}},\ }\href@noop {} {\emph {\bibinfo {title} {Classical {{Field
  Theory}}}}}\ (\bibinfo  {publisher} {Wiley},\ \bibinfo {year}
  {1976})\BibitemShut {NoStop}%
\bibitem [{\citenamefont {Berry}\ and\ \citenamefont
  {Dennis}(2001)}]{Berry2001}%
  \BibitemOpen
  \bibfield  {author} {\bibinfo {author} {\bibfnamefont {M.~V.}\ \bibnamefont
  {Berry}}\ and\ \bibinfo {author} {\bibfnamefont {M.~R.}\ \bibnamefont
  {Dennis}},\ }\bibfield  {title} {\bibinfo {title} {{Polarization
  singularities in isotropic random vector waves}},\ }\href
  {https://doi.org/10.1098/rspa.2000.0660} {\bibfield  {journal} {\bibinfo
  {journal} {Proc. R. Soc. Lond. A.}\ }\textbf {\bibinfo {volume} {457}},\
  \bibinfo {pages} {141} (\bibinfo {year} {2001})}\BibitemShut {NoStop}%
\bibitem [{\citenamefont {Bliokh}\ \emph {et~al.}(2014)\citenamefont {Bliokh},
  \citenamefont {Bekshaev},\ and\ \citenamefont {Nori}}]{Bliokh2014NC}%
  \BibitemOpen
  \bibfield  {author} {\bibinfo {author} {\bibfnamefont {K.~Y.}\ \bibnamefont
  {Bliokh}}, \bibinfo {author} {\bibfnamefont {A.~Y.}\ \bibnamefont
  {Bekshaev}},\ and\ \bibinfo {author} {\bibfnamefont {F.}~\bibnamefont
  {Nori}},\ }\bibfield  {title} {\bibinfo {title} {{Extraordinary momentum and
  spin in evanescent waves}},\ }\href {https://doi.org/10.1038/ncomms4300}
  {\bibfield  {journal} {\bibinfo  {journal} {Nat. Commun.}\ }\textbf {\bibinfo
  {volume} {5}},\ \bibinfo {pages} {3300} (\bibinfo {year} {2014})}\BibitemShut
  {NoStop}%
\bibitem [{\citenamefont {Toftul}\ \emph {et~al.}(2019)\citenamefont {Toftul},
  \citenamefont {Bliokh}, \citenamefont {Petrov},\ and\ \citenamefont
  {Nori}}]{Toftul2019PRL}%
  \BibitemOpen
  \bibfield  {author} {\bibinfo {author} {\bibfnamefont {I.~D.}\ \bibnamefont
  {Toftul}}, \bibinfo {author} {\bibfnamefont {K.~Y.}\ \bibnamefont {Bliokh}},
  \bibinfo {author} {\bibfnamefont {M.~I.}\ \bibnamefont {Petrov}},\ and\
  \bibinfo {author} {\bibfnamefont {F.}~\bibnamefont {Nori}},\ }\bibfield
  {title} {\bibinfo {title} {{Acoustic Radiation Force and Torque on Small
  Particles as Measures of the Canonical Momentum and Spin Densities}},\ }\href
  {https://doi.org/10.1103/PhysRevLett.123.183901} {\bibfield  {journal}
  {\bibinfo  {journal} {Phys. Rev. Lett.}\ }\textbf {\bibinfo {volume} {123}},\
  \bibinfo {pages} {183901} (\bibinfo {year} {2019})}\BibitemShut {NoStop}%
\bibitem [{\citenamefont {Wang}\ \emph {et~al.}(2025)\citenamefont {Wang},
  \citenamefont {Che}, \citenamefont {Cheng}, \citenamefont {Tong},
  \citenamefont {Shi}, \citenamefont {Shen}, \citenamefont {Bliokh},\ and\
  \citenamefont {Zi}}]{Wang2025N}%
  \BibitemOpen
  \bibfield  {author} {\bibinfo {author} {\bibfnamefont {B.}~\bibnamefont
  {Wang}}, \bibinfo {author} {\bibfnamefont {Z.}~\bibnamefont {Che}}, \bibinfo
  {author} {\bibfnamefont {C.}~\bibnamefont {Cheng}}, \bibinfo {author}
  {\bibfnamefont {C.}~\bibnamefont {Tong}}, \bibinfo {author} {\bibfnamefont
  {L.}~\bibnamefont {Shi}}, \bibinfo {author} {\bibfnamefont {Y.}~\bibnamefont
  {Shen}}, \bibinfo {author} {\bibfnamefont {K.~Y.}\ \bibnamefont {Bliokh}},\
  and\ \bibinfo {author} {\bibfnamefont {J.}~\bibnamefont {Zi}},\ }\bibfield
  {title} {\bibinfo {title} {{Topological water-wave structures manipulating
  particles}},\ }\href {https://doi.org/10.1038/s41586-024-08384-y} {\bibfield
  {journal} {\bibinfo  {journal} {Nature}\ }\textbf {\bibinfo {volume} {638}},\
  \bibinfo {pages} {394} (\bibinfo {year} {2025})}\BibitemShut {NoStop}%
\bibitem [{\citenamefont
  {Jackson}(1999)}]{jackson1998ClassicalElectrodynamics}%
  \BibitemOpen
  \bibfield  {author} {\bibinfo {author} {\bibfnamefont {J.~D.}\ \bibnamefont
  {Jackson}},\ }\href@noop {} {\emph {\bibinfo {title} {Classical
  {{Electrodynamics}}}}}\ (\bibinfo  {publisher} {Wiley},\ \bibinfo {year}
  {1999})\BibitemShut {NoStop}%
\bibitem [{\citenamefont {Landau}\ and\ \citenamefont
  {Lifshitz}(2013)}]{Landau_fluid}%
  \BibitemOpen
  \bibfield  {author} {\bibinfo {author} {\bibfnamefont {L.~D.}\ \bibnamefont
  {Landau}}\ and\ \bibinfo {author} {\bibfnamefont {E.~M.}\ \bibnamefont
  {Lifshitz}},\ }\href@noop {} {\emph {\bibinfo {title} {Fluid Mechanics}}}\
  (\bibinfo  {publisher} {Pergamon},\ \bibinfo {address} {Oxford},\ \bibinfo
  {year} {2013})\BibitemShut {NoStop}%
\bibitem [{\citenamefont {Auld}(1973)}]{Auld_book}%
  \BibitemOpen
  \bibfield  {author} {\bibinfo {author} {\bibfnamefont {B.~A.}\ \bibnamefont
  {Auld}},\ }\href@noop {} {\emph {\bibinfo {title} {{Acoustic Fields and Waves
  in Solids. Volume I}}}}\ (\bibinfo  {publisher} {Wiley},\ \bibinfo {address}
  {Hoboken},\ \bibinfo {year} {1973})\BibitemShut {NoStop}%
\bibitem [{\citenamefont {Belinfante}(1940)}]{Belinfante1940}%
  \BibitemOpen
  \bibfield  {author} {\bibinfo {author} {\bibfnamefont {F.~J.}\ \bibnamefont
  {Belinfante}},\ }\bibfield  {title} {\bibinfo {title} {{On the current and
  the density of the electric charge, the energy, the linear momentum and the
  angular momentum of arbitrary fields}},\ }\href
  {https://doi.org/10.1016/S0031-8914(40)90091-X} {\bibfield  {journal}
  {\bibinfo  {journal} {Physica}\ }\textbf {\bibinfo {volume} {7}},\ \bibinfo
  {pages} {449} (\bibinfo {year} {1940})}\BibitemShut {NoStop}%
\bibitem [{\citenamefont {Berry}(2009)}]{Berry2009}%
  \BibitemOpen
  \bibfield  {author} {\bibinfo {author} {\bibfnamefont {M.~V.}\ \bibnamefont
  {Berry}},\ }\bibfield  {title} {\bibinfo {title} {{Optical currents}},\
  }\href {https://doi.org/10.1088/1464-4258/11/9/094001} {\bibfield  {journal}
  {\bibinfo  {journal} {J. Opt. A: Pure Appl. Opt.}\ }\textbf {\bibinfo
  {volume} {11}},\ \bibinfo {pages} {094001} (\bibinfo {year}
  {2009})}\BibitemShut {NoStop}%
\bibitem [{\citenamefont {Markovich}\ and\ \citenamefont
  {Lubensky}(2021)}]{Markovich2021PRL}%
  \BibitemOpen
  \bibfield  {author} {\bibinfo {author} {\bibfnamefont {T.}~\bibnamefont
  {Markovich}}\ and\ \bibinfo {author} {\bibfnamefont {T.~C.}\ \bibnamefont
  {Lubensky}},\ }\bibfield  {title} {\bibinfo {title} {{Odd Viscosity in Active
  Matter: Microscopic Origin and 3D Effects}},\ }\href
  {https://doi.org/10.1103/PhysRevLett.127.048001} {\bibfield  {journal}
  {\bibinfo  {journal} {Phys. Rev. Lett.}\ }\textbf {\bibinfo {volume} {127}},\
  \bibinfo {pages} {048001} (\bibinfo {year} {2021})}\BibitemShut {NoStop}%
\bibitem [{\citenamefont {Mita}(2000)}]{Mita2000AJP}%
  \BibitemOpen
  \bibfield  {author} {\bibinfo {author} {\bibfnamefont {K.}~\bibnamefont
  {Mita}},\ }\bibfield  {title} {\bibinfo {title} {{Virtual probability current
  associated with the spin}},\ }\href {https://doi.org/10.1119/1.19421}
  {\bibfield  {journal} {\bibinfo  {journal} {Am. J. Phys.}\ }\textbf {\bibinfo
  {volume} {68}},\ \bibinfo {pages} {259} (\bibinfo {year} {2000})}\BibitemShut
  {NoStop}%
\bibitem [{\citenamefont {Antognozzi}\ \emph {et~al.}(2016)\citenamefont
  {Antognozzi}, \citenamefont {Bermingham}, \citenamefont {Harniman},
  \citenamefont {Simpson}, \citenamefont {Senior}, \citenamefont {Hayward},
  \citenamefont {Hoerber}, \citenamefont {Dennis}, \citenamefont {Bekshaev},
  \citenamefont {Bliokh},\ and\ \citenamefont {Nori}}]{Antognozzi2016NP}%
  \BibitemOpen
  \bibfield  {author} {\bibinfo {author} {\bibfnamefont {M.}~\bibnamefont
  {Antognozzi}}, \bibinfo {author} {\bibfnamefont {C.~R.}\ \bibnamefont
  {Bermingham}}, \bibinfo {author} {\bibfnamefont {R.~L.}\ \bibnamefont
  {Harniman}}, \bibinfo {author} {\bibfnamefont {S.}~\bibnamefont {Simpson}},
  \bibinfo {author} {\bibfnamefont {J.}~\bibnamefont {Senior}}, \bibinfo
  {author} {\bibfnamefont {R.}~\bibnamefont {Hayward}}, \bibinfo {author}
  {\bibfnamefont {H.}~\bibnamefont {Hoerber}}, \bibinfo {author} {\bibfnamefont
  {M.~R.}\ \bibnamefont {Dennis}}, \bibinfo {author} {\bibfnamefont {A.~Y.}\
  \bibnamefont {Bekshaev}}, \bibinfo {author} {\bibfnamefont {K.~Y.}\
  \bibnamefont {Bliokh}},\ and\ \bibinfo {author} {\bibfnamefont
  {F.}~\bibnamefont {Nori}},\ }\bibfield  {title} {\bibinfo {title} {Direct
  measurements of the extraordinary optical momentum and transverse
  spin-dependent force using a nano-cantilever},\ }\href
  {https://doi.org/10.1038/nphys3732} {\bibfield  {journal} {\bibinfo
  {journal} {Nat. Phys.}\ }\textbf {\bibinfo {volume} {12}},\ \bibinfo {pages}
  {731} (\bibinfo {year} {2016})}\BibitemShut {NoStop}%
\bibitem [{\citenamefont {Ghosh}\ \emph {et~al.}(2024)\citenamefont {Ghosh},
  \citenamefont {Daniel}, \citenamefont {Gorzkowski}, \citenamefont {Bekshaev},
  \citenamefont {Lapkiewicz},\ and\ \citenamefont {Bliokh}}]{Ghosh2024JOSA}%
  \BibitemOpen
  \bibfield  {author} {\bibinfo {author} {\bibfnamefont {B.}~\bibnamefont
  {Ghosh}}, \bibinfo {author} {\bibfnamefont {A.}~\bibnamefont {Daniel}},
  \bibinfo {author} {\bibfnamefont {B.}~\bibnamefont {Gorzkowski}}, \bibinfo
  {author} {\bibfnamefont {A.~Y.}\ \bibnamefont {Bekshaev}}, \bibinfo {author}
  {\bibfnamefont {R.}~\bibnamefont {Lapkiewicz}},\ and\ \bibinfo {author}
  {\bibfnamefont {K.~Y.}\ \bibnamefont {Bliokh}},\ }\bibfield  {title}
  {\bibinfo {title} {{Canonical and Poynting currents in propagation and
  diffraction of structured light: tutorial}},\ }\href
  {https://doi.org/10.1364/JOSAB.522393} {\bibfield  {journal} {\bibinfo
  {journal} {J. Opt. Soc. Am. B}\ }\textbf {\bibinfo {volume} {41}},\ \bibinfo
  {pages} {1276} (\bibinfo {year} {2024})}\BibitemShut {NoStop}%
\bibitem [{\citenamefont {Bekshaev}\ \emph {et~al.}(2015)\citenamefont
  {Bekshaev}, \citenamefont {Bliokh},\ and\ \citenamefont
  {Nori}}]{Bekshaev2015PRX}%
  \BibitemOpen
  \bibfield  {author} {\bibinfo {author} {\bibfnamefont {A.~Y.}\ \bibnamefont
  {Bekshaev}}, \bibinfo {author} {\bibfnamefont {K.~Y.}\ \bibnamefont
  {Bliokh}},\ and\ \bibinfo {author} {\bibfnamefont {F.}~\bibnamefont {Nori}},\
  }\bibfield  {title} {\bibinfo {title} {Transverse spin and momentum in
  two-wave interference},\ }\href {https://doi.org/10.1103/PhysRevX.5.011039}
  {\bibfield  {journal} {\bibinfo  {journal} {Phys. Rev. X}\ }\textbf {\bibinfo
  {volume} {5}},\ \bibinfo {pages} {011039} (\bibinfo {year}
  {2015})}\BibitemShut {NoStop}%
\bibitem [{\citenamefont {Yurchenko}(2002)}]{Yurchenko2002AJP}%
  \BibitemOpen
  \bibfield  {author} {\bibinfo {author} {\bibfnamefont {V.~B.}\ \bibnamefont
  {Yurchenko}},\ }\bibfield  {title} {\bibinfo {title} {{Answer to Question
  {\#}79. Does plane wave not carry a spin?}},\ }\href
  {https://doi.org/10.1119/1.1463741} {\bibfield  {journal} {\bibinfo
  {journal} {Am. J. Phys.}\ }\textbf {\bibinfo {volume} {70}},\ \bibinfo
  {pages} {568} (\bibinfo {year} {2002})}\BibitemShut {NoStop}%
\bibitem [{\citenamefont {Bliokh}\ and\ \citenamefont
  {Nori}(2019{\natexlab{b}})}]{Bliokh2019PRL}%
  \BibitemOpen
  \bibfield  {author} {\bibinfo {author} {\bibfnamefont {K.~Y.}\ \bibnamefont
  {Bliokh}}\ and\ \bibinfo {author} {\bibfnamefont {F.}~\bibnamefont {Nori}},\
  }\bibfield  {title} {\bibinfo {title} {{Klein-Gordon Representation of
  Acoustic Waves and Topological Origin of Surface Acoustic Modes}},\ }\href
  {https://doi.org/10.1103/PhysRevLett.123.054301} {\bibfield  {journal}
  {\bibinfo  {journal} {Phys. Rev. Lett.}\ }\textbf {\bibinfo {volume} {123}},\
  \bibinfo {pages} {054301} (\bibinfo {year} {2019}{\natexlab{b}})}\BibitemShut
  {NoStop}%
\bibitem [{\citenamefont {Burns}\ \emph {et~al.}(2020)\citenamefont {Burns},
  \citenamefont {Bliokh}, \citenamefont {Nori},\ and\ \citenamefont
  {Dressel}}]{Burns2020}%
  \BibitemOpen
  \bibfield  {author} {\bibinfo {author} {\bibfnamefont {L.}~\bibnamefont
  {Burns}}, \bibinfo {author} {\bibfnamefont {K.~Y.}\ \bibnamefont {Bliokh}},
  \bibinfo {author} {\bibfnamefont {F.}~\bibnamefont {Nori}},\ and\ \bibinfo
  {author} {\bibfnamefont {J.}~\bibnamefont {Dressel}},\ }\bibfield  {title}
  {\bibinfo {title} {{Acoustic versus electromagnetic field theory: scalar,
  vector, spinor representations and the emergence of acoustic spin}},\ }\href
  {https://doi.org/10.1088/1367-2630/ab7f91} {\bibfield  {journal} {\bibinfo
  {journal} {New J. Phys.}\ }\textbf {\bibinfo {volume} {22}},\ \bibinfo
  {pages} {053050} (\bibinfo {year} {2020})}\BibitemShut {NoStop}%
\bibitem [{\citenamefont {Gaponov}\ and\ \citenamefont
  {Miller}(1958)}]{Gaponov1958}%
  \BibitemOpen
  \bibfield  {author} {\bibinfo {author} {\bibfnamefont {A.~V.}\ \bibnamefont
  {Gaponov}}\ and\ \bibinfo {author} {\bibfnamefont {M.~A.}\ \bibnamefont
  {Miller}},\ }\bibfield  {title} {\bibinfo {title} {{Potential wells for
  charged particles in a high-frequency electromagnetic field}},\ }\href
  {http://www.jetp.ras.ru/cgi-bin/dn/e_007_01_0168.pdf} {\bibfield  {journal}
  {\bibinfo  {journal} {Sov. Phys. JETP}\ }\textbf {\bibinfo {volume} {7}},\
  \bibinfo {pages} {168} (\bibinfo {year} {1958})}\BibitemShut {NoStop}%
\bibitem [{\citenamefont {Bliokh}\ \emph {et~al.}(2013)\citenamefont {Bliokh},
  \citenamefont {Bekshaev},\ and\ \citenamefont {Nori}}]{Bliokh2013NJP}%
  \BibitemOpen
  \bibfield  {author} {\bibinfo {author} {\bibfnamefont {K.~Y.}\ \bibnamefont
  {Bliokh}}, \bibinfo {author} {\bibfnamefont {A.~Y.}\ \bibnamefont
  {Bekshaev}},\ and\ \bibinfo {author} {\bibfnamefont {F.}~\bibnamefont
  {Nori}},\ }\bibfield  {title} {\bibinfo {title} {{Dual electromagnetism:
  helicity, spin, momentum and angular momentum}},\ }\href
  {https://doi.org/10.1088/1367-2630/15/3/033026} {\bibfield  {journal}
  {\bibinfo  {journal} {New J. Phys.}\ }\textbf {\bibinfo {volume} {15}},\
  \bibinfo {pages} {033026} (\bibinfo {year} {2013})},\ \bibinfo {note}
  {{Corrigendum: New J. Phys. {\bf 18}, 089503 (2016)}}\BibitemShut {NoStop}%
\bibitem [{\citenamefont {Aiello}\ \emph {et~al.}(2015)\citenamefont {Aiello},
  \citenamefont {Banzer}, \citenamefont {Neugebauer},\ and\ \citenamefont
  {Leuchs}}]{Aiello2015NP}%
  \BibitemOpen
  \bibfield  {author} {\bibinfo {author} {\bibfnamefont {A.}~\bibnamefont
  {Aiello}}, \bibinfo {author} {\bibfnamefont {P.}~\bibnamefont {Banzer}},
  \bibinfo {author} {\bibfnamefont {M.}~\bibnamefont {Neugebauer}},\ and\
  \bibinfo {author} {\bibfnamefont {G.}~\bibnamefont {Leuchs}},\ }\bibfield
  {title} {\bibinfo {title} {{From transverse angular momentum to photonic
  wheels}},\ }\href {https://doi.org/10.1038/nphoton.2015.203} {\bibfield
  {journal} {\bibinfo  {journal} {Nat. Photonics}\ }\textbf {\bibinfo {volume}
  {9}},\ \bibinfo {pages} {789} (\bibinfo {year} {2015})}\BibitemShut {NoStop}%
\bibitem [{\citenamefont {Cameron}\ \emph {et~al.}(2012)\citenamefont
  {Cameron}, \citenamefont {Barnett},\ and\ \citenamefont {Yao}}]{Cameron2012}%
  \BibitemOpen
  \bibfield  {author} {\bibinfo {author} {\bibfnamefont {R.~P.}\ \bibnamefont
  {Cameron}}, \bibinfo {author} {\bibfnamefont {S.~M.}\ \bibnamefont
  {Barnett}},\ and\ \bibinfo {author} {\bibfnamefont {A.~M.}\ \bibnamefont
  {Yao}},\ }\bibfield  {title} {\bibinfo {title} {{Optical helicity, optical
  spin and related quantities in electromagnetic theory}},\ }\href
  {https://doi.org/10.1088/1367-2630/14/5/053050} {\bibfield  {journal}
  {\bibinfo  {journal} {New J. Phys.}\ }\textbf {\bibinfo {volume} {14}},\
  \bibinfo {pages} {053050} (\bibinfo {year} {2012})}\BibitemShut {NoStop}%
\bibitem [{\citenamefont {Bliokh}\ \emph
  {et~al.}(2017{\natexlab{a}})\citenamefont {Bliokh}, \citenamefont
  {Bekshaev},\ and\ \citenamefont {Nori}}]{Bliokh2017NJP}%
  \BibitemOpen
  \bibfield  {author} {\bibinfo {author} {\bibfnamefont {K.~Y.}\ \bibnamefont
  {Bliokh}}, \bibinfo {author} {\bibfnamefont {A.~Y.}\ \bibnamefont
  {Bekshaev}},\ and\ \bibinfo {author} {\bibfnamefont {F.}~\bibnamefont
  {Nori}},\ }\bibfield  {title} {\bibinfo {title} {Optical momentum and angular
  momentum in complex media: {{From}} the {{Abraham}}-{{Minkowski}} debate to
  unusual properties of surface plasmon-polaritons},\ }\href
  {https://doi.org/10.1088/1367-2630/aa8913} {\bibfield  {journal} {\bibinfo
  {journal} {New J. Phys.}\ }\textbf {\bibinfo {volume} {19}},\ \bibinfo
  {pages} {123014} (\bibinfo {year} {2017}{\natexlab{a}})}\BibitemShut
  {NoStop}%
\bibitem [{\citenamefont {Cameron}\ and\ \citenamefont
  {Barnett}(2012)}]{Cameron2012NJP_II}%
  \BibitemOpen
  \bibfield  {author} {\bibinfo {author} {\bibfnamefont {R.~P.}\ \bibnamefont
  {Cameron}}\ and\ \bibinfo {author} {\bibfnamefont {S.~M.}\ \bibnamefont
  {Barnett}},\ }\bibfield  {title} {\bibinfo {title} {{Electric-magnetic
  symmetry and Noether's theorem}},\ }\href
  {https://doi.org/10.1088/1367-2630/14/12/123019} {\bibfield  {journal}
  {\bibinfo  {journal} {New J. Phys.}\ }\textbf {\bibinfo {volume} {14}},\
  \bibinfo {pages} {123019} (\bibinfo {year} {2012})}\BibitemShut {NoStop}%
\bibitem [{\citenamefont {Neugebauer}\ \emph {et~al.}(2015)\citenamefont
  {Neugebauer}, \citenamefont {Bauer}, \citenamefont {Aiello},\ and\
  \citenamefont {Banzer}}]{Neugebauer2015PRL}%
  \BibitemOpen
  \bibfield  {author} {\bibinfo {author} {\bibfnamefont {M.}~\bibnamefont
  {Neugebauer}}, \bibinfo {author} {\bibfnamefont {T.}~\bibnamefont {Bauer}},
  \bibinfo {author} {\bibfnamefont {A.}~\bibnamefont {Aiello}},\ and\ \bibinfo
  {author} {\bibfnamefont {P.}~\bibnamefont {Banzer}},\ }\bibfield  {title}
  {\bibinfo {title} {{Measuring the Transverse Spin Density of Light}},\ }\href
  {https://doi.org/10.1103/PhysRevLett.114.063901} {\bibfield  {journal}
  {\bibinfo  {journal} {Phys. Rev. Lett.}\ }\textbf {\bibinfo {volume} {114}},\
  \bibinfo {pages} {063901} (\bibinfo {year} {2015})}\BibitemShut {NoStop}%
\bibitem [{\citenamefont {Neugebauer}\ \emph {et~al.}(2018)\citenamefont
  {Neugebauer}, \citenamefont {Eismann}, \citenamefont {Bauer},\ and\
  \citenamefont {Banzer}}]{Neugebauer2018PRX}%
  \BibitemOpen
  \bibfield  {author} {\bibinfo {author} {\bibfnamefont {M.}~\bibnamefont
  {Neugebauer}}, \bibinfo {author} {\bibfnamefont {J.~S.}\ \bibnamefont
  {Eismann}}, \bibinfo {author} {\bibfnamefont {T.}~\bibnamefont {Bauer}},\
  and\ \bibinfo {author} {\bibfnamefont {P.}~\bibnamefont {Banzer}},\
  }\bibfield  {title} {\bibinfo {title} {{Magnetic and Electric Transverse Spin
  Density of Spatially Confined Light}},\ }\href
  {https://doi.org/10.1103/PhysRevX.8.021042} {\bibfield  {journal} {\bibinfo
  {journal} {Phys. Rev. X}\ }\textbf {\bibinfo {volume} {8}},\ \bibinfo {pages}
  {021042} (\bibinfo {year} {2018})}\BibitemShut {NoStop}%
\bibitem [{\citenamefont {Landau}\ and\ \citenamefont
  {Lifshitz}(1986)}]{LL_elasticity}%
  \BibitemOpen
  \bibfield  {author} {\bibinfo {author} {\bibfnamefont {L.~D.}\ \bibnamefont
  {Landau}}\ and\ \bibinfo {author} {\bibfnamefont {E.~M.}\ \bibnamefont
  {Lifshitz}},\ }\href@noop {} {\emph {\bibinfo {title} {{Theory of
  Elasticity}}}}\ (\bibinfo  {publisher} {Butterworth-Heinemann},\ \bibinfo
  {address} {Oxford},\ \bibinfo {year} {1986})\BibitemShut {NoStop}%
\bibitem [{\citenamefont {Long}\ \emph {et~al.}(2018)\citenamefont {Long},
  \citenamefont {Ren},\ and\ \citenamefont {Chen}}]{Long2018PNAS}%
  \BibitemOpen
  \bibfield  {author} {\bibinfo {author} {\bibfnamefont {Y.}~\bibnamefont
  {Long}}, \bibinfo {author} {\bibfnamefont {J.}~\bibnamefont {Ren}},\ and\
  \bibinfo {author} {\bibfnamefont {H.}~\bibnamefont {Chen}},\ }\bibfield
  {title} {\bibinfo {title} {{Intrinsic spin of elastic waves}},\ }\href
  {https://doi.org/10.1073/pnas.1808534115} {\bibfield  {journal} {\bibinfo
  {journal} {Proc. Natl. Acad. Sci. U.S.A.}\ }\textbf {\bibinfo {volume}
  {115}},\ \bibinfo {pages} {9951} (\bibinfo {year} {2018})}\BibitemShut
  {NoStop}%
\bibitem [{\citenamefont {Yang}\ \emph {et~al.}(2023)\citenamefont {Yang},
  \citenamefont {Zhang}, \citenamefont {Zhao}, \citenamefont {Gao},
  \citenamefont {Yuan}, \citenamefont {Long}, \citenamefont {Pan},
  \citenamefont {Chen}, \citenamefont {Nori}, \citenamefont {Bliokh},
  \citenamefont {Zhong},\ and\ \citenamefont {Ren}}]{Yang2023PRL}%
  \BibitemOpen
  \bibfield  {author} {\bibinfo {author} {\bibfnamefont {C.}~\bibnamefont
  {Yang}}, \bibinfo {author} {\bibfnamefont {D.}~\bibnamefont {Zhang}},
  \bibinfo {author} {\bibfnamefont {J.}~\bibnamefont {Zhao}}, \bibinfo {author}
  {\bibfnamefont {W.}~\bibnamefont {Gao}}, \bibinfo {author} {\bibfnamefont
  {W.}~\bibnamefont {Yuan}}, \bibinfo {author} {\bibfnamefont {Y.}~\bibnamefont
  {Long}}, \bibinfo {author} {\bibfnamefont {Y.}~\bibnamefont {Pan}}, \bibinfo
  {author} {\bibfnamefont {H.}~\bibnamefont {Chen}}, \bibinfo {author}
  {\bibfnamefont {F.}~\bibnamefont {Nori}}, \bibinfo {author} {\bibfnamefont
  {K.~Y.}\ \bibnamefont {Bliokh}}, \bibinfo {author} {\bibfnamefont
  {Z.}~\bibnamefont {Zhong}},\ and\ \bibinfo {author} {\bibfnamefont
  {J.}~\bibnamefont {Ren}},\ }\bibfield  {title} {\bibinfo {title} {{Hybrid
  Spin and Anomalous Spin-Momentum Locking in Surface Elastic Waves}},\ }\href
  {https://doi.org/10.1103/PhysRevLett.131.136102} {\bibfield  {journal}
  {\bibinfo  {journal} {Phys. Rev. Lett.}\ }\textbf {\bibinfo {volume} {131}},\
  \bibinfo {pages} {136102} (\bibinfo {year} {2023})}\BibitemShut {NoStop}%
\bibitem [{\citenamefont {Akhiezer}\ \emph {et~al.}(1975)\citenamefont
  {Akhiezer}, \citenamefont {Akhiezer}, \citenamefont {Polovin}, \citenamefont
  {Sitenko},\ and\ \citenamefont {Stepanov}}]{Akhiezer_plasma}%
  \BibitemOpen
  \bibfield  {author} {\bibinfo {author} {\bibfnamefont {A.~I.}\ \bibnamefont
  {Akhiezer}}, \bibinfo {author} {\bibfnamefont {I.~A.}\ \bibnamefont
  {Akhiezer}}, \bibinfo {author} {\bibfnamefont {R.~V.}\ \bibnamefont
  {Polovin}}, \bibinfo {author} {\bibfnamefont {A.~G.}\ \bibnamefont
  {Sitenko}},\ and\ \bibinfo {author} {\bibfnamefont {K.~N.}\ \bibnamefont
  {Stepanov}},\ }\href@noop {} {\emph {\bibinfo {title} {{Plasma
  Electrodynamics}}}}\ (\bibinfo  {publisher} {Pergamon},\ \bibinfo {address}
  {Oxford},\ \bibinfo {year} {1975})\BibitemShut {NoStop}%
\bibitem [{\citenamefont {Krall}\ and\ \citenamefont
  {Trivelpiece}(1973)}]{Krall_book}%
  \BibitemOpen
  \bibfield  {author} {\bibinfo {author} {\bibfnamefont {N.~A.}\ \bibnamefont
  {Krall}}\ and\ \bibinfo {author} {\bibfnamefont {A.~W.}\ \bibnamefont
  {Trivelpiece}},\ }\href@noop {} {\emph {\bibinfo {title} {{Principles of
  Plasma Physics}}}}\ (\bibinfo  {publisher} {McGraw-Hill},\ \bibinfo {address}
  {Maidenhead},\ \bibinfo {year} {1973})\BibitemShut {NoStop}%
\bibitem [{\citenamefont {Pfeifer}\ \emph {et~al.}(2007)\citenamefont
  {Pfeifer}, \citenamefont {Nieminen}, \citenamefont {Heckenberg},\ and\
  \citenamefont {Rubinsztein-Dunlop}}]{Pfeifer2007}%
  \BibitemOpen
  \bibfield  {author} {\bibinfo {author} {\bibfnamefont {R.~N.~C.}\
  \bibnamefont {Pfeifer}}, \bibinfo {author} {\bibfnamefont {T.~A.}\
  \bibnamefont {Nieminen}}, \bibinfo {author} {\bibfnamefont {N.~R.}\
  \bibnamefont {Heckenberg}},\ and\ \bibinfo {author} {\bibfnamefont
  {H.}~\bibnamefont {Rubinsztein-Dunlop}},\ }\bibfield  {title} {\bibinfo
  {title} {{Colloquium: Momentum of an electromagnetic wave in dielectric
  media}},\ }\href {https://doi.org/10.1103/RevModPhys.79.1197} {\bibfield
  {journal} {\bibinfo  {journal} {Rev. Mod. Phys.}\ }\textbf {\bibinfo {volume}
  {79}},\ \bibinfo {pages} {1197} (\bibinfo {year} {2007})}\BibitemShut
  {NoStop}%
\bibitem [{\citenamefont {Milonni}\ and\ \citenamefont
  {Boyd}(2010)}]{Milonni2010AOP}%
  \BibitemOpen
  \bibfield  {author} {\bibinfo {author} {\bibfnamefont {P.~W.}\ \bibnamefont
  {Milonni}}\ and\ \bibinfo {author} {\bibfnamefont {R.~W.}\ \bibnamefont
  {Boyd}},\ }\bibfield  {title} {\bibinfo {title} {{Momentum of Light in a
  Dielectric Medium}},\ }\href {https://doi.org/10.1364/AOP.2.000519}
  {\bibfield  {journal} {\bibinfo  {journal} {Adv. Opt. Photonics}\ }\textbf
  {\bibinfo {volume} {2}},\ \bibinfo {pages} {519} (\bibinfo {year}
  {2010})}\BibitemShut {NoStop}%
\bibitem [{\citenamefont {Barnett}\ and\ \citenamefont
  {Loudon}(2010)}]{Barnett2010PTRS}%
  \BibitemOpen
  \bibfield  {author} {\bibinfo {author} {\bibfnamefont {S.~M.}\ \bibnamefont
  {Barnett}}\ and\ \bibinfo {author} {\bibfnamefont {R.}~\bibnamefont
  {Loudon}},\ }\bibfield  {title} {\bibinfo {title} {The enigma of optical
  momentum in a medium},\ }\href {https://doi.org/10.1098/rsta.2009.0207}
  {\bibfield  {journal} {\bibinfo  {journal} {Phil. Trans. R. Soc. A}\ }\textbf
  {\bibinfo {volume} {368}},\ \bibinfo {pages} {927} (\bibinfo {year}
  {2010})}\BibitemShut {NoStop}%
\bibitem [{\citenamefont {Partanen}\ \emph {et~al.}(2017)\citenamefont
  {Partanen}, \citenamefont {H{\ifmmode\ddot{a}\else\"{a}\fi}yrynen},
  \citenamefont {Oksanen},\ and\ \citenamefont {Tulkki}}]{Partanen2017PRA}%
  \BibitemOpen
  \bibfield  {author} {\bibinfo {author} {\bibfnamefont {M.}~\bibnamefont
  {Partanen}}, \bibinfo {author} {\bibfnamefont {T.}~\bibnamefont
  {H{\ifmmode\ddot{a}\else\"{a}\fi}yrynen}}, \bibinfo {author} {\bibfnamefont
  {J.}~\bibnamefont {Oksanen}},\ and\ \bibinfo {author} {\bibfnamefont
  {J.}~\bibnamefont {Tulkki}},\ }\bibfield  {title} {\bibinfo {title} {{Photon
  mass drag and the momentum of light in a medium}},\ }\href
  {https://doi.org/10.1103/PhysRevA.95.063850} {\bibfield  {journal} {\bibinfo
  {journal} {Phys. Rev. A}\ }\textbf {\bibinfo {volume} {95}},\ \bibinfo
  {pages} {063850} (\bibinfo {year} {2017})}\BibitemShut {NoStop}%
\bibitem [{\citenamefont {Philbin}(2011)}]{Philbin2011PRA}%
  \BibitemOpen
  \bibfield  {author} {\bibinfo {author} {\bibfnamefont {T.~G.}\ \bibnamefont
  {Philbin}},\ }\bibfield  {title} {\bibinfo {title} {{Electromagnetic energy
  momentum in dispersive media}},\ }\href
  {https://doi.org/10.1103/PhysRevA.83.013823} {\bibfield  {journal} {\bibinfo
  {journal} {Phys. Rev. A}\ }\textbf {\bibinfo {volume} {83}},\ \bibinfo
  {pages} {013823} (\bibinfo {year} {2011})},\ \bibinfo {note} {{Erratum: Phys.
  Rev. A {\bf 85}, 059902 (2012)}}\BibitemShut {NoStop}%
\bibitem [{\citenamefont {Feynman}\ \emph {et~al.}(2011)\citenamefont
  {Feynman}, \citenamefont {Leighton},\ and\ \citenamefont
  {Sands}}]{Feynman_I}%
  \BibitemOpen
  \bibfield  {author} {\bibinfo {author} {\bibfnamefont {R.~P.}\ \bibnamefont
  {Feynman}}, \bibinfo {author} {\bibfnamefont {R.}~\bibnamefont {Leighton}},\
  and\ \bibinfo {author} {\bibfnamefont {M.}~\bibnamefont {Sands}},\
  }\href@noop {} {\emph {\bibinfo {title} {{The Feynman Lectures on Physics,
  Vol. I}}}}\ (\bibinfo  {publisher} {Basic Books},\ \bibinfo {address} {New
  York},\ \bibinfo {year} {2011})\BibitemShut {NoStop}%
\bibitem [{\citenamefont {Huang}\ \emph {et~al.}(2011)\citenamefont {Huang},
  \citenamefont {Law},\ and\ \citenamefont {Huang}}]{Huang2011OE}%
  \BibitemOpen
  \bibfield  {author} {\bibinfo {author} {\bibfnamefont {G.}~\bibnamefont
  {Huang}}, \bibinfo {author} {\bibfnamefont {A.~W.-K.}\ \bibnamefont {Law}},\
  and\ \bibinfo {author} {\bibfnamefont {Z.}~\bibnamefont {Huang}},\ }\bibfield
   {title} {\bibinfo {title} {{Wave-induced drift of small floating objects in
  regular waves}},\ }\href {https://doi.org/10.1016/j.oceaneng.2010.12.015}
  {\bibfield  {journal} {\bibinfo  {journal} {Ocean Eng.}\ }\textbf {\bibinfo
  {volume} {38}},\ \bibinfo {pages} {712} (\bibinfo {year} {2011})}\BibitemShut
  {NoStop}%
\bibitem [{\citenamefont {Huang}\ \emph {et~al.}(2016)\citenamefont {Huang},
  \citenamefont {Huang},\ and\ \citenamefont {Law}}]{Huang2016JEM}%
  \BibitemOpen
  \bibfield  {author} {\bibinfo {author} {\bibfnamefont {G.}~\bibnamefont
  {Huang}}, \bibinfo {author} {\bibfnamefont {Z.~H.}\ \bibnamefont {Huang}},\
  and\ \bibinfo {author} {\bibfnamefont {A.~W.~K.}\ \bibnamefont {Law}},\
  }\bibfield  {title} {\bibinfo {title} {{Analytical Study on Drift of Small
  Floating Objects under Regular Waves}},\ }\href
  {https://doi.org/10.1061/(ASCE)EM.1943-7889.0001067} {\bibfield  {journal}
  {\bibinfo  {journal} {J. Eng. Mech.}\ }\textbf {\bibinfo {volume} {142}},\
  \bibinfo {pages} {06016002} (\bibinfo {year} {2016})}\BibitemShut {NoStop}%
\bibitem [{\citenamefont {Alsina}\ \emph {et~al.}(2020)\citenamefont {Alsina},
  \citenamefont {Jongedijk},\ and\ \citenamefont {van
  Sebille}}]{Alsina2020JGRO}%
  \BibitemOpen
  \bibfield  {author} {\bibinfo {author} {\bibfnamefont {J.~M.}\ \bibnamefont
  {Alsina}}, \bibinfo {author} {\bibfnamefont {C.~E.}\ \bibnamefont
  {Jongedijk}},\ and\ \bibinfo {author} {\bibfnamefont {E.}~\bibnamefont {van
  Sebille}},\ }\bibfield  {title} {\bibinfo {title} {{Laboratory Measurements
  of the Wave-Induced Motion of Plastic Particles: Influence of Wave Period,
  Plastic Size and Plastic Density}},\ }\href
  {https://doi.org/10.1029/2020JC016294} {\bibfield  {journal} {\bibinfo
  {journal} {J. Geophys. Res. Oceans}\ }\textbf {\bibinfo {volume} {125}},\
  \bibinfo {pages} {e2020JC016294} (\bibinfo {year} {2020})}\BibitemShut
  {NoStop}%
\bibitem [{\citenamefont {Calvert}\ \emph {et~al.}(2021)\citenamefont
  {Calvert}, \citenamefont {McAllister}, \citenamefont {Whittaker},
  \citenamefont {Raby}, \citenamefont {Borthwick},\ and\ \citenamefont {van~den
  Bremer}}]{Calvert2021JFM}%
  \BibitemOpen
  \bibfield  {author} {\bibinfo {author} {\bibfnamefont {R.}~\bibnamefont
  {Calvert}}, \bibinfo {author} {\bibfnamefont {M.~L.}\ \bibnamefont
  {McAllister}}, \bibinfo {author} {\bibfnamefont {C.}~\bibnamefont
  {Whittaker}}, \bibinfo {author} {\bibfnamefont {A.}~\bibnamefont {Raby}},
  \bibinfo {author} {\bibfnamefont {A.~G.~L.}\ \bibnamefont {Borthwick}},\ and\
  \bibinfo {author} {\bibfnamefont {T.~S.}\ \bibnamefont {van~den Bremer}},\
  }\bibfield  {title} {\bibinfo {title} {{A mechanism for the increased
  wave-induced drift of floating marine litter}},\ }\href
  {https://doi.org/10.1017/jfm.2021.72} {\bibfield  {journal} {\bibinfo
  {journal} {J. Fluid Mech.}\ }\textbf {\bibinfo {volume} {915}},\ \bibinfo
  {pages} {A73} (\bibinfo {year} {2021})}\BibitemShut {NoStop}%
\bibitem [{\citenamefont {Leader}\ and\ \citenamefont
  {Lorc{\ifmmode\acute{e}\else\'{e}\fi}}(2014)}]{Leader2014PR}%
  \BibitemOpen
  \bibfield  {author} {\bibinfo {author} {\bibfnamefont {E.}~\bibnamefont
  {Leader}}\ and\ \bibinfo {author} {\bibfnamefont {C.}~\bibnamefont
  {Lorc{\ifmmode\acute{e}\else\'{e}\fi}}},\ }\bibfield  {title} {\bibinfo
  {title} {{The angular momentum controversy: What{'}s it all about and does it
  matter?}},\ }\href {https://doi.org/10.1016/j.physrep.2014.02.010} {\bibfield
   {journal} {\bibinfo  {journal} {Phys. Rep.}\ }\textbf {\bibinfo {volume}
  {541}},\ \bibinfo {pages} {163} (\bibinfo {year} {2014})}\BibitemShut
  {NoStop}%
\bibitem [{\citenamefont {Bliokh}\ \emph
  {et~al.}(2017{\natexlab{b}})\citenamefont {Bliokh}, \citenamefont {Ivanov},
  \citenamefont {Guzzinati}, \citenamefont {Clark}, \citenamefont {Van~Boxem},
  \citenamefont
  {B{\ifmmode\acute{e}\else\'{e}\fi}ch{\ifmmode\acute{e}\else\'{e}\fi}},
  \citenamefont {Juchtmans}, \citenamefont {Alonso}, \citenamefont
  {Schattschneider}, \citenamefont {Nori},\ and\ \citenamefont
  {Verbeeck}}]{Bliokh2017PR}%
  \BibitemOpen
  \bibfield  {author} {\bibinfo {author} {\bibfnamefont {K.~Y.}\ \bibnamefont
  {Bliokh}}, \bibinfo {author} {\bibfnamefont {I.~P.}\ \bibnamefont {Ivanov}},
  \bibinfo {author} {\bibfnamefont {G.}~\bibnamefont {Guzzinati}}, \bibinfo
  {author} {\bibfnamefont {L.}~\bibnamefont {Clark}}, \bibinfo {author}
  {\bibfnamefont {R.}~\bibnamefont {Van~Boxem}}, \bibinfo {author}
  {\bibfnamefont {A.}~\bibnamefont
  {B{\ifmmode\acute{e}\else\'{e}\fi}ch{\ifmmode\acute{e}\else\'{e}\fi}}},
  \bibinfo {author} {\bibfnamefont {R.}~\bibnamefont {Juchtmans}}, \bibinfo
  {author} {\bibfnamefont {M.~A.}\ \bibnamefont {Alonso}}, \bibinfo {author}
  {\bibfnamefont {P.}~\bibnamefont {Schattschneider}}, \bibinfo {author}
  {\bibfnamefont {F.}~\bibnamefont {Nori}},\ and\ \bibinfo {author}
  {\bibfnamefont {J.}~\bibnamefont {Verbeeck}},\ }\bibfield  {title} {\bibinfo
  {title} {{Theory and applications of free-electron vortex states}},\ }\href
  {https://doi.org/10.1016/j.physrep.2017.05.006} {\bibfield  {journal}
  {\bibinfo  {journal} {Phys. Rep.}\ }\textbf {\bibinfo {volume} {690}},\
  \bibinfo {pages} {1} (\bibinfo {year} {2017}{\natexlab{b}})}\BibitemShut
  {NoStop}%
\bibitem [{\citenamefont {Larocque}\ \emph {et~al.}(2018)\citenamefont
  {Larocque}, \citenamefont {Kaminer}, \citenamefont {Grillo}, \citenamefont
  {Leuchs}, \citenamefont {Padgett}, \citenamefont {Boyd}, \citenamefont
  {Segev},\ and\ \citenamefont {Karimi}}]{Larocque2018CP}%
  \BibitemOpen
  \bibfield  {author} {\bibinfo {author} {\bibfnamefont {H.}~\bibnamefont
  {Larocque}}, \bibinfo {author} {\bibfnamefont {I.}~\bibnamefont {Kaminer}},
  \bibinfo {author} {\bibfnamefont {V.}~\bibnamefont {Grillo}}, \bibinfo
  {author} {\bibfnamefont {G.}~\bibnamefont {Leuchs}}, \bibinfo {author}
  {\bibfnamefont {M.~J.}\ \bibnamefont {Padgett}}, \bibinfo {author}
  {\bibfnamefont {R.~W.}\ \bibnamefont {Boyd}}, \bibinfo {author}
  {\bibfnamefont {M.}~\bibnamefont {Segev}},\ and\ \bibinfo {author}
  {\bibfnamefont {E.}~\bibnamefont {Karimi}},\ }\bibfield  {title} {\bibinfo
  {title} {{{`}Twisted{'} electrons}},\ }\href
  {https://www.tandfonline.com/doi/full/10.1080/00107514.2017.1418046}
  {\bibfield  {journal} {\bibinfo  {journal} {Contemp. Phys.}\ }\textbf
  {\bibinfo {volume} {59}},\ \bibinfo {pages} {126} (\bibinfo {year}
  {2018})}\BibitemShut {NoStop}%
\bibitem [{\citenamefont {Zhang}\ and\ \citenamefont
  {Niu}(2014)}]{Zhang2014PRL}%
  \BibitemOpen
  \bibfield  {author} {\bibinfo {author} {\bibfnamefont {L.}~\bibnamefont
  {Zhang}}\ and\ \bibinfo {author} {\bibfnamefont {Q.}~\bibnamefont {Niu}},\
  }\bibfield  {title} {\bibinfo {title} {{Angular Momentum of Phonons and the
  Einstein--de Haas Effect}},\ }\href
  {https://doi.org/10.1103/PhysRevLett.112.085503} {\bibfield  {journal}
  {\bibinfo  {journal} {Phys. Rev. Lett.}\ }\textbf {\bibinfo {volume} {112}},\
  \bibinfo {pages} {085503} (\bibinfo {year} {2014})}\BibitemShut {NoStop}%
\bibitem [{\citenamefont {Zhu}\ \emph {et~al.}(2018)\citenamefont {Zhu},
  \citenamefont {Yi}, \citenamefont {Li}, \citenamefont {Xiao}, \citenamefont
  {Zhang}, \citenamefont {Yang}, \citenamefont {Kaindl}, \citenamefont {Li},
  \citenamefont {Wang},\ and\ \citenamefont {Zhang}}]{Zhu2018S}%
  \BibitemOpen
  \bibfield  {author} {\bibinfo {author} {\bibfnamefont {H.}~\bibnamefont
  {Zhu}}, \bibinfo {author} {\bibfnamefont {J.}~\bibnamefont {Yi}}, \bibinfo
  {author} {\bibfnamefont {M.-Y.}\ \bibnamefont {Li}}, \bibinfo {author}
  {\bibfnamefont {J.}~\bibnamefont {Xiao}}, \bibinfo {author} {\bibfnamefont
  {L.}~\bibnamefont {Zhang}}, \bibinfo {author} {\bibfnamefont {C.-W.}\
  \bibnamefont {Yang}}, \bibinfo {author} {\bibfnamefont {R.~A.}\ \bibnamefont
  {Kaindl}}, \bibinfo {author} {\bibfnamefont {L.-J.}\ \bibnamefont {Li}},
  \bibinfo {author} {\bibfnamefont {Y.}~\bibnamefont {Wang}},\ and\ \bibinfo
  {author} {\bibfnamefont {X.}~\bibnamefont {Zhang}},\ }\bibfield  {title}
  {\bibinfo {title} {{Observation of chiral phonons}},\ }\href
  {https://doi.org/10.1126/science.aar2711} {\bibfield  {journal} {\bibinfo
  {journal} {Science}\ }\textbf {\bibinfo {volume} {359}},\ \bibinfo {pages}
  {579} (\bibinfo {year} {2018})}\BibitemShut {NoStop}%
\bibitem [{\citenamefont {Ishito}\ \emph {et~al.}(2023)\citenamefont {Ishito},
  \citenamefont {Mao}, \citenamefont {Kousaka}, \citenamefont {Togawa},
  \citenamefont {Iwasaki}, \citenamefont {Zhang}, \citenamefont {Murakami},
  \citenamefont {Kishine},\ and\ \citenamefont {Satoh}}]{Ishito2023NP}%
  \BibitemOpen
  \bibfield  {author} {\bibinfo {author} {\bibfnamefont {K.}~\bibnamefont
  {Ishito}}, \bibinfo {author} {\bibfnamefont {H.}~\bibnamefont {Mao}},
  \bibinfo {author} {\bibfnamefont {Y.}~\bibnamefont {Kousaka}}, \bibinfo
  {author} {\bibfnamefont {Y.}~\bibnamefont {Togawa}}, \bibinfo {author}
  {\bibfnamefont {S.}~\bibnamefont {Iwasaki}}, \bibinfo {author} {\bibfnamefont
  {T.}~\bibnamefont {Zhang}}, \bibinfo {author} {\bibfnamefont
  {S.}~\bibnamefont {Murakami}}, \bibinfo {author} {\bibfnamefont {J.-I.}\
  \bibnamefont {Kishine}},\ and\ \bibinfo {author} {\bibfnamefont
  {T.}~\bibnamefont {Satoh}},\ }\bibfield  {title} {\bibinfo {title} {{Truly
  chiral phonons in {$\alpha$}-HgS}},\ }\href
  {https://doi.org/10.1038/s41567-022-01790-x} {\bibfield  {journal} {\bibinfo
  {journal} {Nat. Phys.}\ }\textbf {\bibinfo {volume} {19}},\ \bibinfo {pages}
  {35} (\bibinfo {year} {2023})}\BibitemShut {NoStop}%
\bibitem [{\citenamefont {Tauchert}\ \emph {et~al.}(2022)\citenamefont
  {Tauchert}, \citenamefont {Volkov}, \citenamefont {Ehberger}, \citenamefont
  {Kazenwadel}, \citenamefont {Evers}, \citenamefont {Lange}, \citenamefont
  {Donges}, \citenamefont {Book}, \citenamefont {Kreuzpaintner}, \citenamefont
  {Nowak},\ and\ \citenamefont {Baum}}]{Tauchert2022N}%
  \BibitemOpen
  \bibfield  {author} {\bibinfo {author} {\bibfnamefont {S.~R.}\ \bibnamefont
  {Tauchert}}, \bibinfo {author} {\bibfnamefont {M.}~\bibnamefont {Volkov}},
  \bibinfo {author} {\bibfnamefont {D.}~\bibnamefont {Ehberger}}, \bibinfo
  {author} {\bibfnamefont {D.}~\bibnamefont {Kazenwadel}}, \bibinfo {author}
  {\bibfnamefont {M.}~\bibnamefont {Evers}}, \bibinfo {author} {\bibfnamefont
  {H.}~\bibnamefont {Lange}}, \bibinfo {author} {\bibfnamefont
  {A.}~\bibnamefont {Donges}}, \bibinfo {author} {\bibfnamefont
  {A.}~\bibnamefont {Book}}, \bibinfo {author} {\bibfnamefont {W.}~\bibnamefont
  {Kreuzpaintner}}, \bibinfo {author} {\bibfnamefont {U.}~\bibnamefont
  {Nowak}},\ and\ \bibinfo {author} {\bibfnamefont {P.}~\bibnamefont {Baum}},\
  }\bibfield  {title} {\bibinfo {title} {{Polarized phonons carry angular
  momentum in ultrafast demagnetization}},\ }\href
  {https://doi.org/10.1038/s41586-021-04306-4} {\bibfield  {journal} {\bibinfo
  {journal} {Nature}\ }\textbf {\bibinfo {volume} {602}},\ \bibinfo {pages}
  {73} (\bibinfo {year} {2022})}\BibitemShut {NoStop}%
\bibitem [{\citenamefont {Rozenman}\ \emph {et~al.}(2019)\citenamefont
  {Rozenman}, \citenamefont {Fu}, \citenamefont {Arie},\ and\ \citenamefont
  {Shemer}}]{Rozenman2019F}%
  \BibitemOpen
  \bibfield  {author} {\bibinfo {author} {\bibfnamefont {G.~G.}\ \bibnamefont
  {Rozenman}}, \bibinfo {author} {\bibfnamefont {S.}~\bibnamefont {Fu}},
  \bibinfo {author} {\bibfnamefont {A.}~\bibnamefont {Arie}},\ and\ \bibinfo
  {author} {\bibfnamefont {L.}~\bibnamefont {Shemer}},\ }\bibfield  {title}
  {\bibinfo {title} {{Quantum Mechanical and Optical Analogies in Surface
  Gravity Water Waves}},\ }\href {https://doi.org/10.3390/fluids4020096}
  {\bibfield  {journal} {\bibinfo  {journal} {Fluids}\ }\textbf {\bibinfo
  {volume} {4}},\ \bibinfo {pages} {96} (\bibinfo {year} {2019})}\BibitemShut
  {NoStop}%
\bibitem [{\citenamefont {Zhu}\ \emph {et~al.}(2024)\citenamefont {Zhu},
  \citenamefont {Zhao}, \citenamefont {Han}, \citenamefont {Zi}, \citenamefont
  {Hu},\ and\ \citenamefont {Chen}}]{Zhu2024NRP}%
  \BibitemOpen
  \bibfield  {author} {\bibinfo {author} {\bibfnamefont {S.}~\bibnamefont
  {Zhu}}, \bibinfo {author} {\bibfnamefont {X.}~\bibnamefont {Zhao}}, \bibinfo
  {author} {\bibfnamefont {L.}~\bibnamefont {Han}}, \bibinfo {author}
  {\bibfnamefont {J.}~\bibnamefont {Zi}}, \bibinfo {author} {\bibfnamefont
  {X.}~\bibnamefont {Hu}},\ and\ \bibinfo {author} {\bibfnamefont
  {H.}~\bibnamefont {Chen}},\ }\bibfield  {title} {\bibinfo {title}
  {{Controlling water waves with artificial structures}},\ }\href
  {https://doi.org/10.1038/s42254-024-00701-8} {\bibfield  {journal} {\bibinfo
  {journal} {Nat. Rev. Phys.}\ }\textbf {\bibinfo {volume} {6}},\ \bibinfo
  {pages} {231} (\bibinfo {year} {2024})}\BibitemShut {NoStop}%
\bibitem [{\citenamefont {Valet}\ \emph
  {et~al.}(2025{\natexlab{a}})\citenamefont {Valet}, \citenamefont {Yamamoto},
  \citenamefont {Pigeau}, \citenamefont {de~Loubens},\ and\ \citenamefont
  {Klein}}]{Valet2025_I}%
  \BibitemOpen
  \bibfield  {author} {\bibinfo {author} {\bibfnamefont {T.}~\bibnamefont
  {Valet}}, \bibinfo {author} {\bibfnamefont {K.}~\bibnamefont {Yamamoto}},
  \bibinfo {author} {\bibfnamefont {B.}~\bibnamefont {Pigeau}}, \bibinfo
  {author} {\bibfnamefont {G.}~\bibnamefont {de~Loubens}},\ and\ \bibinfo
  {author} {\bibfnamefont {O.}~\bibnamefont {Klein}},\ }\bibfield  {title}
  {\bibinfo {title} {{The Orbital Angular Momentum of Azimuthal Spin-Waves}},\
  }\bibfield  {journal} {\bibinfo  {journal} {arXiv}\ }\href
  {https://doi.org/10.48550/arXiv.2503.06556} {10.48550/arXiv.2503.06556}
  (\bibinfo {year} {2025}{\natexlab{a}}),\ \Eprint
  {https://arxiv.org/abs/2503.06556} {2503.06556} \BibitemShut {NoStop}%
\bibitem [{\citenamefont {Valet}\ \emph
  {et~al.}(2025{\natexlab{b}})\citenamefont {Valet}, \citenamefont {Yamamoto},
  \citenamefont {Pigeau}, \citenamefont {de~Loubens},\ and\ \citenamefont
  {Klein}}]{Valet2025_II}%
  \BibitemOpen
  \bibfield  {author} {\bibinfo {author} {\bibfnamefont {T.}~\bibnamefont
  {Valet}}, \bibinfo {author} {\bibfnamefont {K.}~\bibnamefont {Yamamoto}},
  \bibinfo {author} {\bibfnamefont {B.}~\bibnamefont {Pigeau}}, \bibinfo
  {author} {\bibfnamefont {G.}~\bibnamefont {de~Loubens}},\ and\ \bibinfo
  {author} {\bibfnamefont {O.}~\bibnamefont {Klein}},\ }\bibfield  {title}
  {\bibinfo {title} {{Field Theory of Linear Spin-Waves in Finite Textured
  Ferromagnets}},\ }\bibfield  {journal} {\bibinfo  {journal} {arXiv}\ }\href
  {https://doi.org/10.48550/arXiv.2503.06557} {10.48550/arXiv.2503.06557}
  (\bibinfo {year} {2025}{\natexlab{b}}),\ \Eprint
  {https://arxiv.org/abs/2503.06557} {2503.06557} \BibitemShut {NoStop}%
\bibitem [{\citenamefont {Shen}\ \emph {et~al.}(2023)\citenamefont {Shen},
  \citenamefont {Zhan}, \citenamefont {Wright}, \citenamefont
  {Christodoulides}, \citenamefont {Wise}, \citenamefont {Willner},
  \citenamefont {Zou}, \citenamefont {Zhao}, \citenamefont {Porras},
  \citenamefont {Chong}, \citenamefont {Wan}, \citenamefont {Bliokh},
  \citenamefont {Liao}, \citenamefont {Hern{\'a}ndez-Garc{\'\i}a},
  \citenamefont {Murnane}, \citenamefont {Yessenov}, \citenamefont {Abouraddy},
  \citenamefont {Wong}, \citenamefont {Go}, \citenamefont {Kumar},
  \citenamefont {Guo}, \citenamefont {Fan}, \citenamefont {Papasimakis},
  \citenamefont {Zheludev}, \citenamefont {Chen}, \citenamefont {Zhu},
  \citenamefont {Agrawal}, \citenamefont {Mounaix}, \citenamefont {Fontaine},
  \citenamefont {Carpenter}, \citenamefont {Jolly}, \citenamefont {Dorrer},
  \citenamefont {Alonso}, \citenamefont {Lopez-Quintas}, \citenamefont
  {L{\'o}pez-Ripa}, \citenamefont {Sola}, \citenamefont {Huang}, \citenamefont
  {Zhang}, \citenamefont {Ruan}, \citenamefont {Dorrah}, \citenamefont
  {Capasso},\ and\ \citenamefont {Forbes}}]{Shen2023JO}%
  \BibitemOpen
  \bibfield  {author} {\bibinfo {author} {\bibfnamefont {Y.}~\bibnamefont
  {Shen}}, \bibinfo {author} {\bibfnamefont {Q.}~\bibnamefont {Zhan}}, \bibinfo
  {author} {\bibfnamefont {L.~G.}\ \bibnamefont {Wright}}, \bibinfo {author}
  {\bibfnamefont {D.~N.}\ \bibnamefont {Christodoulides}}, \bibinfo {author}
  {\bibfnamefont {F.~W.}\ \bibnamefont {Wise}}, \bibinfo {author}
  {\bibfnamefont {A.~E.}\ \bibnamefont {Willner}}, \bibinfo {author}
  {\bibfnamefont {K.-H.}\ \bibnamefont {Zou}}, \bibinfo {author} {\bibfnamefont
  {Z.}~\bibnamefont {Zhao}}, \bibinfo {author} {\bibfnamefont {M.~A.}\
  \bibnamefont {Porras}}, \bibinfo {author} {\bibfnamefont {A.}~\bibnamefont
  {Chong}}, \bibinfo {author} {\bibfnamefont {C.}~\bibnamefont {Wan}}, \bibinfo
  {author} {\bibfnamefont {K.~Y.}\ \bibnamefont {Bliokh}}, \bibinfo {author}
  {\bibfnamefont {C.-T.}\ \bibnamefont {Liao}}, \bibinfo {author}
  {\bibfnamefont {C.}~\bibnamefont {Hern{\'a}ndez-Garc{\'\i}a}}, \bibinfo
  {author} {\bibfnamefont {M.}~\bibnamefont {Murnane}}, \bibinfo {author}
  {\bibfnamefont {M.}~\bibnamefont {Yessenov}}, \bibinfo {author}
  {\bibfnamefont {A.~F.}\ \bibnamefont {Abouraddy}}, \bibinfo {author}
  {\bibfnamefont {L.~J.}\ \bibnamefont {Wong}}, \bibinfo {author}
  {\bibfnamefont {M.}~\bibnamefont {Go}}, \bibinfo {author} {\bibfnamefont
  {S.}~\bibnamefont {Kumar}}, \bibinfo {author} {\bibfnamefont
  {C.}~\bibnamefont {Guo}}, \bibinfo {author} {\bibfnamefont {S.}~\bibnamefont
  {Fan}}, \bibinfo {author} {\bibfnamefont {N.}~\bibnamefont {Papasimakis}},
  \bibinfo {author} {\bibfnamefont {N.~I.}\ \bibnamefont {Zheludev}}, \bibinfo
  {author} {\bibfnamefont {L.}~\bibnamefont {Chen}}, \bibinfo {author}
  {\bibfnamefont {W.}~\bibnamefont {Zhu}}, \bibinfo {author} {\bibfnamefont
  {A.}~\bibnamefont {Agrawal}}, \bibinfo {author} {\bibfnamefont
  {M.}~\bibnamefont {Mounaix}}, \bibinfo {author} {\bibfnamefont {N.~K.}\
  \bibnamefont {Fontaine}}, \bibinfo {author} {\bibfnamefont {J.}~\bibnamefont
  {Carpenter}}, \bibinfo {author} {\bibfnamefont {S.~W.}\ \bibnamefont
  {Jolly}}, \bibinfo {author} {\bibfnamefont {C.}~\bibnamefont {Dorrer}},
  \bibinfo {author} {\bibfnamefont {B.}~\bibnamefont {Alonso}}, \bibinfo
  {author} {\bibfnamefont {I.}~\bibnamefont {Lopez-Quintas}}, \bibinfo {author}
  {\bibfnamefont {M.}~\bibnamefont {L{\'o}pez-Ripa}}, \bibinfo {author}
  {\bibfnamefont {{\'I}.~J.}\ \bibnamefont {Sola}}, \bibinfo {author}
  {\bibfnamefont {J.}~\bibnamefont {Huang}}, \bibinfo {author} {\bibfnamefont
  {H.}~\bibnamefont {Zhang}}, \bibinfo {author} {\bibfnamefont
  {Z.}~\bibnamefont {Ruan}}, \bibinfo {author} {\bibfnamefont {A.~H.}\
  \bibnamefont {Dorrah}}, \bibinfo {author} {\bibfnamefont {F.}~\bibnamefont
  {Capasso}},\ and\ \bibinfo {author} {\bibfnamefont {A.}~\bibnamefont
  {Forbes}},\ }\bibfield  {title} {\bibinfo {title} {Roadmap on spatiotemporal
  light fields},\ }\href {https://doi.org/10.1088/2040-8986/ace4dc} {\bibfield
  {journal} {\bibinfo  {journal} {J. Opt.}\ }\textbf {\bibinfo {volume} {25}},\
  \bibinfo {pages} {093001} (\bibinfo {year} {2023})}\BibitemShut {NoStop}%
\end{thebibliography}%

\end{document}